\documentclass[conference]{IEEEtran}
\usepackage{hyperref}
\newif\iflong
\longtrue

\usepackage[T1]{fontenc}
\usepackage{url}

\usepackage{changepage}

\usepackage{flushend} 

\usepackage{graphicx}
\usepackage{grffile} 
\newcommand{\includegraphicsifexists}[2][,]{\IfFileExists{#2}{\includegraphics[#1]{#2}}{}}

\usepackage{tikz}
\usetikzlibrary{shapes,arrows,positioning,calc}

\usepackage{boxedminipage}
{\smallskip
\noindent
\let\emph=\textbf
\begin{boxedminipage}{\columnwidth}\begin{center}\em}%
{\end{center}\end{boxedminipage}%
\smallskip
}

\usepackage{paralist}

\usepackage{rotating}

\usepackage[scaled=0.81]{beramono} 

\usepackage[english]{babel}
\usepackage{calc}
\usepackage{cite}
\usepackage[fleqn,cmex10]{amsmath}
\usepackage{array}
\usepackage[font={scriptsize,normal},skip=5pt]{caption}
\usepackage[tight,scriptsize]{subfig}
\usepackage{fixltx2e}
\usepackage{textcomp}
\usepackage{xcolor}
\definecolor{darkblue}{rgb}{0,0,0.4}
\definecolor{lightblue}{RGB}{220,230,240}
\definecolor{lightgreen}{rgb}{0.47,0.66,0.53}
\definecolor{lightgray}{rgb}{0.8,0.8,0.8}

\definecolor{colorstrong}{HTML}{33A02C}
\definecolor{colormedium}{HTML}{A6611A}
\definecolor{colorweak}{HTML}{1F78B4}

\definecolor{colorstrong-main}{HTML}{000000}
\definecolor{colormedium-main}{HTML}{000000}
\definecolor{colorweak-main}{HTML}{000000}

\usepackage{colortbl}
\usepackage{enumerate}

\newcommand{\fakepar}[1]{\textbf{#1.}}
\newcommand{\rqpar}[1]{\noindent\textbf{#1}}

\newenvironment{boxcenter}[1][\topsep]
  {\setlength{\topsep}{#1}\par\kern\topsep\centering}
  {\par\kern\topsep}

\usepackage{fancybox}
\newcommand{\takehome}[1]{\begin{boxcenter}[1pt]
\ovalbox{
    \begin{minipage}{0.87\columnwidth}
      {\it #1}
    \end{minipage}
}\end{boxcenter}
}

\makeatletter
\renewcommand*{\thetable}{\arabic{table}}
\renewcommand*{\thefigure}{\arabic{figure}}
\let\c@table\c@figure
\makeatother

\usepackage{xspace}
\newcommand{\rc}{Rosetta\ Code\xspace}
\newcommand{\sgn}{\operatorname{sgn}}

\newcommand{\vanish}[1]{{\tiny\raisebox{0.8pt}{\textless}\,}\ifcase#1\relax$\varepsilon$\else 10\raisebox{1pt}{\scalebox{0.8}{-#1}}\fi}
\newcommand{\pc}[1]{#1\%}

\usepackage{mathptmx}

\usepackage{array}
\newcolumntype{L}[1]{>{\raggedright\let\\\tabularnewline}m{#1}}
\newcolumntype{C}[1]{>{\centering\let\\\tabularnewline}m{#1}}
\newcolumntype{R}[1]{>{\raggedleft\let\\\tabularnewline}m{#1}}

\begin{document}

\title{A Comparative Study of Programming Languages in~Rosetta Code}

\author{%
  \IEEEauthorblockN{Sebastian Nanz $\quad\cdot\quad$ Carlo A.\ Furia}
  \IEEEauthorblockA{Chair of Software Engineering, Department of Computer Science,  ETH Zurich, Switzerland\\
  \url{firstname.lastname@inf.ethz.ch}
  }
}

\maketitle

\begin{abstract}
Sometimes debates on programming languages are more religious than scientific.
Questions about which language is more succinct or efficient, or makes developers more productive are discussed with fervor, and their answers are too often based on anecdotes and unsubstantiated beliefs.
In this study, we use the largely untapped research potential of Rosetta Code, a code repository of solutions to common programming tasks in various languages, which offers a large data set for analysis. Our study is based on 7'087 solution programs corresponding to 745 tasks in 8 widely used languages representing the major programming paradigms (procedural: C and Go; object-oriented: C\# and Java; functional: F\# and Haskell; scripting: Python and Ruby).
Our statistical analysis reveals, most notably, that: functional and scripting languages are more concise than procedural and object-oriented languages; C is hard to beat when it comes to raw speed on large inputs, but performance differences over inputs of moderate size are less pronounced and allow even interpreted languages to be competitive; compiled strongly-typed languages, where more defects can be caught at compile time, are less prone to runtime failures than interpreted or weakly-typed languages.
We discuss implications of these results for developers, language designers, and educators.
\end{abstract}


\section{Introduction}
\label{sec:introduction}

\emph{What is the best programming language for\ldots$\!$?}
Questions about programming languages and the properties of their programs are asked often but well-founded answers are not easily available.
From an engineering viewpoint, the design of a programming language is the result of multiple trade-offs that achieve certain desirable properties (such as speed) at the expense of others (such as simplicity). 
%
Technical aspects are, however, hardly ever the only relevant concerns when it comes to choosing a programming language. 
Factors as heterogeneous as a strong supporting community, similarity to other widespread languages, or availability of libraries are often instrumental in deciding a language's popularity and how it is used in the wild~\cite{Meyerovich:2013:EAP:2509136.2509515}.
If we want to reliably answer questions about properties of programming languages, we have to analyze, empirically, the artifacts programmers write in those languages.
Answers grounded in empirical evidence can be valuable in helping language users and designers make informed choices. 

To control for the many factors that may affect the properties of programs, some empirical studies of programming languages~\cite{Prechelt:2000:ECS:619056.621567,DBLP:conf/oopsla/Hanenberg10,SS13-quorum,nanz-west-silveira:2013:benchmarking} have performed \emph{controlled experiments} where human subjects (typically students) in highly controlled environments solve small programming tasks in different languages.
Such controlled experiments provide the most reliable data about the impact of certain programming language features such as syntax and typing, but they are also necessarily limited in scope and generalizability by the number and types of tasks solved, and by the use of novice programmers as subjects. Real-world programming also develops over far more time than that allotted for short exam-like programming assignments; and produces programs that change features and improve quality over multiple development iterations.

At the opposite end of the spectrum, empirical studies based on analyzing programs in \emph{public repositories} such as GitHub~\cite{Bhattacharya:2011:APL:1985793.1985817,okur:2012:libraries,GitHubstudy-fse2014} can count on large amounts of mature code improved by experienced developers over substantial time spans.
Such set-ups are suitable for studies of defect proneness and code evolution, but they also greatly complicate analyses that require directly comparable data across different languages: projects in code repositories target disparate categories of software, and even those in the same category (such as ``web browsers'') often differ broadly in features, design, and style, and hence cannot be considered to be implementing minor variants of the same task.

The study presented in this paper explores a middle ground between highly controlled but small programming assignments and large but incomparable software projects: programs in \rc.
The \rc repository~\cite{rosettacode} collects solutions, written in hundreds of different languages, to an open collection of over 700 programming \emph{tasks}. 
Most tasks are quite detailed descriptions of problems that go beyond simple programming assignments, from sorting algorithms to pattern matching and from numerical analysis to GUI programming. 
Solutions to the same task in different languages are thus significant samples of what each programming language can achieve and are directly comparable.
The community of contributors to \rc (nearly 25'000 users at the time of writing) includes programmers that scrutinize, revise, and improve each other's solutions.

Our study analyzes 7'087 solution programs to 745 tasks in 8 widely used languages representing the major programming paradigms (procedural: C and Go; object-oriented: C\# and Java; functional: F\# and Haskell; scripting: Python and Ruby).
The study's research questions target various program features including conciseness, size of executables, running time, memory usage, and failure proneness.
A quantitative statistical analysis, cross-checked for consistency against a careful inspection of plotted data, reveals the following main findings about the programming languages we analyzed:
\begin{itemize}
\item Functional and scripting languages enable writing more concise code than procedural and object-oriented languages.
\item Languages that compile into bytecode produce smaller executables than those that compile into native machine code.
\item C is hard to beat when it comes to raw speed on large inputs. 
  Go is the runner-up, also in terms of economical memory usage.
\item In contrast, performance differences between languages shrink over inputs of moderate size, where languages with a lightweight runtime may be competitive even if they are interpreted.
\item Compiled strongly-typed languages, where more defects can be caught at compile time, are less prone to runtime failures than interpreted or weakly-typed languages.
\end{itemize}

Beyond the findings specific to the programs and programming languages that we analyzed, our study paves the way for follow-up research that can benefit from the wealth of data in \rc and generalize our findings to other domains. To this end, \autoref{sec:implications} discusses some practical implications of these findings for developers, language designers, and educators, whose choices about programming languages can increasingly rely on a growing fact base built on complementary sources.

The bulk of the paper describes the design of our empirical study (\autoref{sec:methodology}), and its research questions and overall results (\autoref{sec:results}).
We refer to a detailed technical report~\cite{rosettacode-extended} for the complete fine-grain details of the measures, statistics, and plots.
To support repetition and replication studies, we also make the complete data available online\footnote{\url{https://bitbucket.org/nanzs/rosettacodedata}}, together with the scripts we wrote to produce and analyze it.

\section{Methodology}
\label{sec:methodology}

\subsection{The \rc repository} \label{sec:rc-description}

\rc~\cite{rosettacode} is a code repository with a wiki interface. This study is based on a repository's snapshot taken on 24 June 2014\footnote{Cloned into our Git repository\footnotemark[1] using a modified version of the Perl module RosettaCode-0.0.5 available from \url{http://cpan.org/}.}; henceforth ``\rc'' denotes this snapshot.

\rc is organized in 745 \emph{tasks}.
Each task is a natural language description of a computational problem or theme, such as the bubble sort algorithm or reading the JSON data format.
Contributors can provide \emph{solutions} to tasks in their favorite programming languages, or revise  already available solutions.
\rc features 379 languages (with at least one solution per language) for a total of 49'305 solutions and 3'513'262 lines (total lines of program files).
A solution consists of a piece of code, which ideally should accurately follow a task's description and be self-contained (including test inputs); that is, the code should compile and execute in a proper environment without modifications. 

\iflong
Tasks significantly differ in the detail, prescriptiveness, and generality of their descriptions.
The most detailed ones, such as ``Bubble sort'', consist of well-defined algorithms, described informally and in pseudo-code, and include tests (input/output pairs) to demonstrate solutions.
Other tasks are much vaguer and only give a general theme, which may be inapplicable to some languages or admit widely different solutions.
For instance, task ``Memory allocation'' just asks to ``show how to explicitly allocate and deallocate blocks of memory''.
\fi

\subsection{Task selection} \label{sec:task-selection}
\iflong
Whereas even vague task descriptions may prompt well-written solutions, our
\fi
study requires comparable solutions to clearly-defined tasks.
To identify them, we categorized tasks, based on their description, according to whether they are suitable for lines-of-code analysis (\textsc{loc}), compilation (\textsc{comp}), and execution (\textsc{exec}); 
$T_C$ denotes the set of tasks in a category $C$.
Categories are increasingly restrictive: lines-of-code analysis only includes tasks sufficiently well-defined that their solutions can be considered minor variants of a unique problem; compilation further requires that tasks demand complete solutions rather than sketches or snippets; execution further requires that tasks include meaningful inputs and algorithmic components (typically, as opposed to data-structure and interface definitions).
As \autoref{tab:task-select} shows, many tasks are too vague to be used in the study, but the differences between the tasks in the three categories are limited.

\begin{table}[!h]
\begin{center}
\begin{tabular}{c r|rrr|rr}
 & \textsc{all} & \textsc{loc} & \textsc{comp} & \textsc{exec} & \textsc{perf} & \textsc{scal} \\ \hline
\textsc{\# tasks} &
 745 & 454 & 452 & 436 & 50 & 46 \\ \hline
\end{tabular}
\end{center}
\caption{Classification and selection of \rc tasks.}
\label{tab:task-select}
\end{table}

Most tasks do not describe sufficiently precise and varied inputs to be usable in an analysis of runtime performance.
For instance, some tasks are computationally trivial, and hence do not determine measurable resource usage when running\iflong; others do not give specific inputs to be tested, and hence solutions may run on incomparable inputs; others still are well-defined but their performance without interactive input is immaterial, such as in the case of graphic animation tasks\fi.
To identify tasks that can be meaningfully used in analyses of performance, we introduced two additional categories (\textsc{perf} and \textsc{scal}) of tasks suitable for performance comparisons: \textsc{perf} describes ``everyday'' workloads that are not necessarily very resource intensive, but whose descriptions include well-defined inputs that can be consistently used in every solution; in contrast, \textsc{scal} describes ``computing-intensive'' workloads with inputs that can easily be scaled up to substantial size and require well-engineered solutions. 
\iflong
For example, sorting algorithms are computing-intensive tasks working on large input lists; ``Cholesky matrix decomposition'' is an ``everyday'' task working on two test input matrices that can be decomposed quickly. 
\fi
 The corresponding sets $T_{\text{\textsc{perf}}}$ and $T_{\text{\textsc{scal}}}$ are disjoint subsets of the execution tasks $T_{\text{\textsc{exec}}}$; \autoref{tab:task-select} gives their size.

\subsection{Language selection}
\label{sec:language-selection}

\rc includes solutions in 379 languages.
Analyzing all of them is not worth the huge effort, given that many languages are not used in practice or cover only few tasks.
To find a representative and significant subset, we rank languages according to a combination of their rankings in \rc and in the TIOBE index~\cite{tiobe}.
A language's \rc ranking is based on the number of tasks for which at least one solution in that language exists: the larger the number of tasks the higher the ranking; \autoref{tab:rosettacode-rank} lists the top-20 languages (\textsc{lang}) in the \rc ranking (\textsc{rosetta}) with the number of tasks they implement (\textsc{\# tasks}).
The TIOBE programming community index~\cite{tiobe} is a long-standing, monthly-published language popularity ranking based on hits in various search engines; \autoref{tab:tiobe-rank} lists the top-20 languages in the TIOBE index with their TIOBE score (\textsc{tiobe}).

\newcommand{\rch}{\rowcolor{lightblue}}
\newcommand{\rank}[1]{\##1}
\begin{table}[htb]
  \centering
  \begin{tabular}{R{1cm}L{3cm}R{1cm}R{1cm}}
\textsc{rosetta} & \textsc{lang}   & \textsc{\# tasks} & \textsc{tiobe} \\ \hline
\rch  \rank{1} & Tcl         &     718 & \rank{43} \\
      \rank{2} & Racket      &     706 & --\footnotemark[3] \\
\rch  \rank{3} & {\bf Python}&     675 & \rank{8}\\
      \rank{4} & Perl 6      &     644 & -- \\
\rch  \rank{5} & {\bf Ruby  }&     635 & \rank{14} \\
      \rank{6} & J           &     630 & -- \\
\rch  \rank{7} & {\bf C     }&     630 & \rank{1} \\
\rch  \rank{8} & D           &     622 & \rank{50} \\
\rch  \rank{9} & Go          &     617 & \rank{30} \\
     \rank{10} & PicoLisp    &     605 & -- \\
\rch \rank{11} & {\bf Perl  }&     601 & \rank{11} \\
\rch \rank{12} & Ada         &     582 & \rank{29} \\
     \rank{13} & Mathematica &     580 & -- \\
     \rank{14} & REXX        &     566 & -- \\
\rch \rank{15} & Haskell     &     553 & \rank{38} \\
     \rank{16} & AutoHotkey  &     536 & -- \\
\rch \rank{17} & {\bf Java  }&     534 & \rank{2}\\
     \rank{18} & BBC BASIC   &     515 & -- \\
     \rank{19} & Icon        &     473 & -- \\
     \rank{20} & OCaml       &     471 & -- \\
    \end{tabular}
  \vspace{1ex}
  \caption{\rc ranking: top 20.}
  \label{tab:rosettacode-rank}
\end{table}

\renewcommand{\rch}{\rowcolor{lightblue}}
\begin{table}[htb]
  \centering
  \begin{tabular}{R{1cm}L{3cm}R{1cm}R{1cm}}
   \textsc{tiobe}  & \textsc{lang}      & \textsc{\# tasks} & \textsc{rosetta}   \\ \hline
\rch  \rank{1} & {\bf C                   } & 630 & \rank{7}\\
\rch  \rank{2} & {\bf Java                } & 534 & \rank{17} \\
      \rank{3} & {\bf Objective-C         } & 136 & \rank{72} \\
\rch  \rank{4} & {\bf C++                 } & 461 & \rank{22} \\
      \rank{5} & {\bf (Visual) Basic      } &  34 & \rank{145} \\
\rch  \rank{6} & {\bf C\#                 } & 463 & \rank{21} \\
\rch  \rank{7} & {\bf PHP                 } & 324 & \rank{36} \\
\rch  \rank{8} & {\bf Python              } & 675 & \rank{3}\\
\rch  \rank{9} & {\bf JavaScript          } & 371 & \rank{28} \\
     \rank{10} & {\bf Transact-SQL        } &   4 & \rank{266} \\
\rch \rank{11} & {\bf Perl                } & 601 & \rank{11} \\
     \rank{12} & {\bf Visual Basic .NET   } & 104 & \rank{81} \\
\rch \rank{13} & {\bf F\#                 } & 341 & \rank{33} \\
\rch \rank{14} & {\bf Ruby                } & 635 & \rank{5 }\\
     \rank{15} & {\bf ActionScript        } & 113 & \rank{77} \\
     \rank{16} & {\bf Swift               } &   --\footnotemark[4] &     -- \\
     \rank{17} & {\bf Delphi/Object Pascal} & 219 & \rank{53} \\
     \rank{18} & {\bf Lisp                } &   --\footnotemark[5] &     -- \\
\rch \rank{19} & {\bf MATLAB              } & 305 & \rank{40} \\
     \rank{20} & {\bf Assembly            } &   --\footnotemark[5] &     -- \\
  \end{tabular}
  \vspace{1ex}
  \caption{TIOBE index ranking: top 20.}
  \label{tab:tiobe-rank}
\end{table}

A language $\ell$ must satisfy two criteria to be included in our study:
\begin{enumerate}[C1.]
\item \label{crit:tiobe} $\ell$ ranks in the top-50 positions in the TIOBE index;
\item \label{crit:rc} $\ell$ implements at least one third ($\approx 250$) of the \rc tasks. 
\end{enumerate}
Criterion C\ref{crit:tiobe} selects widely-used, popular languages.
Criterion C\ref{crit:rc} selects languages that can be compared on a substantial number of tasks, conducing to statistically significant results. 
Languages in \autoref{tab:rosettacode-rank} that fulfill criterion C\ref{crit:tiobe} are shaded (the top-20 in TIOBE are in bold); and so are languages in \autoref{tab:tiobe-rank} that fulfill criterion C\ref{crit:rc}.
\iflong
A comparison of the two tables indicates that some popular languages are underrepresented in \rc, such as Objective-C, (Visual) Basic, and Transact-SQL; conversely, some languages popular in \rc have a low TIOBE ranking, such as Tcl, Racket, and Perl 6.
\fi



Twenty-four languages satisfy both criteria. We assign scores to them, based on the following rules: 
\begin{enumerate}[R1.]
\item \label{crit:tioberank} A language $\ell$ receives a TIOBE score $\tau_{\ell} = 1$ iff it is in the top-20 in TIOBE (\autoref{tab:tiobe-rank}); otherwise, $\tau_{\ell} = 2$.
\item \label{crit:rcrank} A language $\ell$ receives a \rc score $\rho_{\ell}$ corresponding to its ranking in \rc (first column in \autoref{tab:rosettacode-rank}).
\end{enumerate}
Using these scores, languages are ranked in increasing lexicographic order of $(\tau_{\ell}, \rho_{\ell})$. This ranking method sticks to the same rationale as C\ref{crit:tiobe} (prefer popular languages) and C\ref{crit:rc} (ensure a statistically significant base for analysis), and helps mitigate the role played by languages that are ``hyped'' in either the TIOBE or the \rc ranking.

To cover the most popular programming paradigms, we partition languages in four categories: procedural, object-oriented, functional, scripting.
Two languages (R and MATLAB) mainly are special-purpose; hence we drop them.
In each category, we rank languages using our ranking method and pick the top two languages. 
\autoref{tab:rosetta-code-popularity} shows the overall ranking; the shaded rows contain the eight languages selected for the study.

\footnotetext[3]{No rank means that the language is not in the top-50 in the TIOBE index.}
\footnotetext[4]{Not represented in \rc.}
\footnotetext[5]{Only represented in \rc in dialect versions.}
\addtocounter{footnote}{3}

\newcommand{\cc}{}
\newcommand{\mc}[2]{\multicolumn{2}{#1}{{\textsc{#2}}}}
\renewcommand{\rank}[1]{#1}
\begin{table}[htb]
\setlength{\tabcolsep}{2.3pt}
  \centering
  \begin{tabular}{lr|lr|lr|lr}
    \mc{c|}{procedural} & \mc{c|}{object-oriented} & \mc{c|}{functional} & \mc{c}{scripting}        \\ 
    $\ell$ & $(\tau_\ell, \rho_{\ell})$ & $\ell$ & $(\tau_\ell, \rho_{\ell})$ & $\ell$ & $(\tau_\ell, \rho_{\ell})$ & $\ell$ & $(\tau_\ell, \rho_{\ell})$ \\
    \hline
\rch {\bf C}   & (1,\rank{7})  & {\bf Java} & (1,\rank{17}) & {\bf F\#}   & (1,\rank{8}) & {\bf Python}     & (1,\rank{3}) \\
\rch Go        & (2,\rank{9})  & {\bf C\#}  & (1,\rank{21}) & Haskell     & (2,\rank{15}) & {\bf Ruby}       & (1,\rank{5}) \\
     Ada       & (2,\rank{12}) & {\bf C++}  & (1,\rank{22}) & Common Lisp\hspace{-2ex} & (2,\rank{23}) & {\bf Perl}       & (1,\rank{11}) \\
     PL/I      & (2,\rank{30}) & D          & (2,\rank{50}) & Scala       & (2,\rank{25}) & {\bf JavaScript}\hspace{-2ex} & (1,\rank{28}) \\
     Fortran\hspace{-2ex}   & (2,\rank{39}) &            &               & Erlang      & (2,\rank{26}) & {\bf PHP}        & (1,\rank{36}) \\
               &               &            &               & Scheme      & (2,\rank{47}) & Tcl              & (2,\rank{1})\\
               &               &            &               &             &               & Lua              & (2,\rank{35})
  \end{tabular}
  \caption{Combined ranking: the top-2 languages in each category are selected for the study.}
  \label{tab:rosetta-code-popularity}
\end{table}

\subsection{Experimental setup}  \label{sec:setup}

\rc collects solution files by task and language.
The following table details the total size of the data considered in our experiments (\textsc{lines} are total lines of program files).
\begin{center}
\begin{scriptsize}
\setlength{\tabcolsep}{2.5pt}
\begin{tabular}{l rrrrrrrr|r}
&                       C  &  C\#  & F\# & Go  & Haskell & Java & Python & Ruby & \textsc{all} \\
\hline
\textsc{tasks}
  & 630 & 463 & 341 & 617 & 553 & 534 & 675 & 635 & 745 \\
\textsc{files}
  & 989 & 640 & 426 & 869 & 980 & 837 & 1'319 & 1'027 & 7'087  \\
\textsc{lines}
  & 44'643 &  21'295 &   6'473 &  36'395 &  14'426 &  27'891 &  27'223 &  19'419 & 197'765 \\
\hline
\end{tabular}
\end{scriptsize}
\end{center}

Our experiments measure properties of \rc solutions in various dimensions\iflong: source-code features (such as lines of code), compilation features (such as size of executables), and runtime features (such as execution time)\fi.
\iflong
Correspondingly, we have to
\fi
 perform the following actions for each solution file \verb|f| of every task $t$ in each language $\ell$:
\begin{itemize}
\item \textbf{Merge:} if \verb|f| depends on other files (for example, an application consisting of two classes in two different files), make them available in the same location where \verb|f| is; $F$ denotes the resulting self-contained collection of source files that correspond to one solution of $t$ in $\ell$.
\item \textbf{Patch:} if $F$ has errors that prevent correct compilation or execution (for example, a library is used but not imported), correct $F$ as needed.
\item \textbf{LOC:} measure source-code features of $F$.
\item \textbf{Compile:} compile $F$ into native code (C, Go, and Haskell) or bytecode (C\#, F\#, Java, Python); \emph{executable} denotes the files produced by compilation.\footnote{For Ruby, which does not produce compiled code of any kind, this step is replaced by a syntax check of $F$.} Measure compilation features.
\item \textbf{Run:} run the executable and measure runtime features.
\end{itemize}

\iflong
Actions \textit{merge} and \textit{patch} are solution-specific and are required for the actions that follow.
In contrast, \textit{LOC}, \textit{compile}, and \textit{run} are only language-specific and produce the actual experimental data.
To automate executing the actions to the extent possible, we built a system of scripts that we now describe in some detail.
\fi

\fakepar{Merge}
We stored the information necessary for this step in the form of makefiles---one for every task that requires merging, that is such that there is no one-to-one correspondence between source-code files and solutions.
\iflong
A makefile has one target for every task solution $F$, and a default \verb|all| target that builds all solution targets for the current task.
Each target's recipe calls a placeholder script \verb|comp|, passing to it the list of input files that constitute the solution together with other necessary solution-specific compilation files (for example, library flags for the linker).
\fi
We wrote the makefiles after attempting a compilation with default options for all solution files, each compiled in isolation: we inspected all failed compilation attempts and provided makefiles whenever necessary.

\fakepar{Patch}
We stored the information necessary for this step in the form of diffs---one for every solution file that requires correction.
We wrote the diffs after attempting a compilation with the makefiles\iflong: we inspected all failed compilation attempts, and wrote diffs whenever necessary\fi.
\iflong
Some corrections could not be expressed as diffs because they involved renaming or splitting files (for example, some C files include both declarations and definitions, but the former should go in separate header files); we implemented these corrections by adding shell commands directly in the makefiles.
\fi

An important decision was \emph{what to patch}.
We want to have as many compiled solutions as possible, but we also do not want to alter the \rc data before measuring it.
We did not fix errors that had to do with functional correctness or very solution-specific features.
We did fix simple errors: missing library inclusions, omitted variable declarations, and typos.
These guidelines try to replicate the moves of a user who would like to reuse \rc solutions but may not be fluent with the languages.
As a result of following this process, we have a reasonably high confidence that patched solutions are correct implementations of the tasks.

Diffs play an additional role for tasks for performance analysis ($T_{\text{\textsc{perf}}}$ and $T_{\text{\textsc{scal}}}$ in \autoref{sec:task-selection}).
Solutions to these tasks must not only be correct but also run on the same inputs (tasks $T_{\text{\textsc{perf}}}$) and on the same ``large'' inputs (tasks $T_{\text{\textsc{scal}}}$).
We checked all solutions to tasks $T_{\text{\textsc{perf}}}$ and patched them when necessary to ensure they work on comparable inputs, but we did not change the inputs themselves from those suggested in the task descriptions.
In contrast, we inspected all solutions to tasks $T_{\text{\textsc{scal}}}$ and patched them by supplying task-specific inputs that are computationally demanding.
\iflong
A significant example of computing-intensive tasks were the sorting algorithms, which we patched to build and sort large integer arrays (generated on the fly using a linear congruential generator function with fixed seed).
The input size was chosen after a few trials so as to be feasible for most languages within a timeout of 3 minutes; for example, the sorting algorithms deal with arrays of size from $3 \cdot 10^4$ elements for quadratic-time algorithms to $2 \cdot 10^6$ elements for linear-time algorithms.
\fi

\fakepar{LOC}
For each language $\ell$, we wrote a script $\ell\texttt{\_loc}$ that inputs a list of files, calls \verb|cloc|\footnote{\url{http://cloc.sourceforge.net/}} on them to count the lines of code, and logs the results.

\fakepar{Compile}
For each language $\ell$, we wrote a script $\ell\texttt{\_compile}$ that inputs a list of files and compilation flags, calls the appropriate compiler on them, and logs the results.
The following table shows the compiler versions used for each language, as well as the optimization flags.
We tried to select a stable compiler version complete with matching standard libraries, and the best optimization level among those that are not too aggressive or involve rigid or extreme trade-offs.
\begin{center}
\begin{scriptsize}
\begin{tabular}{llrl}
\textsc{lang} & \textsc{compiler} & \textsc{version}     & \multicolumn{1}{l}{\textsc{flags}} \\
\hline
C          &   \verb|gcc| (GNU)        &  4.6.3        &  \verb|-O2|  \\
C\#        &   \verb|mcs| (Mono 3.2.1)       &  3.2.1.0        &  \verb|-optimize| \\
F\#        &   \verb|fsharpc| (Mono 3.2.1)   & 3.1    &  \verb|-O| \\
Go         &   \verb|go|               & 1.3           &  \\
Haskell    &   \verb|ghc|              & 7.4.1         &  \verb|-O2| \\
Java       &   \verb|javac| (OracleJDK 8)  & 1.8.0\_11     &  \\
Python     &   \verb|python| (CPython) & 2.7.3/3.2.3   &  \\
Ruby       &   \verb|ruby|             & 2.1.2        &  \verb|-c|
\end{tabular}
\end{scriptsize}
\end{center}
\verb|C_compile| tries to detect the C dialect (\verb|gnu90|, \verb|C99|, \ldots) until compilation succeeds.
\verb|Java_compile| looks for names of public classes in each source file and renames the files to match the class names (as required by the Java compiler).
\verb|Python_compile| tries to detect the version of Python (2.x or 3.x) until compilation succeeds.
\verb|Ruby_compile| only performs a syntax check of the source (flag \verb|-c|), since Ruby has no (standard) stand-alone compilation.

\fakepar{Run}
For each language $\ell$, we wrote a script $\ell\texttt{\_run}$ that inputs an executable name, executes it, and logs the results.
Native executables are executed directly, whereas bytecode is executed using the appropriate virtual machines.
To have reliable performance measurements, the scripts: repeat each execution 6 times; discard the timing of the first execution (to fairly accommodate bytecode languages that load virtual machines from disk\iflong: it is only in the first execution that the virtual machine is loaded from disk, with corresponding possibly significant one-time overhead; in the successive executions the virtual machine is read from cache, with only limited overhead\fi); check that the 5 remaining execution times are within one standard deviation of the mean; log the mean execution time. 
If an execution does not terminate within a time-out of 3 minutes it is forcefully terminated.

\fakepar{Overall process}
\iflong
A Python script orchestrates the whole experiment.
\fi
For every language $\ell$, for every task $t$, for each action $\texttt{act} \in \{\texttt{loc}, \texttt{compile}, \texttt{run}\}$:
\begin{enumerate}
\item if patches exist for any solution of $t$ in $\ell$, apply them;
\item if no makefile exists for task $t$ in $\ell$, call script $\ell\texttt{\_act}$ directly on each solution file \verb|f| of $t$;
\item if a makefile exists, invoke it and pass $\ell\texttt{\_act}$ as command \iflong \verb|comp|\fi to be used; the makefile defines the self-contained collection of source files $F$ on which the script works.
\end{enumerate}
\iflong
Since the command-line interface of the $\ell\texttt{\_loc}$, $\ell\texttt{\_compile}$, and $\ell\texttt{\_run}$ scripts is uniform, the same makefiles work as recipes for all actions \verb|act|.
\fi

\subsection{Experiments} \label{sec:experiments}

The experiments ran on a Ubuntu 12.04 LTS 64bit GNU/Linux box with Intel Quad Core2 CPU at 2.40~GHz and 4~GB of RAM.
At the end of the experiments, we extracted all logged data for statistical analysis using R.

\subsection{Statistical analysis} \label{sec:statistics}

The statistical analysis targets pairwise comparisons between languages. Each comparison uses a different metric $M$ 
\iflong
including lines of code (conciseness), size of the executable (native or bytecode), CPU time, maximum RAM usage (i.e., maximum resident set size), number of page faults, and number of runtime failures. Metrics are \emph{normalized} as we detail below.
\fi
Let $\ell$ be a programming language, $t$ a task, and $M$ a metric.
$\ell_M(t)$ denotes the vector of measures of $M$, one for each solution to task $t$ in language $\ell$.
$\ell_M(t)$ may be empty if there are no solutions to task $t$ in $\ell$.
The comparison of languages $X$ and $Y$ based on $M$ works as follows.
Consider a subset $T$ of the tasks such that, for every $t \in T$, both $X$ and $Y$ have at least one solution to $t$.
\iflong
$T$ may be further restricted based on a measure-dependent \emph{criterion}; for example, to check conciseness, we may choose to only consider a task $t$ if both $X$ and $Y$ have at least one solution that compiles without errors (solutions that do not satisfy the criterion are discarded).
\fi

Following this procedure, each $T$ determines two data vectors $x_M^\alpha$ and $y_M^\alpha$, for the two languages $X$ and $Y$, by aggregating the measures per task using an \emph{aggregation function} $\alpha$; as aggregation functions, we normally consider both minimum and mean.
For each task $t \in T$, the $t$-th component of the two vectors $x_M^\alpha$ and $y_M^\alpha$ is:
\[
\begin{split}
x_M^\alpha(t) & =     \alpha(X_M(t))/\nu_M(t, X, Y)\,, \\
y_M^\alpha(t) & =     \alpha(Y_M(t))/\nu_M(t, X, Y)\,,
\end{split}
\]
where $\nu_M(t, X, Y)$ is a normalization factor defined as:
\[
\nu_M(t, X, Y) =
\begin{cases}
\min\left( X_M(t)Y_M(t) \right)  
   &  \text{if } \min(X_M(t)Y_M(t)) > 0\,, \\
1  & \text{otherwise}\,,
\end{cases}
\]
where juxtaposing vectors denotes concatenating them.
\iflong
Thus, the normalization factor is the smallest value of metric $M$ measured across all solutions of $t$ in $X$ and in $Y$ if such a value is positive; otherwise, when the minimum is zero, the normalization factor is one.
\fi
This definition ensures that $x_M^\alpha(t)$ and $y_M^\alpha(t)$ are well-defined even when a minimum of zero occurs due to the limited precision of some measures such as running time.

As statistical test, we normally\footnote{Failure analysis (RQ5) uses the $U$ test, as described there.} use the Wilcoxon signed-rank test, a paired non-parametric difference test which assesses whether the mean ranks of $x_M^\alpha$ and of $y_M^\alpha$ differ. We display the test results in a table, under column labeled with language $X$ at row labeled with language $Y$, and include various measures:

  \begin{enumerate}
  \item The $p$-value, which estimates the probability that chance can explain the differences between $x_M^\alpha$ and $y_M^\alpha$. Intuitively, if $p$ is small  
it means that there is a high chance that $X$ and $Y$ exhibit a genuinely different behavior with respect to metric $M$.

  \item The effect size, computed as Cohen's $d$, defined as the standardized mean difference:
$d = (\overline{x_M^\alpha} - \overline{y_M^\alpha})/{s}$, 
where $\overline{V}$ is the mean of a vector $V$, and $s$ is the pooled standard deviation of the data.
For statistically significant differences, $d$ estimates how large the difference is.

\item The signed ratio
\[
R = \sgn(\widetilde{x_M^\alpha} - \widetilde{y_M^\alpha})\,\frac{\max(\widetilde{x_M^\alpha}, \widetilde{y_M^\alpha})}{\min(\widetilde{x_M^\alpha}, \widetilde{y_M^\alpha})}
\]
of the largest median to the smallest median (where $\widetilde{V}$ is the median of a vector $V$), which gives an unstandardized measure of the difference between the two medians.\footnote{The definition of $R$ uses median as average to lessen the influence of outliers.} 
Sign and absolute value of $R$ have direct interpretations whenever the difference between $X$ and $Y$ is significant: if $M$ is such that ``smaller is better'' (for instance, running time), then a \emph{positive} sign $\sgn(\widetilde{x_M^\alpha} - \widetilde{y_M^\alpha})$ indicates that the average solution in language $Y$ is \emph{better} (smaller) with respect to $M$ than the average solution in language $X$; the absolute value of $R$ indicates \emph{how many times} $X$ is larger than $Y$ on average.
   \end{enumerate}

Throughout the paper, we will say that language $X$: \emph{is significantly different} from language $Y$, if $p < 0.01$; and that it \emph{tends to be different} from $Y$ if $0.01 \leq p < 0.05$. We will say that the effect size is: \emph{vanishing} if $d < 0.05$; \emph{small} if $0.05 \leq d < 0.3$; \emph{medium} if $0.3 \leq d < 0.7$; and \emph{large} if $d \geq 0.7$.

\subsection{Visualizations of language comparisons} \label{sec:visualizations}

Each results table is accompanied by a \emph{language relationship graph}, which helps visualize the results of the the pairwise language relationships.
In such graphs, nodes correspond to programming languages.
Two nodes $\ell_1$ and $\ell_2$ are arranged so that their \emph{horizontal} distance is \emph{roughly} proportional to the absolute value of ratio $R$ for the two languages; an exact proportional display is not possible in general, as the pairwise ordering of languages may not be 
a total order. Vertical distances are chosen only to improve readability and carry no meaning.

A \emph{solid} arrow is drawn from node $X$ to $Y$ if language $Y$ is significantly better than language $X$ in the given metric, and a \emph{dashed} arrow if $Y$ tends to be better than $X$ (using the terminology from \autoref{sec:statistics}).
To improve the visual layout, edges that express an ordered pair that is subsumed by others are omitted, that is if $X \rightarrow W \rightarrow Y$ the edge from $X$ to $Y$ is omitted. The thickness of arrows is proportional to the effect size; 
if the effect is vanishing, no arrow is drawn.

\section{Results}
\label{sec:results}

\rqpar{RQ1. Which programming languages make for more concise code?}

To answer this question, we measure the non-blank non-comment \emph{lines of code} of solutions of tasks $T_{\textsc{loc}}$ marked for lines of code count that compile without errors. The requirement of successful compilation ensures that only syntactically correct programs are considered to measure conciseness. To check the impact of this requirement, we also compared these results with a measurement including all solutions (whether they compile or not), obtaining qualitatively similar results. 

For all research questions but RQ5, we considered both minimum and mean as aggregation functions (\autoref{sec:statistics}). For brevity, the presentation describes results for only one of them (typically the minimum). 
For lines of code measurements, aggregating by minimum means that we consider, for each task, the shortest solution available in the language.

\autoref{tab:loc_exit0_min-main} shows the results of the pairwise comparison, where $p$ is the $p$-value, $d$ the effect size, and $R$ the ratio, as described in \autoref{sec:statistics}. In the table, $\varepsilon$ denotes the smallest positive floating-point value representable in R. 

\begin{table}[ht]
\setlength{\tabcolsep}{4pt}
\begin{center}
{\scriptsize
\begin{tabular}{cc|rrrrrrr}
  \hline
\textsc{lang} & \textsc{} & \multicolumn{1}{c}{C} & \multicolumn{1}{c}{C\#} & \multicolumn{1}{c}{F\#} & \multicolumn{1}{c}{Go} & \multicolumn{1}{c}{Haskell} & \multicolumn{1}{c}{Java} & \multicolumn{1}{c}{Python} \\ 
  \hline
C\# & $p$ & 0.543 &  &  &  &  &  &  \\ 
   & $d$ & 0.004 &  &  &  &  &  &  \\ 
   & $R$ & -1.1 &  &  &  &  &  &  \\ 
   \hline
F\# & $p$ & \textcolor{colorstrong-main}{\vanish{0}} & \textcolor{colorstrong-main}{\vanish{0}} &  &  &  &  &  \\ 
   & $d$ & \textcolor{colorstrong-main}{0.735} & \textcolor{colorstrong-main}{0.945} &  &  &  &  &  \\ 
   & $R$ & 2.5 & 2.6 &  &  &  &  &  \\ 
   \hline
Go & $p$ & 0.377 & 0.082 & \textcolor{colorstrong-main}{\vanish{29}} &  &  &  &  \\ 
   & $d$ & \textcolor{colorweak-main}{0.155} & \textcolor{colorweak-main}{0.083} & \textcolor{colormedium-main}{0.640} &  &  &  &  \\ 
   & $R$ & 1.0 & 1.0 & -2.5 &  &  &  &  \\ 
   \hline
Haskell & $p$ & \textcolor{colorstrong-main}{\vanish{0}} & \textcolor{colorstrong-main}{\vanish{0}} & 0.168 & \textcolor{colorstrong-main}{\vanish{0}} &  &  &  \\ 
   & $d$ & \textcolor{colorstrong-main}{1.071} & \textcolor{colorstrong-main}{1.286} & \textcolor{colorweak-main}{0.085} & \textcolor{colorstrong-main}{1.255} &  &  &  \\ 
   & $R$ & 2.9 & 2.7 & 1.3 & 2.9 &  &  &  \\ 
   \hline
Java & $p$ & \textcolor{colorweak-main}{0.026} & \textcolor{colorstrong-main}{\vanish{4}} & \textcolor{colorstrong-main}{\vanish{25}} & \textcolor{colorweak-main}{0.026} & \textcolor{colorstrong-main}{\vanish{32}} &  &  \\ 
   & $d$ & \textcolor{colorweak-main}{0.262} & \textcolor{colormedium-main}{0.319} & \textcolor{colorstrong-main}{0.753} & \textcolor{colorweak-main}{0.148} & \textcolor{colorstrong-main}{1.052} &  &  \\ 
   & $R$ & 1.0 & 1.1 & -2.3 & 1.0 & -2.9 &  &  \\ 
   \hline
Python & $p$ & \textcolor{colorstrong-main}{\vanish{0}} & \textcolor{colorstrong-main}{\vanish{0}} & \textcolor{colorstrong-main}{\vanish{4}} & \textcolor{colorstrong-main}{\vanish{0}} & \textcolor{colorweak-main}{0.021} & \textcolor{colorstrong-main}{\vanish{0}} &  \\ 
   & $d$ & \textcolor{colorstrong-main}{0.951} & \textcolor{colorstrong-main}{1.114} & \textcolor{colormedium-main}{0.359} & \textcolor{colorstrong-main}{0.816} & \textcolor{colorweak-main}{0.209} & \textcolor{colorstrong-main}{0.938} &  \\ 
   & $R$ & 2.9 & 3.6 & 1.6 & 3.0 & 1.2 & 2.8 &  \\ 
   \hline
Ruby & $p$ & \textcolor{colorstrong-main}{\vanish{0}} & \textcolor{colorstrong-main}{\vanish{0}} & \textcolor{colorweak-main}{0.013} & \textcolor{colorstrong-main}{\vanish{0}} & 0.764 & \textcolor{colorstrong-main}{\vanish{0}} & \textcolor{colorweak-main}{0.015} \\ 
   & $d$ & \textcolor{colormedium-main}{0.558} & \textcolor{colorstrong-main}{0.882} & \textcolor{colorweak-main}{0.103} & \textcolor{colorstrong-main}{0.742} & \textcolor{colorweak-main}{0.107} & \textcolor{colorstrong-main}{0.763} & 0.020 \\ 
   & $R$ & 2.5 & 2.7 & 1.2 & 2.5 & -1.2 & 2.2 & -1.3 \\ 
   \hline
\end{tabular}
}
\caption{Comparison of lines of code (by minimum).} 
\label{tab:loc_exit0_min-main}
\end{center}
\end{table}

\begin{figure}[htb]
\centering
  \begin{tikzpicture}[
  lang/.style={draw=none,font=\footnotesize,inner sep=1pt,outer sep=1pt},
  align=center, xscale=2.0, yscale=0.2
  ]
\node [lang] (C) at (-3,3) {C};
\node [lang] (C-sharp) at (-3.6,6) {C\#};
\node [lang] (F-Sharp) at (-1.2,3) {F\#};
\node [lang] (Go) at (-3,9) {Go};
\node [lang] (Haskell) at (-0.8,9) {Haskell};
\node [lang] (Java) at (-2.9,6) {Java};
\node [lang] (Python) at (0,6) {Python};
\node [lang] (Ruby) at (-0.9,6) {Ruby};
\draw [->,densely dotted,thin](C) edge (Java);
\draw [->,,thick](C-sharp) edge (Java);
\draw [->,,very thick](Java) edge (F-Sharp);
\draw [->,,thick](F-Sharp) edge (Python);
\draw [->,densely dotted,thin](F-Sharp) edge (Ruby);
\draw [->,densely dotted,thin](Go) edge (Java);
\draw [->,,very thick](Java) edge (Haskell);
\draw [->,densely dotted,thin](Haskell) edge (Python);
  \end{tikzpicture}
\caption{Comparison of lines of code (by minimum).}
\label{fig:loc_exit0_min_normrank_network-main}
\end{figure}
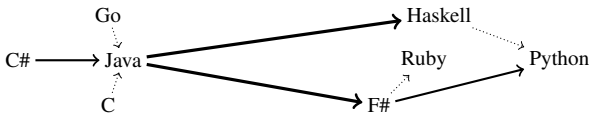

\autoref{fig:loc_exit0_min_normrank_network-main} shows the corresponding language relationship graph; remember that arrows point to the more concise languages, thickness denotes larger effects, and horizontal distances are roughly proportional to average differences. 
Languages are clearly divided into two groups: functional and scripting languages tend to provide the most concise code, whereas procedural and object-oriented languages are significantly more verbose. 
The absolute difference between the two groups is major; for instance, Java programs are on average 2.2--2.9 times longer than programs in functional and scripting languages.

Within the two groups, differences are less pronounced. Among the scripting languages, and among the functional languages, no statistically significant differences exist. Python tends to be the most concise, even against functional languages (1.2--1.6 times shorter on average). Among procedural and object-oriented languages, Java tends to be slightly more concise, with small to medium effect sizes. 

\takehome{Functional and scripting languages provide significantly more concise code than procedural and object-oriented languages.}

\rqpar{RQ2. Which programming languages compile into smaller executables?}

To answer this question, we measure the \emph{size of the executables} of solutions of tasks $T_{\textsc{comp}}$ marked for compilation that compile without errors. We consider both native-code executables (C, Go, and Haskell) and bytecode executables (C\#, F\#, Java, Python). Ruby's standard programming environment does not offer compilation to bytecode and Ruby programs are therefore not included in the measurements for RQ2. 

\begin{figure}[htb]
\centering
  \begin{tikzpicture}[
  lang/.style={draw=none,font=\footnotesize,inner sep=1pt,outer sep=1pt},
  align=center, xscale=0.015, yscale=0.18
  ]
\node [lang] (C) at (-3,0) {C};
\node [lang] (C-sharp) at (-1.4,6) {C\#};
\node [lang] (F-Sharp) at (-2.0,3) {F\#};
\node [lang] (Go) at (-453,0) {Go};
\node [lang] (Haskell) at (-338,0) {Haskell};
\node [lang] (Java) at (-1.3,9) {Java};
\node [lang] (Python) at (0,12) {Python};
\draw [->,,very thick](C) edge (F-Sharp);
\draw [->,,very thick](Haskell) edge (C);
\draw [->,,very thick](F-Sharp) edge (C-sharp);
\draw [->,,thick](C-sharp) edge (Java);
\draw [->,,very thick](Go) edge (Haskell);
\draw [->,,thick](Java) edge (Python);
  \end{tikzpicture}
\caption{Comparison of size of executables (by minimum).}
\label{fig:binsize_exit0_min_normrank_network-main}
\end{figure}
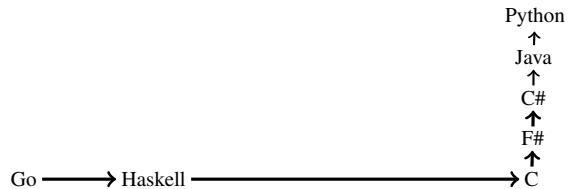

\autoref{tab:binsize_exit0_min-main} shows the results of the statistical analysis, and \autoref{fig:binsize_exit0_min_normrank_network-main} the corresponding language relationship graph. 

\begin{table}[ht]
\setlength{\tabcolsep}{4pt}
\begin{center}
{\scriptsize
\begin{tabular}{cc|rrrrrrr}
  \hline
\textsc{lang} & \textsc{} & \multicolumn{1}{c}{C} & \multicolumn{1}{c}{C\#} & \multicolumn{1}{c}{F\#} & \multicolumn{1}{c}{Go} & \multicolumn{1}{c}{Haskell} & \multicolumn{1}{c}{Java} \\ 
  \hline
C\# & $p$ & \textcolor{colorstrong-main}{\vanish{0}} &  &  &  &  &  \\ 
   & $d$ & \textcolor{colorstrong-main}{2.669} &  &  &  &  &  \\ 
   & $R$ & 2.1 &  &  &  &  &  \\ 
   \hline
F\# & $p$ & \textcolor{colorstrong-main}{\vanish{0}} & \textcolor{colorstrong-main}{\vanish{15}} &  &  &  &  \\ 
   & $d$ & \textcolor{colorstrong-main}{1.395} & \textcolor{colorstrong-main}{1.267} &  &  &  &  \\ 
   & $R$ & 1.4 & -1.5 &  &  &  &  \\ 
   \hline
Go & $p$ & \textcolor{colorstrong-main}{\vanish{52}} & \textcolor{colorstrong-main}{\vanish{39}} & \textcolor{colorstrong-main}{\vanish{31}} &  &  &  \\ 
   & $d$ & \textcolor{colorstrong-main}{3.639} & \textcolor{colorstrong-main}{2.312} & \textcolor{colorstrong-main}{2.403} &  &  &  \\ 
   & $R$ & -153.3 & -340.7 & -217.8 &  &  &  \\ 
   \hline
Haskell & $p$ & \textcolor{colorstrong-main}{\vanish{45}} & \textcolor{colorstrong-main}{\vanish{35}} & \textcolor{colorstrong-main}{\vanish{29}} & \textcolor{colorstrong-main}{\vanish{0}} &  &  \\ 
   & $d$ & \textcolor{colorstrong-main}{2.469} & \textcolor{colorstrong-main}{2.224} & \textcolor{colorstrong-main}{2.544} & \textcolor{colorstrong-main}{1.071} &  &  \\ 
   & $R$ & -111.2 & -240.3 & -150.8 & 1.3 &  &  \\ 
   \hline
Java & $p$ & \textcolor{colorstrong-main}{\vanish{0}} & \textcolor{colorstrong-main}{\vanish{4}} & \textcolor{colorstrong-main}{\vanish{0}} & \textcolor{colorstrong-main}{\vanish{0}} & \textcolor{colorstrong-main}{\vanish{0}} &  \\ 
   & $d$ & \textcolor{colorstrong-main}{3.148} & \textcolor{colormedium-main}{0.364} & \textcolor{colorstrong-main}{1.680} & \textcolor{colorstrong-main}{3.121} & \textcolor{colorstrong-main}{1.591} &  \\ 
   & $R$ & 2.3 & 1.1 & 1.7 & 341.4 & 263.8 &  \\ 
   \hline
Python & $p$ & \textcolor{colorstrong-main}{\vanish{0}} & \textcolor{colorstrong-main}{\vanish{15}} & \textcolor{colorstrong-main}{\vanish{0}} & \textcolor{colorstrong-main}{\vanish{0}} & \textcolor{colorstrong-main}{\vanish{0}} & \textcolor{colorstrong-main}{\vanish{5}} \\ 
   & $d$ & \textcolor{colorstrong-main}{5.686} & \textcolor{colorstrong-main}{0.899} & \textcolor{colorstrong-main}{1.517} & \textcolor{colorstrong-main}{3.430} & \textcolor{colorstrong-main}{1.676} & \textcolor{colormedium-main}{0.395} \\ 
   & $R$ & 2.9 & 1.3 & 2.0 & 452.7 & 338.3 & 1.3 \\ 
   \hline
\end{tabular}
}
\caption{Comparison of size of executables (by minimum).}
\label{tab:binsize_exit0_min-main}
\end{center}
\end{table}

It is apparent that measuring executable sizes determines a total order of languages, with Go producing the largest and Python the smallest executables. Based on this order, three consecutive groups naturally emerge: Go and Haskell compile to native and have ``large'' executables; F\#, C\#, Java, and Python compile to bytecode and have ``small'' executables; C compiles to native but with size close to bytecode executables.

Size of bytecode does not differ much across languages: F\#, C\#, and Java executables are, on average, only 1.3--2.0 times larger than Python's. 
The differences between sizes of native executables are more spectacular, with Go's and Haskell's being on average 153.3 and 111.2 times larger than C's. This is largely a result of Go and Haskell using static linking by default, as opposed to \verb|gcc| defaulting to dynamic linking whenever possible. With dynamic linking, C produces very compact binaries, which are on average a mere 2.9 times larger than Python's bytecode. C was compiled with level \verb|-O2| optimization, which should be a reasonable middle ground: binaries tend to be larger under more aggressive speed optimizations, and smaller under executable size optimizations (flag \verb|-Os|). 

\takehome{Languages that compile into bytecode have significantly smaller executables than those that compile into native machine code.}

\rqpar{RQ3. Which programming languages have better running-time performance?}

To answer this question, we measure the \emph{running time} of solutions of tasks $T_{\textsc{scal}}$ marked for running time measurements on computing-intensive workloads that run without errors or timeout (set to 3 minutes). 
As discussed in \autoref{sec:task-selection} and \autoref{sec:setup}, we manually patched solutions to tasks in $T_{\textsc{scal}}$ to ensure that they work on the same inputs of substantial size. This ensures that---as is crucial for running time measurements---all solutions used in these experiments run on the very same inputs. 

\begin{table}[htb]
\begin{center}
{\scriptsize
\setlength{\tabcolsep}{2pt}
  \begin{tabular}{rll}
   & \textsc{name} & \textsc{input} \\ \hline
 1 & 9 billion names of God the integer & $n = 10^5$ \\
 2--3 & Anagrams & $100\ \times$ \verb|unixdict.txt| (20.6 MB) \\
 4 & Arbitrary-precision integers & $5^{4^{3^{2}}}$ \\
 5 & Combinations & $25 \choose 10$\\
 6 & Count in factors & $n = 10^6$ \\
 7 & Cut a rectangle & $10 \times 10$ rectangle \\
 8 & Extensible prime generator & $10^7$th prime \\
 9 & Find largest left truncatable prime & $10^7$th prime \\
10 & Hamming numbers & $10^7$th Hamming number \\
11 & Happy numbers & $10^6$th Happy number \\
12 & Hofstadter Q sequence & \# flips up to $10^5$th term  \\
13--16 & Knapsack problem/{[\emph{all versions}]} & from task description \\
17 & Ludic numbers & from task description  \\
18 & LZW compression & $100\ \times$ \verb|unixdict.txt| (20.6 MB) \\
19 & Man or boy test & $n = 16$ \\
20 & N-queens problem & $n = 13$ \\
21 & Perfect numbers & first $5$ perfect numbers  \\
22 & Pythagorean triples & perimeter $< 10^8$ \\
23 & Self-referential sequence & $n = 10^6$ \\
24 & Semordnilap & $100\ \times$ \verb|unixdict.txt| \\
25 & Sequence of non-squares & non-squares $< 10^6$ \\
26--34 & Sorting algorithms/[\emph{quadratic}] & $n \simeq 10^4$ \\
35--41 & Sorting algorithms/[\emph{$n \log n$ and linear}] & $n \simeq 10^6$ \\
42--43 & Text processing/{[\emph{all versions}]}  & from task description (1.2 MB) \\
44 & Topswops & $n = 12$ \\
45 & Towers of Hanoi & $n = 25$  \\
46 & Vampire number & from task description \\
  \end{tabular}
}
  \caption{Computing-intensive tasks.}
  \label{tab:scalability-tasks}
\end{center}
\end{table}

\begin{figure}[htb]
\centering
  \begin{tikzpicture}[
  lang/.style={draw=none,font=\footnotesize,inner sep=1pt,outer sep=1pt},
  align=center, xscale=0.3, yscale=0.4
  ]
\node [lang] (C) at (1,3) {C};
\node [lang] (C-sharp) at (-9.5,3) {C\#};
\node [lang] (F-Sharp) at (-8.5,6.5) {F\#};
\node [lang] (Go) at (-1.6,3) {Go};
\node [lang] (Haskell) at (-14,0) {Haskell};
\node [lang] (Java) at (-5,5) {Java};
\node [lang] (Python) at (-19,1) {Python};
\node [lang] (Ruby) at (-24,5) {Ruby};
\draw [->,,thick](Go) edge (C);
\draw [->,densely dotted,thick](C-sharp) edge ($(Go.west)+(-0pt,-0pt)$);
\draw [->,densely dotted,thick](Python) edge (C-sharp);
\draw [->,,thick](Ruby) edge (C-sharp);
\draw [->,densely dotted,thick,bend left=20](F-Sharp) edge (Go.north);
\draw [->,,very thick,bend right=10](Haskell) edge (Go.south);
\draw [->,densely dotted,thick](Java) edge ($(Go.west)+(-4pt,18pt)$);
\draw [->,,very thick](Python) edge ($(Go.west)+(1pt,-12pt)$);
\draw [->,,very thick](Ruby) edge ($(Go.west)+(-2pt,8pt)$);
\draw [->,densely dotted,thick](Ruby) edge (Java);
  \end{tikzpicture}
\caption{Comparison of running time (by minimum) for computing-intensive tasks.}
\label{fig:scalability_exit0_min_normrank_network-main}
\end{figure}
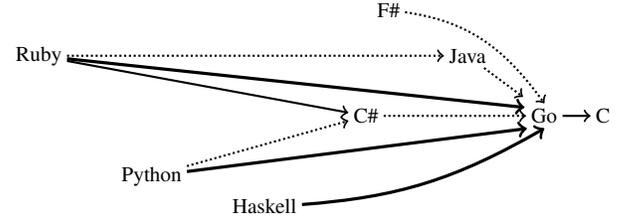

\autoref{tab:scalability-tasks} summarizes the tasks $T_{\textsc{scal}}$ and their inputs. It is a diverse collection which spans from text processing tasks on large input files (``Anagrams'', ``Semordnilap''), to combinatorial puzzles (``N-queens problem'', ``Towers of Hanoi''), to NP-complete problems (``Knapsack problem'') and sorting algorithms of varying complexity. 
We chose inputs sufficiently large to probe the performance of the programs, and to make  input/output overhead negligible w.r.t.\ total running time. 
%
\autoref{tab:scalability_exit0_min-main} shows the results of the statistical analysis, and \autoref{fig:scalability_exit0_min_normrank_network-main} the corresponding language relationship graph.

\begin{table}[htb]
\setlength{\tabcolsep}{4pt}
\begin{center}
{\scriptsize
\begin{tabular}{cc|rrrrrrr}
  \hline
\textsc{lang} & \textsc{} & \multicolumn{1}{c}{C} & \multicolumn{1}{c}{C\#} & \multicolumn{1}{c}{F\#} & \multicolumn{1}{c}{Go} & \multicolumn{1}{c}{Haskell} & \multicolumn{1}{c}{Java} & \multicolumn{1}{c}{Python} \\ 
  \hline
C\# & $p$ & \textcolor{colorstrong-main}{0.001} &  &  &  &  &  &  \\ 
   & $d$ & \textcolor{colormedium-main}{0.328} &  &  &  &  &  &  \\ 
   & $R$ & -7.5 &  &  &  &  &  &  \\ 
   \hline
F\# & $p$ & \textcolor{colorweak-main}{0.012} & 0.075 &  &  &  &  &  \\ 
   & $d$ & \textcolor{colormedium-main}{0.453} & \textcolor{colormedium-main}{0.650} &  &  &  &  &  \\ 
   & $R$ & -8.6 & -2.6 &  &  &  &  &  \\ 
   \hline
Go & $p$ & \textcolor{colorstrong-main}{\vanish{4}} & \textcolor{colorweak-main}{0.020} & \textcolor{colorweak-main}{0.016} &  &  &  &  \\ 
   & $d$ & \textcolor{colormedium-main}{0.453} & \textcolor{colormedium-main}{0.338} & \textcolor{colormedium-main}{0.578} &  &  &  &  \\ 
   & $R$ & -1.6 & 6.3 & 5.6 &  &  &  &  \\ 
   \hline
Haskell & $p$ & \textcolor{colorstrong-main}{\vanish{4}} & 0.084 & 0.929 & \textcolor{colorstrong-main}{\vanish{3}} &  &  &  \\ 
   & $d$ & \textcolor{colorstrong-main}{0.895} & \textcolor{colorweak-main}{0.208} & \textcolor{colormedium-main}{0.424} & \textcolor{colorstrong-main}{0.705} &  &  &  \\ 
   & $R$ & -24.5 & -2.9 & -1.6 & -13.1 &  &  &  \\ 
   \hline
Java & $p$ & \textcolor{colorstrong-main}{\vanish{4}} & 0.661 & 0.158 & \textcolor{colorweak-main}{0.0135} & 0.098 &  &  \\ 
   & $d$ & \textcolor{colormedium-main}{0.374} & \textcolor{colormedium-main}{0.364} & \textcolor{colormedium-main}{0.469} & \textcolor{colormedium-main}{0.563} & \textcolor{colormedium-main}{0.424} &  &  \\ 
   & $R$ & -3.2 & 1.8 & 4.9 & -2.4 & 6.6 &  &  \\ 
   \hline
Python & $p$ & \textcolor{colorstrong-main}{\vanish{5}} & \textcolor{colorweak-main}{0.033} & 0.938 & \textcolor{colorstrong-main}{\vanish{3}} & 0.894 & 0.082 &  \\ 
   & $d$ & \textcolor{colorstrong-main}{0.704} & \textcolor{colormedium-main}{0.336} & \textcolor{colormedium-main}{0.318} & \textcolor{colorstrong-main}{0.703} & \textcolor{colormedium-main}{0.386} & \textcolor{colorweak-main}{0.182} &  \\ 
   & $R$ & -29.8 & -4.3 & -1.1 & -15.5 & 1.1 & -9.2 &  \\ 
   \hline
Ruby & $p$ & \textcolor{colorstrong-main}{\vanish{3}} & \textcolor{colorstrong-main}{0.004} & 0.754 & \textcolor{colorstrong-main}{\vanish{3}} & 0.360 & \textcolor{colorweak-main}{0.013} & 0.055 \\ 
   & $d$ & \textcolor{colorstrong-main}{0.999} & \textcolor{colormedium-main}{0.358} & \textcolor{colorweak-main}{0.113} & \textcolor{colorstrong-main}{0.984} & \textcolor{colorweak-main}{0.250} & \textcolor{colorweak-main}{0.204} & 0.020 \\ 
   & $R$ & -34.1 & -6.2 & -1.4 & -18.5 & -1.4 & -7.9 & -1.6 \\ 
   \hline
\end{tabular}
}
\caption{Comparison of running time (by minimum) for computing-intensive tasks.}
\label{tab:scalability_exit0_min-main}
\end{center}
\end{table}

C is unchallenged over the computing-intensive tasks $T_{\textsc{scal}}$. Go is the runner-up, but significantly slower with medium effect size: the average Go program is 1.6 times slower than the average C program. Programs in other languages are much slower than Go programs, with medium to large effect size (2.4--18.5 times slower than Go on average).

\takehome{C is king on computing-intensive workloads. Go is the runner-up. Other languages, with object-oriented or functional features, incur further performance losses.}

The results on the tasks $T_{\textsc{scal}}$ clearly identified the procedural languages---C in particular---as the fastest.
However, the raw speed demonstrated on those tasks represents challenging conditions that are relatively infrequent in the many classes of applications that are not algorithmically intensive.
To find out performance differences under other conditions, we measure running time on the tasks $T_{\textsc{perf}}$, which are still clearly defined and run on the same inputs, but are not markedly computationally intensive and do not naturally scale to large instances.
Examples of such tasks are checksum algorithms (Luhn's credit card validation), string manipulation tasks (reversing the space-separated words in a string), and standard system library accesses (securing a temporary file).

The results, which we only discuss in the text for brevity, are definitely more mixed than those related to tasks $T_{\textsc{scal}}$, which is what one could expect given that we are now looking into modest running times in absolute value, where every language has at least decent performance.
First of all, C loses its undisputed supremacy, as it is not significantly faster than Go and Haskell---even though, when differences are statistically significant, C remains ahead of the other languages.
The procedural languages and Haskell collectively emerge as the fastest in tasks $T_{\textsc{perf}}$; none of them sticks out as \emph{the} fastest because the differences among them are insignificant and may sensitively depend on the tasks that each language implements in \rc.
Among the other languages (C\#, F\#, Java, Python, and Ruby), Python emerges as the fastest.
Overall, we confirm that the distinction between $T_{\textsc{perf}}$ and $T_{\textsc{scal}}$ tasks---which we dub ``everyday'' and ``computing-intensive''---is quite important to understand performance differences among languages.
On tasks $T_{\textsc{perf}}$, languages with an agile runtime, such as the scripting languages, or with natively efficient operations on lists and string, such as Haskell, may turn out to be efficient in practice.

\takehome{The distinction between ``everyday'' and ``computing-intensive'' workloads is important when assessing running-time performance. On ``everyday'' workloads, languages may be able to compete successfully regardless of their programming paradigm.} 

\rqpar{RQ4. Which programming languages use memory more efficiently?}

\begin{figure}[htb]
\centering
  \begin{tikzpicture}[
  lang/.style={draw=none,font=\footnotesize,inner sep=1pt,outer sep=1pt},
  align=center, xscale=0.5, yscale=0.3
  ]
  \node [lang] (C) at (0,0) {C};
  \node [lang] (Go) at (-1.8,0) {Go};
  \node [lang] (C-sharp) at (-5.5,0) {C\#};
  \node [lang] (F-Sharp) at (-9,3) {F\#};
  \node [lang] (Python) at (-9,0) {Python};
  \node [lang] (Ruby) at (-9,-3) {Ruby};
  \node [lang] (Java) at (-6.5,3) {Java};
  \node [lang] (Haskell) at (-13,3) {Haskell};
 
  \draw [->,,thin](Go) edge (C);
  \draw [->,densely dotted,thin](F-Sharp) edge (Python);
  \draw [->,,thick](C-sharp) edge (Go);
  \draw [->,,very thick](Java) edge (C-sharp);
  \draw [->,,thick](Ruby) edge (C-sharp);
  \draw [->,,thick,bend left](Haskell) edge (Java);
  \draw [->,densely dotted,thick](Haskell) edge (Ruby);
  \draw [->,,thin](Python) edge (Java);
  \draw [->,densely dotted,thin](Python) edge (Ruby);
\end{tikzpicture}
\caption{Comparison of maximum RAM used (by minimum).}
\label{fig:maxram_exit0_min_normrank_network-main}
\end{figure}
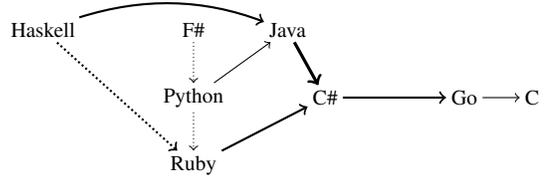

To answer this question, we measure the maximum RAM usage (i.e., maximum resident set size) of solutions of tasks $T_{\textsc{scal}}$ marked for comparison on computing-intensive tasks that run without errors or timeout; this measure includes the memory footprint of the runtime environments. 
%
\autoref{tab:maxram_exit0_min-main} shows the results of the statistical analysis, and \autoref{fig:maxram_exit0_min_normrank_network-main} the corresponding language relationship graph.

\begin{table}[ht]
\setlength{\tabcolsep}{4pt}
\begin{center}
{\scriptsize
\begin{tabular}{cc|rrrrrrr}
  \hline
\textsc{lang} & \textsc{} & \multicolumn{1}{c}{C} & \multicolumn{1}{c}{C\#} & \multicolumn{1}{c}{F\#} & \multicolumn{1}{c}{Go} & \multicolumn{1}{c}{Haskell} & \multicolumn{1}{c}{Java} & \multicolumn{1}{c}{Python} \\ 
  \hline
C\# & $p$ & \textcolor{colorstrong-main}{\vanish{4}} &  &  &  &  &  &  \\ 
   & $d$ & \textcolor{colorstrong-main}{2.022} &  &  &  &  &  &  \\ 
   & $R$ & -2.5 &  &  &  &  &  &  \\ 
   \hline
F\# & $p$ & \textcolor{colorstrong-main}{0.006} & \textcolor{colorstrong-main}{0.010} &  &  &  &  &  \\ 
   & $d$ & \textcolor{colorstrong-main}{0.761} & \textcolor{colorstrong-main}{1.045} &  &  &  &  &  \\ 
   & $R$ & -4.5 & -5.2 &  &  &  &  &  \\ 
   \hline
Go & $p$ & \textcolor{colorstrong-main}{\vanish{3}} & \textcolor{colorstrong-main}{\vanish{4}} & \textcolor{colorstrong-main}{0.006} &  &  &  &  \\ 
   & $d$ & \textcolor{colorweak-main}{0.064} & \textcolor{colormedium-main}{0.391} & \textcolor{colorstrong-main}{0.788} &  &  &  &  \\ 
   & $R$ & -1.8 & 4.1 & 3.5 &  &  &  &  \\ 
   \hline
Haskell & $p$ & \textcolor{colorstrong-main}{\vanish{3}} & 0.841 & 0.062 & \textcolor{colorstrong-main}{\vanish{3}} &  &  &  \\ 
   & $d$ & \textcolor{colorweak-main}{0.287} & \textcolor{colorweak-main}{0.123} & \textcolor{colormedium-main}{0.614} & \textcolor{colormedium-main}{0.314} &  &  &  \\ 
   & $R$ & -14.2 & -3.5 & 2.5 & -5.6 &  &  &  \\ 
   \hline
Java & $p$ & \textcolor{colorstrong-main}{\vanish{5}} & \textcolor{colorstrong-main}{\vanish{4}} & 0.331 & \textcolor{colorstrong-main}{\vanish{5}} & \textcolor{colorstrong-main}{0.007} &  &  \\ 
   & $d$ & \textcolor{colorstrong-main}{0.890} & \textcolor{colorstrong-main}{1.427} & \textcolor{colorweak-main}{0.278} & \textcolor{colormedium-main}{0.527} & \textcolor{colormedium-main}{0.617} &  &  \\ 
   & $R$ & -5.4 & -2.4 & 1.4 & -2.6 & 2.1 &  &  \\ 
   \hline
Python & $p$ & \textcolor{colorstrong-main}{\vanish{5}} & 0.351 & \textcolor{colorweak-main}{0.034} & \textcolor{colorstrong-main}{\vanish{4}} & 0.992 & \textcolor{colorstrong-main}{0.006} &  \\ 
   & $d$ & \textcolor{colormedium-main}{0.330} & \textcolor{colormedium-main}{0.445} & \textcolor{colorweak-main}{0.096} & \textcolor{colormedium-main}{0.417} & 0.010 & \textcolor{colorweak-main}{0.206} &  \\ 
   & $R$ & -5.0 & -2.2 & 1.0 & -2.7 & 2.9 & -1.1 &  \\ 
   \hline
Ruby & $p$ & \textcolor{colorstrong-main}{\vanish{5}} & \textcolor{colorstrong-main}{0.002} & 0.530 & \textcolor{colorstrong-main}{\vanish{4}} & \textcolor{colorweak-main}{0.049} & 0.222 & \textcolor{colorweak-main}{0.036} \\ 
   & $d$ & \textcolor{colormedium-main}{0.403} & \textcolor{colormedium-main}{0.525} & \textcolor{colorweak-main}{0.242} & \textcolor{colormedium-main}{0.531} & \textcolor{colormedium-main}{0.301} & \textcolor{colormedium-main}{0.301} & \textcolor{colorweak-main}{0.061} \\ 
   & $R$ & -5.0 & -5.0 & 1.5 & -2.5 & 1.8 & -1.1 & 1.0 \\ 
   \hline
\end{tabular}
}
\caption{Comparison of maximum RAM used (by minimum).}
\label{tab:maxram_exit0_min-main}
\end{center}
\end{table}

C and Go clearly emerge as the languages that make the most economical usage of RAM.
Go's frugal memory usage (on average only 1.8 times higher than C) is remarkable, given that its runtime includes garbage collection.
In contrast, all other languages use considerably more memory (2.5--14.2 times on average over either C or Go), which is justifiable in light of their bulkier runtimes, supporting not only garbage collection but also features such as dynamic binding (C\# and Java), lazy evaluation, pattern matching (Haskell and F\#), dynamic typing, and reflection (Python and Ruby).

Differences between languages in the same category (procedural, scripting, and functional) are generally small or insignificant.
The exception are object-oriented languages, where Java uses significantly more RAM than C\# (on average, 2.4 times more).
Among the other languages, Haskell emerges as the least memory-efficient, although some differences are insignificant.

While maximum RAM usage is a major indication of the efficiency of memory usage, modern architectures include many-layered memory hierarchies whose influence on performance is multi-faceted.
To complement the data about maximum RAM and refine our understanding of memory usage, we also measured \emph{average} RAM usage and number of \emph{page faults}.
Average RAM tends to be practically zero in all tasks but very few; correspondingly, the statistics are inconclusive as they are based on tiny samples.
By contrast, the data about page faults 
clearly partitions the languages in two classes: the functional languages trigger significantly more page faults than all other languages; in fact, the only statistically significant differences are those involving F\# or Haskell, whereas programs in other languages hardly ever trigger a single page fault.
The difference in page faults between Haskell programs and F\# programs is insignificant.
The page faults recorded in our experiments indicate that functional languages exhibit significant non-locality of reference.
The overall impact of this phenomenon probably depends on a machine's architecture; RQ3, however, showed that functional languages are generally competitive in terms of running-time performance, so that their non-local behavior might just denote a particular instance of the space vs.\ time trade-off.

\takehome{Procedural languages use significantly less memory than other languages. Functional languages make distinctly non-local memory accesses.}

\rqpar{RQ5. Which programming languages are less failure prone?}

To answer this question, we measure \emph{runtime failures} of solutions of tasks $T_{\textsc{exec}}$ marked for execution that compile without errors or timeout.
We exclude programs that time out because whether a timeout is indicative of failure depends on the task: for example, interactive applications will time out in our setup waiting for user input, but this should not be recorded as failure.
Thus, a terminating program \emph{fails} if it returns an exit code other than $0$ (for example, when throwing an uncaught exception).
The measure of failures is ordinal and not normalized: $\ell_f$ denotes a vector of binary values, one for each solution in language $\ell$ where we measure runtime failures; a value in $\ell_f$ is $1$ iff the corresponding program fails and it is $0$ if it does not fail.

Data about failures differs from that used to answer the other research questions in that we cannot aggregate it by task, since failures in different solutions, even for the same task, are in general unrelated.
Therefore, we use the Mann-Whitney $U$ test, an unpaired non-parametric ordinal test which can be applied to compare samples of different size.
For two languages $X$ and $Y$, the $U$ test assesses whether the two samples $X_f$ and $Y_f$ of binary values representing failures are likely to come from the same population.

\begin{table}[!hb]
\centering
\scriptsize
\setlength{\tabcolsep}{5pt}
\begin{tabular}{c rrrrrrrr}
&                C  &  C\#  & F\# & Go  & Haskell & Java & Python & Ruby \\
\hline
\# ran solutions &  391  & 246   & 215  & 389  & 376     & 297  & 675    & 516 \\
\% no error &   \pc{87}  & \pc{93}  & \pc{89} & \pc{98} & \pc{93}    & \pc{85} & \pc{85}   & \pc{86}  \\
\hline
\end{tabular}
\caption{Number of solutions that ran without timeout, and their percentage that ran without errors.}
\label{tab:failure-stats-main}
\end{table}

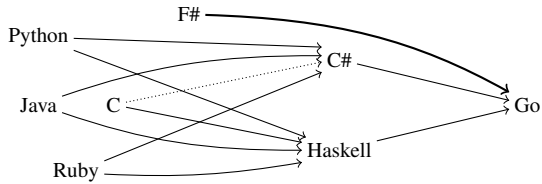
\begin{figure}[htb]
\centering
  \begin{tikzpicture}[
  lang/.style={draw=none,font=\footnotesize,inner sep=1pt,outer sep=1pt},
  align=center, xscale=0.5, yscale=0.3
  ]
  \node [lang] (Go) at (-2,0) {Go};
  \node [lang] (C-sharp) at (-7,2) {C\#};
  \node [lang] (Haskell) at (-7,-2) {Haskell};
  \node [lang] (F-Sharp) at (-11,4) {F\#};
  \node [lang] (C) at (-13,0) {C};
  \node [lang] (Ruby) at (-14,-3) {Ruby};
  \node [lang] (Java) at (-15,0) {Java};
  \node [lang] (Python) at (-15,3) {Python};

  \draw [->,thin](C-sharp) edge (Go);
  \draw [->,thin](Haskell) edge (Go);
  \draw [->,densely dotted,thin](C) edge ($(C-sharp.west)+(0,-3pt)$);
  \draw [->,thin](C) edge (Haskell);
  \draw [->,thin,bend left=15](Java) edge ($(C-sharp.west)+(0,5pt)$);
  \draw [->,thin,bend right=15](Java) edge (Haskell);
  \draw [->,thin](Ruby) edge (C-sharp.south west);
  \draw [->,thin,bend right=10](Ruby) edge (Haskell.south west);
  
  \draw [->,thin](Python) edge (C-sharp.north west);
  \draw [->,thin](Python) edge (Haskell);
  \draw [->,thick,bend left=20](F-Sharp) edge (Go.north west);
  \end{tikzpicture}
\caption{Comparisons of runtime failure proneness.}
\label{fig:rq5-faults-network}
\end{figure}

\autoref{tab:rq5-failures} shows the results of the tests; we do not report unstandardized measures of difference, such as $R$ in the previous tables, since they would be uninformative on ordinal data.
\autoref{fig:rq5-faults-network} is the corresponding language relationship graph. 
Horizontal distances are proportional to the fraction of solutions that run without errors (last row of \autoref{tab:failure-stats-main}).

\begin{table}[ht]
\setlength{\tabcolsep}{4pt}
\begin{center}
{\scriptsize
\begin{tabular}{cc|rrrrrrr}
  \hline
\textsc{lang} & \textsc{} & \multicolumn{1}{c}{C} & \multicolumn{1}{c}{C\#} & \multicolumn{1}{c}{F\#} & \multicolumn{1}{c}{Go} & \multicolumn{1}{c}{Haskell} & \multicolumn{1}{c}{Java} & \multicolumn{1}{c}{Python} \\ 
\hline
C\# & $p$ & \textcolor{colorweak-main}{0.037} &  &  &  &  &  &  \\ 
   & $d$ & \textcolor{colorweak-main}{0.170} &  &  &  &  &  &  \\ 
   \hline
F\# & $p$ & 0.500 & 0.204 &  &  &  &  &  \\ 
   & $d$ & \textcolor{colorweak-main}{0.057} & \textcolor{colorweak-main}{0.119} &  &  &  &  &  \\ 
   \hline
Go & $p$ & \textcolor{colorstrong-main}{\vanish{7}} & \textcolor{colorstrong-main}{0.011} & \textcolor{colorstrong-main}{\vanish{5}} &  &  &  &  \\ 
   & $d$ & \textcolor{colormedium-main}{0.410} & \textcolor{colorweak-main}{0.267} & \textcolor{colormedium-main}{0.398} &  &  &  &  \\ 
   \hline
Haskell & $p$ & \textcolor{colorstrong-main}{0.006} & 0.748 & 0.083 & \textcolor{colorstrong-main}{0.002} &  &  &  \\ 
   & $d$ & \textcolor{colorweak-main}{0.200} & 0.026 & \textcolor{colorweak-main}{0.148} & \textcolor{colorweak-main}{0.227} &  &  &  \\ 
   \hline
Java & $p$ & 0.386 & \textcolor{colorstrong-main}{0.006} & 0.173 & \textcolor{colorstrong-main}{\vanish{9}} & \textcolor{colorstrong-main}{\vanish{3}} &  &  \\ 
   & $d$ & \textcolor{colorweak-main}{0.067} & \textcolor{colorweak-main}{0.237} & \textcolor{colorweak-main}{0.122} & \textcolor{colormedium-main}{0.496} & \textcolor{colorweak-main}{0.271} &  &  \\ 
   \hline
Python & $p$ & {0.332} & \textcolor{colorstrong-main}{0.003} & {0.141} & \textcolor{colorstrong-main}{\vanish{10}} & \textcolor{colorstrong-main}{\vanish{3}} & {0.952} &  \\ 
   & $d$ & \textcolor{colorweak-main}{0.062} & \textcolor{colorweak-main}{0.222} & \textcolor{colorweak-main}{0.115} & \textcolor{colormedium-main}{0.428} & \textcolor{colorweak-main}{0.250} & {0.004} &  \\ 
   \hline
Ruby & $p$ & 0.589 & \textcolor{colorstrong-main}{0.010} & 0.260 & \textcolor{colorstrong-main}{\vanish{9}} & \textcolor{colorstrong-main}{\vanish{3}} & 0.678 & {0.658} \\ 
   & $d$ & 0.036 & \textcolor{colorweak-main}{0.201} & \textcolor{colorweak-main}{0.091} & \textcolor{colormedium-main}{0.423} & \textcolor{colorweak-main}{0.231} & 0.030 & {0.026} \\ 
   \hline
\end{tabular}
}
\caption{Comparisons of runtime failure proneness.} 
\label{tab:rq5-failures}
\end{center}
\end{table}

\iflong
\begin{table}[ht]
\centering
\scriptsize
\setlength{\tabcolsep}{4pt}
\begin{tabular}{c rrrrrrrr}
&                       C  &  C\#  & F\# & Go  & Haskell & Java & Python & Ruby \\
\hline
\# comp.\ solutions &  524 & 354  & 254  & 497 & 519     & 446   & 775   & 581  \\
\% no error & \pc{85} & \pc{90} & \pc{95} & \pc{89} & \pc{84} & \pc{78} & \pc{100} & \pc{100} \\
\hline
\end{tabular}
\caption{Number of solutions considered for compilation, and their percentage that compiled without errors.}
\label{tab:failure-stats-main-compile}
\end{table}
\fi

Go clearly sticks out as the least failure prone language.
\iflong If we look, in \autoref{tab:failure-stats-main-compile}, at the fraction of solutions that failed to compile, and hence didn't contribute data to failure analysis, Go is not significantly different from other compiled languages. \fi %
Together, these two elements indicate that the Go compiler is particularly good at catching sources of failures at compile time, since only a small fraction of compiled programs fail at runtime.
Go's restricted type system (no inheritance, no overloading, no genericity, no pointer arithmetic) may help make compile-time checks effective.
By contrast, the scripting languages tend to be the most failure prone of the lot.
This is a consequence of Python and Ruby being \emph{interpreted} languages\footnote{Even if Python compiles to bytecode, the translation process only performs syntactic checks (and is not invoked separately normally anyway).}: any syntactically correct program is executed, and hence most errors manifest themselves only at runtime.

There are few major differences among the remaining \emph{compiled} languages, where it is useful to distinguish between weak (C) and strong (the other languages) type systems~\cite[Sec.~3.4.2]{GJ-PLconcepts}.
F\# shows no statistically significant differences with any of C, C\#, and Haskell.
C tends to be more failure prone than C\# and is significantly more failure prone than Haskell; similarly to the explanation behind the interpreted languages' failure proneness, C's weak type system may be responsible for fewer failures being caught at compile time than at runtime.
In fact, the association between weak typing and failure proneness was also found in other studies~\cite{GitHubstudy-fse2014}.
Java is unusual in that it has a strong type system and is compiled, but is significantly more error prone than Haskell and C\#, which also are strongly typed and compiled.
Future work will determine if Java's behavior is spurious or indicative of concrete issues.

\takehome{Compiled strongly-typed languages are significantly less prone to runtime failures than interpreted or weakly-typed languages, since more errors are caught at compile time. Thanks to its simple static type system, Go is the least failure-prone language in our study.}

\section{Discussion}
\label{sec:implications}

The results of our study can help different stakeholders---developers, language designers, and educators---to make better informed choices about language usage and design.

The conciseness of functional and scripting programming languages suggests that the characterizing features of these languages---such as list comprehensions, type polymorphism, dynamic typing, and extensive support for reflection and list and map data structures---provide for great expressiveness.
In times where more and more languages combine elements belonging to different paradigms, language designers can focus on these features to improve the expressiveness and raise the level of abstraction.
For programmers, using a programming language that makes for concise code can help write software with fewer bugs.
In fact, some classic research suggests~\cite{CapersJones,McConnell,LesHatton889} that bug density is largely constant across programming languages---all else being equal~\cite{FentonN99,MohagheghiCKS04}; therefore, shorter programs will tend to have fewer bugs.

The results about executable size are an instance of the ubiquitous space vs.\ time trade-off.
Languages that compile to native can perform more aggressive compile-time optimizations since they produce code that is very close to the actual hardware it will be executed on.
\iflong
In fact, compilers to native tend to have several optimization options, which exercise different trade-offs.
GNU's \verb|gcc|, for instance, has a \verb|-Os| flag that optimizes for executable size instead of speed (but we didn't use this highly specialized optimization in our experiments).
\fi
However, with the ever increasing availability of cheap and compact memory, differences between languages have significant implications only for applications that run on highly constrained hardware such as embedded devices\iflong (where, in fact, bytecode languages are becoming increasingly common)\fi.
Interpreted languages such as Ruby exercise yet another trade-off, where there is no visible binary at all and all optimizations are done at runtime.

No one will be surprised by our results that C dominates other languages in terms of raw speed and efficient memory usage.
Major progresses in compiler technology notwithstanding, higher-level programming languages do incur a noticeable performance loss to accommodate features such as automatic memory management or dynamic typing in their runtimes.
Nevertheless, our results on ``everyday'' workloads showed that pretty much any language can be competitive when it comes to the regular-size inputs that make up the overwhelming majority of programs.
When teaching and developing software, we should then remember that ``most applications do not actually need better performance than Python offers''~\cite[p.~337]{ESR-unix}.

Another interesting lesson emerging from our performance measurements is how Go achieves respectable running times as well as good results in memory usage, thereby distinguishing itself from the pack just as C does (in fact, Go's developers include prominent figures\iflong---Ken Thompson, most notably---\fi{}who were also primarily involved in the development of C).
Go's design choices may be traced back to a careful selection of features that differentiates it from most other language designs (which tend to be more feature-prodigal): while it offers automatic memory management and some dynamic typing, it deliberately omits genericity and inheritance, and offers only a limited support for exceptions.
In our study, we have seen that Go features not only good performance but also a compiler that is quite effective at finding errors at compile time rather than leaving them to leak into runtime failures.
Besides being appealing for certain kinds of software development (Go's concurrency mechanisms, which we didn't consider in this study, may be another feature to consider), Go also shows to language designers that there still is uncharted territory in the programming language landscape\iflong, and innovative solutions could be discovered that are germane to requirements in certain special domains\fi.

Evidence in our, as well as others' (\autoref{sec:related-work}), analysis tends to confirm what advocates of static strong typing have long claimed: that it makes it possible to catch more errors earlier, at compile time.
But the question remains of which process leads to overall higher programmer productivity (or, in a different context, to effective learning): postponing testing and catching as many errors as possible at compile time, or running a prototype as soon as possible while frequently going back to fixing and refactoring?
The traditional knowledge that bugs are more expensive to fix the later they are detected is not an argument against the ``test early'' approach, since testing early may be the quickest way to find an error in the first place.
This is another area where new trade-offs can be explored by selectively---or flexibly~\cite{DBLP:conf/icse/BayneCE11}---combining features\iflong that enhance compilation or execution\fi.

\section{Threats to Validity}
\label{sec:threats-validity}

Threats to \emph{construct validity}---are we measuring the right things?---are quite limited given that our research questions, and the measures we take to answer them, target widespread well-defined features (conciseness, performance, and so on) with straightforward matching measures (lines of code, running time, and so on).
A partial exception is RQ5, which targets the multifaceted notion of failure proneness, but the question and its answer are consistent with related empirical work that approached the same theme from other angles, which reflects positively on the soundness of our constructs. Regarding conciseness, lines of code remains a widely used metric, but it will be interesting to correlate it with other proposed measures of conciseness.

We took great care in the study's design and execution to minimize threats to \emph{internal validity}---are we measuring things right?
We manually inspected all task descriptions to ensure that the study only includes well-defined tasks and comparable solutions.
We also manually inspected, and modified whenever necessary, all solutions used to measure performance, where it is of paramount importance that the same inputs be applied in every case.
To ensure reliable runtime measures (running time, memory usage, and so on), we ran every executable multiple times, checked that each repeated run's deviation from the average is moderate (less than one standard deviation), and based our statistics on the average (mean) behavior.
Data analysis often showed highly statistically significant results, which also reflects favorably on the soundness of the study's data.
Our experimental setup tried to use \iflong \emph{standard} tools with \fi \emph{default} settings; this may limit the scope of our findings, but also helps reduce bias\iflong due to different familiarity with different languages\fi.
Exploring different directions, such as pursuing the best optimizations possible in each language~\cite{nanz-west-silveira:2013:benchmarking}\iflong for each task\fi, is an interesting goal of future work.

A possible threat to \emph{external validity}---do the findings generalize?---has to do with whether the properties of \rc programs are representative of real-world software projects.
On one hand, \rc tasks tend to favor algorithmic problems, and solutions are quite small on average compared to any realistic application or library.
On the other hand, every large project is likely to include a small set of core functionalities whose quality, performance, and reliability significantly influences the whole system's; \rc programs can be indicative of such core functionalities.
In addition, measures of performance are meaningful only on comparable implementations of algorithmic tasks, and hence \rc's algorithmic bias helped provide a solid base for comparison of this aspect (\autoref{sec:task-selection} and RQ3,4).
The size and level of activity of the \rc community mitigates the threat that contributors to \rc are not representative of the skills \iflong and expertise \fi of real-world programmers.
However, to ensure wider generalizability, we plan to analyze other characteristics of the \rc programming community. 
\iflong To sum up, while some quantitative details of our results may vary on different codebases and methodologies, the big, mainly qualitative, picture, is likely robust.\fi

Another potential threat comes from the choice of programming languages.
\autoref{sec:language-selection} describes how we selected languages representative of real-world popularity among major paradigms.
Classifying programming languages into paradigms has become harder in recent times, when multi-paradigm languages are the norm\iflong(many programming languages offer procedures, some form of object system, and even functional features such as closures and list comprehensions)\fi. \iflong\footnote{At the 2012 LASER summer school on ``Innovative languages for software engineering'', Mehdi Jazayeri mentioned the proliferation of multi-paradigm languages as a disincentive to updating his book on programming language concepts~\cite{GJ-PLconcepts}.}\fi
Nonetheless, we maintain that paradigms still significantly influence the way in which programs are written, and it is natural to associate major programming languages to a specific paradigm based on their \rc programs.
For example, even though Python offers classes and other object-oriented features, practically no solutions in \rc use them.
Extending the study to more languages and new paradigms belongs to future work.

\section{Related Work}
\label{sec:related-work}


\iflong
\textbf{Controlled experiments} are a popular approach to language comparisons: study participants program the same tasks in different languages while researchers measure features such as code size and execution or development time.
\fi
Prechelt~\cite{Prechelt:2000:ECS:619056.621567} compares 7 programming languages on a single task in 80 solutions written by students\iflong and other volunteers\fi. \iflong Measures include program size, execution time, memory consumption, and development time. \fi Findings include: the program written in Perl, Python, REXX, or Tcl is ``only half as long'' as written in C, C++, or Java; \iflong performance results are more mixed, but \fi C and C++ are generally faster than Java. The study asks questions similar to ours but is limited by the small sample size. Languages and their compilers have evolved since 2000 (when~\cite{Prechelt:2000:ECS:619056.621567} was published), making the results difficult to compare; however, some tendencies (conciseness of scripting languages, performance-dominance of C) are visible in our study too.
Harrison et al.~\cite{harrison:1996:comparing} compare the code quality of C++ against \iflong the functional language \fi SML's on 12 tasks, finding few significant differences. Our study targets a broader set of research questions (only RQ5 is related to quality). Hanenberg~\cite{DBLP:conf/oopsla/Hanenberg10} conducts a study with 49 students over 27 hours of development time comparing static vs.\ dynamic type systems, finding no significant differences. In contrast to controlled experiments, our approach cannot take development time into account.

\iflong Many recent comparative studies have targeted programming languages for concurrency and parallelism. \fi
%
Studying 15 students on a single problem, Szafron and Schaeffer~\cite{Szafron94experimentallyassessing} identify a message-passing library that is somewhat superior to higher-level parallel programming, even though the latter is more ``usable'' overall. This highlights the difficulty of reconciling results of different metrics. We do not attempt this in our study, as the suitability of a language for certain projects may depend on external factors\iflong that assign different weights to different metrics\fi.
Other studies~\cite{Cantonnet04productivityanalysis,Hochstein:2005:PPP:1105760.1105800,Ebcioglu06experiment,Hochstein:2008:PSC:1412750.1412840} compare parallel programming approaches (UPC, MPI, OpenMP, and X10) using mostly small student populations.
%
%
%
%
In the realm of concurrent programming, 
a study~\cite{rossbach-et-al:2010:transactional} with 237 undergraduate students implementing one program with locks, monitors, or transactions  suggests that transactions leads to the fewest errors. 
In a usability study with 67 students~\cite{nanz-et-al:2011:design}, we find advantages of the SCOOP concurrency model over Java's monitors.
Pankratius et al.~\cite{Pankratius:2012:CFI:2337223.2337238} compare Scala and Java using 13 students and one software engineer working on three tasks. 
They conclude that Scala's functional style leads to more compact code and comparable performance.
To eschew the limitations of classroom studies---based on the unrepresentative performance of novice programmers\iflong\ (for instance, in~\cite{Ebcioglu06experiment}, about a third of the student subjects fail the parallel programming task in that they cannot achieve any speedup)\fi{}---previous work of ours~\cite{nanz-west-silveira:2013:benchmarking,nanz-et-al:2013:examining} compared Chapel, Cilk, Go, and TBB on 96 solutions to 6 tasks that were checked for style and performance by \iflong notable \fi language experts. \iflong \cite{nanz-west-silveira:2013:benchmarking,nanz-et-al:2013:examining} also introduced language dependency diagrams similar to those used in the present paper.\fi 

A common problem with all the aforementioned studies is that they often target few tasks and solutions, and therefore fail to achieve statistical significance or generalizability. The large sample size in our study minimizes these problems.

\iflong
\textbf{Surveys} can help characterize the perception of programming languages. 
\fi
Meyerovich and Rabkin~\cite{Meyerovich:2013:EAP:2509136.2509515} study the reasons behind language adoption. One key finding is that the intrinsic features of a language (such as reliability) are less important for adoption when compared to extrinsic ones such as existing code\iflong, open-source libraries, and previous experience\fi. This puts our study into perspective, and shows that some features we investigate are very important to developers (e.g., performance\iflong as second most important attribute\fi).
Bissyand{\'e} et al.~\cite{Bissyande:2013:PII:2546398.2546404} study similar questions\iflong: the popularity, interoperability, and impact of languages\fi. Their rankings, according to lines of code or usage in projects, may suggest alternatives to the TIOBE ranking we used\iflong for selecting languages\fi. 

\iflong
\textbf{Repository mining}, as we have done in this study, has become a customary approach to answering a variety of questions about programming languages. 
\fi
Bhattacharya and Neamtiu~\cite{Bhattacharya:2011:APL:1985793.1985817} study 4 projects in C and C++ to understand the impact on software quality, finding an advantage in C++.
With similar goals, Ray et al.~\cite{GitHubstudy-fse2014} mine 729 projects in 17 languages from GitHub. They find that strong typing is modestly better than weak typing, and functional languages have an advantage over procedural languages. Our study looks at a broader spectrum of research questions in a more controlled environment, but our results on failures (RQ5) confirm the superiority of statically strongly typed languages.
Other studies investigate specialized features of programming languages. For example, recent studies by us~\cite{EFNPM-FM14-Coat} and others~\cite{writingcontracts} investigate the use of contracts and their interplay with other language features such as inheritance.
Okur and Dig~\cite{okur:2012:libraries} analyze 655 open-source applications with parallel programming to identify adoption trends and usage problems, addressing questions that are orthogonal to ours.

\section{Conclusions}
\label{sec:conclusion}

Programming languages are essential tools for the working computer scientist, and it is no surprise that what is the ``right tool for the job'' can be the subject of intense debates. To put such debates on strong foundations, we must understand how features of different languages relate to each other. Our study revealed differences regarding some of the most frequently discussed language features---conciseness, performance, failure-proneness---and is therefore of value to software developers and language designers.
The key to having highly significant statistical results in our study was the use of a large program chrestomathy: \rc. 
The repository can be a valuable resource also for future programming language research that corroborates, or otherwise complements, our findings.
\iflong
Besides using \rc, researchers can also improve it (by correcting any detected errors) and can increase its research value (by maintaining easily accessible up-to-date statistics). 
\fi

\textbf{Acknowledgments.}
Thanks to \rc's Mike Mol for helpful replies to our questions about the repository. 
Comments by Donald ``Paddy'' McCarthy helped us fix a problem with one of the analysis scripts for Python.
Other comments by readers of Slashdot, Reddit, and Hacker News were also valuable---occasionally even gracious. 
We thank members of the Chair of Software Engineering for their helpful feedback on a draft of this paper.
This work was partially supported by ERC grant CME \#291389.



\iflong
\clearpage
\newpage
\onecolumn
\def\GraphWidth{0.35\textwidth}
\def\AllGraphWidth{0.5\textwidth}

\newcommand{\GraphC}[2]{%
\begin{figure}[!p]
\begin{center}
\begin{tabular}{cc}
\includegraphicsifexists[width=\GraphWidth]{./#1_C-C-sharp.pdf}
&
\includegraphicsifexists[width=\GraphWidth]{./#1_C-F-Sharp.pdf} \\
\includegraphicsifexists[width=\GraphWidth]{./#1_C-Go.pdf}
&
\includegraphicsifexists[width=\GraphWidth]{./#1_C-Haskell.pdf} \\
\includegraphicsifexists[width=\GraphWidth]{./#1_C-Java.pdf}
&
\includegraphicsifexists[width=\GraphWidth]{./#1_C-Python.pdf} \\
\includegraphicsifexists[width=\GraphWidth]{./#1_C-Ruby.pdf}
\end{tabular}
\end{center}
\caption{#2}
\label{fig:#1:C}
\end{figure}%

\begin{figure}[!p]
\begin{center}
\begin{tabular}{cc}
\includegraphicsifexists[width=\GraphWidth]{./#1_C-C-sharp_scatter.pdf}
&
\includegraphicsifexists[width=\GraphWidth]{./#1_C-F-Sharp_scatter.pdf} \\
\includegraphicsifexists[width=\GraphWidth]{./#1_C-Go_scatter.pdf}
&
\includegraphicsifexists[width=\GraphWidth]{./#1_C-Haskell_scatter.pdf} \\
\includegraphicsifexists[width=\GraphWidth]{./#1_C-Java_scatter.pdf}
&
\includegraphicsifexists[width=\GraphWidth]{./#1_C-Python_scatter.pdf} \\
\includegraphicsifexists[width=\GraphWidth]{./#1_C-Ruby_scatter.pdf}
\end{tabular}
\end{center}
\caption{#2}
\label{fig:#1:C-scatter}
\end{figure}%
}

\newcommand{\GraphCsharp}[2]{%
\begin{figure}[!p]
\begin{center}
\begin{tabular}{cc}
\includegraphicsifexists[width=\GraphWidth]{./#1_C-sharp-F-Sharp.pdf}
&
\includegraphicsifexists[width=\GraphWidth]{./#1_C-sharp-Go.pdf} \\
\includegraphicsifexists[width=\GraphWidth]{./#1_C-sharp-Haskell.pdf}
&
\includegraphicsifexists[width=\GraphWidth]{./#1_C-sharp-Java.pdf} \\
\includegraphicsifexists[width=\GraphWidth]{./#1_C-sharp-Python.pdf}
&
\includegraphicsifexists[width=\GraphWidth]{./#1_C-sharp-Ruby.pdf}
\end{tabular}
\end{center}
\caption{#2}
\label{fig:#1:C-sharp}
\end{figure}%

\begin{figure}[!p]
\begin{center}
\begin{tabular}{cc}
\includegraphicsifexists[width=\GraphWidth]{./#1_C-sharp-F-Sharp_scatter.pdf}
&
\includegraphicsifexists[width=\GraphWidth]{./#1_C-sharp-Go_scatter.pdf} \\
\includegraphicsifexists[width=\GraphWidth]{./#1_C-sharp-Haskell_scatter.pdf}
&
\includegraphicsifexists[width=\GraphWidth]{./#1_C-sharp-Java_scatter.pdf} \\
\includegraphicsifexists[width=\GraphWidth]{./#1_C-sharp-Python_scatter.pdf}
&
\includegraphicsifexists[width=\GraphWidth]{./#1_C-sharp-Ruby_scatter.pdf}
\end{tabular}
\end{center}
\caption{#2}
\label{fig:#1:C-sharp-scatter}
\end{figure}%
}

\newcommand{\GraphFSharp}[2]{%
\begin{figure}[!p]
\begin{center}
\begin{tabular}{cc}
\includegraphicsifexists[width=\GraphWidth]{./#1_F-Sharp-Go.pdf}
&
\includegraphicsifexists[width=\GraphWidth]{./#1_F-Sharp-Haskell.pdf} \\
\includegraphicsifexists[width=\GraphWidth]{./#1_F-Sharp-Java.pdf}
&
\includegraphicsifexists[width=\GraphWidth]{./#1_F-Sharp-Python.pdf} \\
\includegraphicsifexists[width=\GraphWidth]{./#1_F-Sharp-Ruby.pdf}
\end{tabular}
\end{center}
\caption{#2}
\label{fig:#1:F-Sharp}
\end{figure}%

\begin{figure}[!p]
\begin{center}
\begin{tabular}{cc}
\includegraphicsifexists[width=\GraphWidth]{./#1_F-Sharp-Go_scatter.pdf}
&
\includegraphicsifexists[width=\GraphWidth]{./#1_F-Sharp-Haskell_scatter.pdf} \\
\includegraphicsifexists[width=\GraphWidth]{./#1_F-Sharp-Java_scatter.pdf}
&
\includegraphicsifexists[width=\GraphWidth]{./#1_F-Sharp-Python_scatter.pdf} \\
\includegraphicsifexists[width=\GraphWidth]{./#1_F-Sharp-Ruby_scatter.pdf}
\end{tabular}
\end{center}
\caption{#2}
\label{fig:#1:F-Sharp-scatter}
\end{figure}%
}

\newcommand{\GraphGo}[2]{%
\begin{figure}[!p]
\begin{center}
\begin{tabular}{cc}
\includegraphicsifexists[width=\GraphWidth]{./#1_Go-Haskell.pdf}
&
\includegraphicsifexists[width=\GraphWidth]{./#1_Go-Java.pdf} \\
\includegraphicsifexists[width=\GraphWidth]{./#1_Go-Python.pdf}
&
\includegraphicsifexists[width=\GraphWidth]{./#1_Go-Ruby.pdf}
\end{tabular}
\end{center}
\caption{#2}
\label{fig:#1:Go}
\end{figure}%

\begin{figure}[!p]
\begin{center}
\begin{tabular}{cc}
\includegraphicsifexists[width=\GraphWidth]{./#1_Go-Haskell_scatter.pdf}
&
\includegraphicsifexists[width=\GraphWidth]{./#1_Go-Java_scatter.pdf} \\
\includegraphicsifexists[width=\GraphWidth]{./#1_Go-Python_scatter.pdf}
&
\includegraphicsifexists[width=\GraphWidth]{./#1_Go-Ruby_scatter.pdf}
\end{tabular}
\end{center}
\caption{#2}
\label{fig:#1:Go-scatter}
\end{figure}%
}

\newcommand{\GraphHaskell}[2]{%
\begin{figure}[!p]
\begin{center}
\begin{tabular}{cc}
\includegraphicsifexists[width=\GraphWidth]{./#1_Haskell-Java.pdf}
&
\includegraphicsifexists[width=\GraphWidth]{./#1_Haskell-Python.pdf} \\
\includegraphicsifexists[width=\GraphWidth]{./#1_Haskell-Ruby.pdf}
\end{tabular}
\end{center}
\caption{#2}
\label{fig:#1:Haskell}
\end{figure}%

\begin{figure}[!p]
\begin{center}
\begin{tabular}{cc}
\includegraphicsifexists[width=\GraphWidth]{./#1_Haskell-Java_scatter.pdf}
&
\includegraphicsifexists[width=\GraphWidth]{./#1_Haskell-Python_scatter.pdf} \\
\includegraphicsifexists[width=\GraphWidth]{./#1_Haskell-Ruby_scatter.pdf}
\end{tabular}
\end{center}
\caption{#2}
\label{fig:#1:Haskell-scatter}
\end{figure}%
}

\newcommand{\GraphJava}[2]{%
\begin{figure}[!p]
\begin{center}
\begin{tabular}{cc}
\includegraphicsifexists[width=\GraphWidth]{./#1_Java-Python.pdf}
&
\includegraphicsifexists[width=\GraphWidth]{./#1_Java-Ruby.pdf}
\end{tabular}
\end{center}
\caption{#2}
\label{fig:#1:Java}
\end{figure}%

\begin{figure}[!p]
\begin{center}
\begin{tabular}{cc}
\includegraphicsifexists[width=\GraphWidth]{./#1_Java-Python_scatter.pdf}
&
\includegraphicsifexists[width=\GraphWidth]{./#1_Java-Ruby_scatter.pdf}
\end{tabular}
\end{center}
\caption{#2}
\label{fig:#1:Java-scatter}
\end{figure}%
}

\newcommand{\GraphPython}[2]{%
\begin{figure}[!p]
\begin{center}
\begin{tabular}{cc}
\includegraphicsifexists[width=\GraphWidth]{./#1_Python-Ruby.pdf}
\end{tabular}
\end{center}
\caption{#2}
\label{fig:#1:Python}
\end{figure}%

\begin{figure}[!p]
\begin{center}
\begin{tabular}{cc}
\includegraphicsifexists[width=\GraphWidth]{./#1_Python-Ruby_scatter.pdf}
\end{tabular}
\end{center}
\caption{#2}
\label{fig:#1:Python-scatter}
\end{figure}%
}

\newcommand{\GraphAll}[2]{%
\begin{figure}[!p]
\begin{center}
\begin{tabular}{cc}
\includegraphicsifexists[width=\AllGraphWidth]{./#1_all.pdf}
\end{tabular}
\end{center}
\caption{#2}
\label{fig:#1:all}
\end{figure}%
}

\newcommand{\GraphMeasure}[2]{%
\GraphC{#1}{#2 (C vs.\ other languages)}
\GraphCsharp{#1}{#2 (C\# vs.\ other languages)}
\GraphFSharp{#1}{#2 (F\# vs.\ other languages)}
\GraphGo{#1}{#2 (Go vs.\ other languages)}
\GraphHaskell{#1}{#2 (Haskell vs.\ other languages)}
\GraphJava{#1}{#2 (Java vs.\ other languages)}
\GraphPython{#1}{#2 (Python  vs.\ other languages)}
\GraphAll{#1}{#2 (all languages)}%
}

\newcommand{\NetworkFig}[3][]{%
\begin{figure}[!h]
\centering
  \begin{tikzpicture}[
  lang/.style={draw=none,font=\footnotesize,inner sep=1pt,outer sep=1pt},
  align=center, xscale=0.4, yscale=0.5
  ]
  \input{./#2_network.tex}
  \end{tikzpicture}
\caption{#3}
\label{fig:#2_network}
\end{figure}

\IfFileExists{./#2_normrank_network.tex}{%
\begin{figure}[!h]
\centering
  \begin{tikzpicture}[
  lang/.style={draw=none,font=\footnotesize,inner sep=1pt,outer sep=1pt},
  align=center, xscale=0.4, yscale=0.5
  ]
  \input{./#2_normrank_network.tex}
  \end{tikzpicture}
\caption{#3 (normalized horizontal distances)}
\label{fig:#2_normrank_network}
\end{figure}
}{}

\IfFileExists{./#1_normrank_network.tex}{%
\begin{figure}[!h]
\centering
  \begin{tikzpicture}[
  lang/.style={draw=none,font=\footnotesize,inner sep=1pt,outer sep=1pt},
  align=center, xscale=0.4, yscale=0.5
  ]
  \input{./#1_normrank_network.tex}
  \end{tikzpicture}
\caption{#3 (normalized horizontal distances)}
\label{fig:#1_normrank_network}
\end{figure}%
}{}
}

\tableofcontents
\listoffigures
\listoftables

\pagebreak

For all data processing we used R version 2.14.1.
The Wilcoxon signed-rank test and the Mann-Whitney $U$-test were performed using package \verb|coin| version 1.0-23, except for the test statistics $W$ and $U$ that were computed using R's standard function \verb|wilcox.test|; Cohen's $d$ calculations were performed using package \verb|lsr| version 0.3.2.

\section{Appendix: Pairwise comparisons} \label{sec:app:pairwise}

Sections~\ref{sec:app:conciseness} to~\ref{sec:app:n-solutions} describe the complete measured, rendered as graphs and tables, for a number of pairwise comparisons between programming languages; the actual graphs and table appear in the remaining parts of this appendix.

Each comparison targets a different metric $M$, including lines of code (conciseness), lines of comments per line of code (comments), binary size (in kilobytes, where binaries may be native or byte code), CPU user time (in seconds, for different sets of \emph{performance} $T_{\text{\textsc{perf}}}$---``everyday'' in the main paper---and \emph{scalability} $T_{\text{\textsc{scal}}}$---``computing-intensive'' in the main paper---tasks), maximum RAM usage (i.e., maximum resident set size, in kilobytes), number of page faults, time outs (with a timeout limit of 3 minutes), and number of \rc solutions for the same task.
Most metrics are \emph{normalized}, as we detail in the subsections.

A metric may also be such that \emph{smaller is better} (such as lines of code: the fewer the more concise a program is) or \emph{larger is better} (such as comments per line of code: the more the more comments are available).
Indeed, comments per line of code and number of solutions per task are ``larger is better'' metrics; all other metrics are ``smaller is better''.
We discuss below how this feature influences how the results should be read.

Let $\ell$ be a programming language, $t$ a task, and $M$ a metric.
$\ell_M(t)$ denotes the vector of measures of $M$, one for each solution to task $t$ in language $\ell$.
$\ell_M(t)$ may be empty if there are no solutions to task $t$ in $\ell$.

Using this notation, the comparison of programming languages $X$ and $Y$ based on $M$ works as follows.
Consider a subset $T$ of the tasks such that, for every $t \in T$, both $X$ and $Y$ have at least one solution to $t$.
$T$ may be further restricted based on a measure-dependent \emph{criterion}, which we describe in the following subsections; for example, \autoref{sec:app:conciseness} only considers a task $t$ if both $X$ and $Y$ have at least one solution that compiles without errors (solutions that do not satisfy the criterion are discarded).
Based on $T$, we build two data vectors $x_M^\alpha$ and $y_M^\alpha$ for the two languages by aggregating metric $M$ per task using an \emph{aggregation function} $\alpha$.

To this end, if $M$ is normalized, the normalization factor $\nu_M(t, X, Y)$ denotes the smallest value of $M$ for all solutions of $t$ in $X$ and in $Y$; otherwise it is just one:
\[
\nu_M(t, X, Y) \quad=\quad
\begin{cases}
\min\left( X_M(t) Y_M(t) \right)  
   &  \text{if }M\text{ is normalized and } \min( X_M(t) Y_M(t)) > 0\,, \\
1  & \text{otherwise}\,,
\end{cases}
\]
where juxtaposing vectors denotes concatenating them.
Note that the normalization factor is one also if $M$ is normalized but the minimum is zero; this is to avoid divisions by zero when normalizing.
(A minimum of zero may occur due to the limited precision of some measures such as running time.)

We are finally ready to define vectors $x_M^\alpha$ and $y_M^\alpha$.
The vectors have the same length $|T| = |x_M^\alpha| = |y_M^\alpha|$ and are ordered by task; thus, $x_M^\alpha(t)$ and $y_M^\alpha(t)$ denote the value in $x_M^\alpha$ and in $y_M^\alpha$ corresponding to task $t$, for $t \in T$:
\[
\begin{split}
x_M^\alpha(t) & =     \alpha(X_M(t)/\nu_M(t, X, Y))\,, \\
y_M^\alpha(t) & =     \alpha(Y_M(t)/\nu_M(t, X, Y))\,.
\end{split}
\]
As aggregation functions, we normally consider both minimum and mean; hence the sets of graphs and tables are often double, one for each aggregation function.
For unstandardized measures, we also define the vectors $\mathsf{x}_M^\alpha$ and $\mathsf{y}_M^\alpha$ by task $t$ as non-normalized counterparts to $x_M^\alpha$ and $y_M^\alpha$:
\[
\begin{split}
\mathsf{x}_M^\alpha(t) & =     \alpha(X_M(t))\,, \\
\mathsf{y}_M^\alpha(t) & =     \alpha(Y_M(t))\,.
\end{split}
\]

The data in $x_M^\alpha$ and $y_M^\alpha$ determines two graphs and a statistical test.
\begin{itemize}

\item One graph includes \emph{line plots} of $x_M^\alpha$ and of $y_M^\alpha$, with the horizontal axis representing task number and the vertical axis representing values of $M$ (possibly normalized).

For example, \autoref{fig:loc_exit0_min:C} includes a graph with normalized values of lines of code aggregated per task by minimum for C and Python.
There you can see that there are close to 350 tasks with at least one solution in both C and Python that compiles successfully; and that there is a task whose shortest solution in C is over 50 times larger (in lines of code) than its shortest solution in Python.

\item Another graph is a \emph{scatter plot} of $x_M^\alpha$ and of $y_M^\alpha$, namely of points with coordinates $(x_M^\alpha(t), y_M^\alpha(t))$ for all available tasks $t \in T$. This graph also includes a linear regression line fitted using the least squares approach. Since axes have the same scales in these graphs, a linear regression line that bisects the graph diagonally at 45$^\circ$ would mean that there is no visible difference in metric $M$ between the two languages.
  Otherwise, if $M$ is such that ``smaller is better'', the flatter or lower the regression line, the better language $Y$ tends to be compared against language $X$ on metric $M$. In fact, a flatter or lower line denotes more points $(v_X, v_Y)$ with $v_Y < v_X$ than the other way round, or more tasks where $Y$ is better (smaller metric).
  Conversely, if $M$ is such that ``larger is better'', the steeper or higher the regression line, the better language $Y$ tends to be compared against language $X$ on metric $M$.

For example, \autoref{fig:loc_exit0_min:C-scatter} includes a graph with normalized values of lines of code aggregated per task by minimum for C and Python.
There you can see that most tasks are such that the shortest solution in C is larger than the shortest solution in Python; the regression line is almost horizontal at ordinate 1.

\item The \emph{statistical test} is a Wilcoxon signed-rank test, a paired non-parametric difference test which assesses whether the mean ranks of $x_M^\alpha$ and of $y_M^\alpha$ differ.
  The test results appear in a table, under column labeled with language $X$ at a row labeled with language $Y$, and includes various statistics:

  \begin{enumerate}
   \item The $p$-value estimates the probability that the differences between $x_M^\alpha$ and $y_M^\alpha$ are due to chance; thus, if $p$ is small (typically at least $p < 0.1$, but preferably $p \ll 0.01$) it means that there is a high chance that $X$ and $Y$ exhibit a genuinely different behavior with respect to metric $M$.
     Significant $p$-values are colored: \textcolor{colorstrong}{highly significant} ($p < 0.01$) and  \textcolor{colorweak}{significant but not highly} so (``tends to be significant'': $0.01 \leq p < 0.05$).

   \item The total sample size $N$ is $|x_M^\alpha| + |x_M^\alpha|$, that is twice the number of tasks considered for metric $M$.

   \item The test statistics $W$ is the absolute value of the sum of the signed ranks (see a description of the test for details).

   \item The related test statistics $Z$ is derivable from $W$.

   \item The effect size, computed as Cohen's $d$, which, for statistically significant differences, gives an idea of how large the difference is. As a rule of thumb, $d < 0.3$ denotes a small effect size, $0.3 \leq d < 0.7$ denotes a medium effect size, and $d \geq 0.7$ denotes a large effect size.
     Non-negligible effect sizes are colored: \textcolor{colorstrong}{large effect size}, \textcolor{colormedium}{medium effect size}, and \textcolor{colorweak}{small} (but non vanishing, that is $> 0.05$) \textcolor{colorweak}{effect size}.

   \item The difference $\Delta = \widetilde{\mathsf{x}_M^\alpha} - \widetilde{\mathsf{y}_M^\alpha}$ of the medians of non-normalized vectors $\mathsf{x}_M^\alpha$ and $\mathsf{x}_M^\alpha$, which gives an unstandardized measure and sign of the size of the overall difference. Namely, if $M$ is such that ``smaller is better'' and the difference between $X$ and $Y$ is significant, a positive $\Delta$ indicates that language $Y$ is on average better (smaller) on $M$ than language $X$.
     Conversely, if $M$ is such that ``larger is better'', a negative $\Delta$ indicates that language $Y$ is on average better (larger) on $M$ than language $X$.

     \item The ratio
\[
R \quad=\quad \sgn(\Delta)\,\frac{\max(\widetilde{\mathsf{x}_M^\alpha}, \widetilde{\mathsf{y}_M^\alpha})}{\min(\widetilde{\mathsf{x}_M^\alpha}, \widetilde{\mathsf{y}_M^\alpha})}
\]
of the largest median to the smallest median, with the same sign as $\Delta$. This is another unstandardized measure and sign of the size of the difference with a more direct interpretation in terms of ratio of average measures.
Note that $\sgn(v) = 1$ if $v \geq 0$; otherwise $\sgn(v) = -1$.
   \end{enumerate}

For example, \autoref{tab:loc_exit0_min} includes a cell comparing C (column header) against Python (row header) for normalized values of lines of code aggregated per task by minimum.
The $p$-value is practically zero, and hence the differences are highly significant.
The effect size is large ($d > 0.9$), and hence the magnitude of the differences is considerable.
Since the metric for conciseness is ``smaller is better'', a positive $\Delta$ indicates that Python is the more concise language on average; the value of $R$ further indicates that the average C solution is over $2.9$ times longer in lines of code than the average Python solution.
These figures quantitatively confirm what we observed in the line and scatter plots.
\end{itemize}

We also include a cumulative line plot with all languages at once, which is only meant as a qualitative visualization.

\subsection{Conciseness} \label{sec:app:conciseness}

The metric for conciseness is non-blank non-comment lines of code, counted using \verb|cloc| version 1.6.2.
The metric is normalized and smaller is better.
As aggregation functions we consider minimum `$\min$' and mean.
The criterion only selects solutions that compile successfully (compilation returns with exit status 0), and only include tasks $T_{\textsc{loc}}$ which we manually marked for lines of code count.

\subsection{Conciseness (all tasks)} \label{sec:app:conciseness-all}

The metric for conciseness on all tasks is non-blank non-comment lines of code, counted using \verb|cloc| version 1.6.2.
The metric is normalized and smaller is better.
As aggregation functions we consider minimum `$\min$' and mean.
The criterion only include tasks $T_{\textsc{loc}}$ which we manually marked for lines of code count (but otherwise includes all solutions, including those that do not compile correctly).

\subsection{Comments} \label{sec:app:comments}

The metric for comments is comment lines of code per non-blank non-comment lines of code, counted using \verb|cloc| version 1.6.2.
The metric is normalized and \emph{larger} is better.
As aggregation functions we consider minimum `$\min$' and mean.
The criterion only selects tasks $T_{\textsc{loc}}$ which we manually marked for lines of code count (but otherwise includes all solutions, including those that do not compile correctly).

\subsection{Binary size} \label{sec:app:binsize}

The metric for binary size is size of binary (in kilobytes), measured using GNU \verb|du| version 8.13.
The metric is normalized and smaller is better.
As aggregation functions we consider minimum `$\min$' and mean.
The criterion only selects solutions that compile successfully (compilation returns with exit status 0 and creates a non-empty binary), and only include tasks $T_{\textsc{comp}}$ which we manually marked for compilation.

The ``binary'' is either native code or byte code, according to the language.
Ruby does not feature in this comparison since it does not generate byte code, and hence the graphs and tables for this metric do not include Ruby.

\subsection{Performance} \label{sec:app:performance}

The metric for performance is CPU user time (in seconds), measured using GNU \verb|time| version 1.7.
The metric is normalized and smaller is better.
As aggregation functions we consider minimum `$\min$' and mean.
The criterion only selects solutions that execute successfully (execution returns with exit status 0), and only include tasks $T_{\textsc{perf}}$ which we manually marked for performance comparison.

We selected the performance tasks based on whether they represent well-defined comparable tasks were measuring performance makes sense, and we ascertained that all solutions used in the analysis indeed implement the task correctly (and the solutions are comparable, that is interpret the task consistently and run on comparable inputs).

\subsection{Scalability} \label{sec:app:scalability}

The metric for scalability is CPU user time (in seconds), measured using GNU \verb|time| version 1.7.
The metric is normalized and smaller is better.
As aggregation functions we consider minimum `$\min$' and mean.
The criterion only selects solutions that execute successfully (execution returns with exit status 0), and only include tasks $T_{\textsc{scal}}$ which we manually marked for scalability comparison.
\autoref{tab:scalability-tasks-appendix} lists the scalability tasks and describes the size $n$ of their inputs in the experiments.

We selected the scalability tasks based on whether they represent well-defined comparable tasks were measuring scalability makes sense.
We ascertained that all solutions used in the analysis indeed implement the task correctly (and the solutions are comparable, that is interpret the task consistently); and we modified the input to all solutions so that they are uniform across languages and represent challenging (or at least non-trivial) input sizes.

\begin{table}[htb]
  \centering
{\scriptsize
\setlength{\tabcolsep}{2pt}
  \begin{tabular}{rll}
   & \textsc{task name} & \textsc{input size} $n$ \\ \hline
 1 & 9 billion names of God the integer & $n = 10^5$ \\
 2 & Anagrams & $100\ \times$ \verb|unixdict.txt| (20.6 MB) \\
 3 & Anagrams/Deranged anagrams & $100\ \times$ \verb|unixdict.txt| (20.6 MB) \\
 4 & Arbitrary-precision integers (included) & $5^{4^{3^{2}}}$ \\
 5 & Combinations & $25 \choose 10$\\
 6 & Count in factors & $n = 10^6$ \\
 7 & Cut a rectangle & $10 \times 10$  \\
 8 & Extensible prime generator & $10^7$th prime \\
 9 & Find largest left truncatable prime in a given base & $10^7$th prime \\
10 & Hamming numbers & $10^7$th Hamming number \\
11 & Happy numbers & $10^6$th Happy number \\
12 & Hofstadter Q sequence & \# flips up to $10^5$th term  \\
13 & Knapsack problem/0-1 & input from Rosetta Code task description \\
14 & Knapsack problem/Bounded & input from Rosetta Code task description\\
15 & Knapsack problem/Continuous & input from Rosetta Code task description\\
16 & Knapsack problem/Unbounded & input from Rosetta Code task description \\
17 & Ludic numbers & input from Rosetta Code task description  \\
18 & LZW compression & $100\ \times$ \verb|unixdict.txt| (20.6 MB) \\
19 & Man or boy test & $n = 16$ \\
20 & N-queens problem & $n = 13$ \\
21 & Perfect numbers & first $5$ perfect numbers  \\
22 & Pythagorean triples & perimeter $< 10^8$ \\
23 & Self-referential sequence & $n = 10^6$ \\
24 & Semordnilap & $100\ \times$ \verb|unixdict.txt| \\
25 & Sequence of non-squares & non-squares $< 10^6$ \\
26 & Sorting algorithms/Bead sort & $n = 10^4$, nonnegative values $< 10^4$  \\
27 & Sorting algorithms/Bubble sort & $n = 3 \cdot 10^4$ \\
28 & Sorting algorithms/Cocktail sort & $n = 3 \cdot 10^4$ \\
29 & Sorting algorithms/Comb sort & $n = 10^6$ \\
30 & Sorting algorithms/Counting sort & $n = 2 \cdot 10^6$, nonnegative values $< 2 \cdot 10^6$ \\
31 & Sorting algorithms/Gnome sort & $n = 3 \cdot 10^4$ \\
32 & Sorting algorithms/Heapsort & $n = 10^6$ \\
33 & Sorting algorithms/Insertion sort & $n = 3 \cdot 10^4$ \\
34 & Sorting algorithms/Merge sort & $n = 10^6$\\
35 & Sorting algorithms/Pancake sort & $n = 3 \cdot 10^4$ \\
36 & Sorting algorithms/Quicksort & $n = 2 \cdot 10^6$ \\
37 & Sorting algorithms/Radix sort & $n = 2 \cdot 10^6$, nonnegative values $< 2 \cdot 10^6$ \\
38 & Sorting algorithms/Selection sort & $n = 3 \cdot 10^4$ \\
39 & Sorting algorithms/Shell sort & $n = 2 \cdot 10^6$ \\
40 & Sorting algorithms/Stooge sort & $n = 3 \cdot 10^3$ \\
41 & Sorting algorithms/Strand sort & $n = 3 \cdot 10^4$ \\
42 & Text processing/1 & input from Rosetta Code task description (1.2 MB) \\
43 & Text processing/2 & input from Rosetta Code task description (1.2 MB) \\
44 & Topswops & $n = 12$ \\
45 & Towers of Hanoi & $n = 25$  \\
46 & Vampire number & input from Rosetta Code task description \\
  \end{tabular}
}
  \caption{Names and input size of scalability tasks}
  \label{tab:scalability-tasks-appendix}
\end{table}

\subsection{Memory usage} \label{sec:app:maxram}

The metric for memory (RAM) usage is maximum resident set size (in kilobytes), measured using GNU \verb|time| version 1.7.
The metric is normalized and smaller is better.
As aggregation functions we consider minimum `$\min$' and mean.
The criterion only selects solutions that execute successfully (execution returns with exit status 0), and only include tasks $T_{\textsc{scal}}$ which we manually marked for scalability comparison.

\subsection{Page faults} \label{sec:app:pfs}

The metric for page faults is number of page faults in an execution, measured using GNU \verb|time| version 1.7.
The metric is normalized and smaller is better.
As aggregation functions we consider minimum `$\min$' and mean.
The criterion only selects solutions that execute successfully (execution returns with exit status 0), and only include tasks $T_{\textsc{scal}}$ which we manually marked for scalability comparison.

A number of tests could not be performed due to languages not generating any page faults (all pairs are ties). 
In those cases, the metric is immaterial.

\subsection{Timeouts} \label{sec:app:timeout}

The metric for timeouts is ordinal and two-valued: a solution receives a value of one if it times out within the allotted time; otherwise it receives a value of zero.
Time outs were detected using GNU \verb|timeout| version 8.13.
The metric is \emph{not} normalized and smaller is better.
As aggregation function we consider maximum `$\max$', corresponding to letting $\ell(t) = 1$ iff all selected solutions to task $t$ in language $\ell$ time out.
The criterion only selects solutions that either execute successfully (execution terminates and returns with exit status 0) or are still running at the timeout limit, and only include tasks $T_{\textsc{scal}}$ which we manually marked for scalability comparison.

The line plots for this metric are actually point plots for better readability.
Also for readability, the majority of tasks with the same value in the languages under comparison correspond to a different color (marked ``all'' in the legends).

\subsection{Solutions per task} \label{sec:app:n-solutions}

The metric for timeouts is a counting metric: each solution receives a value of one.
The metric is \emph{not} normalized and \emph{larger} is better.
As aggregation function we consider the sum; hence each task receives a value corresponding to the number of solutions in \rc for that task in each language.
The criterion only selects tasks $T_{\textsc{comp}}$ which we manually marked for compilation (but otherwise includes all solutions, including those that do not compile correctly).

The line plots for this metric are actually point plots for better readability.
Also for readability, the majority of tasks with the same value in the languages under comparison correspond to a different color (marked ``all'' in the legends).

\subsection{Other comparisons} \label{sec:app:other-comparisons}

\autoref{tab:make_stats_stat}, \autoref{tab:run_stats_stat}, \autoref{tab:misc_stats_make}, and \autoref{tab:misc_stats_run} display the results of additional statistics comparing programming languages.

\subsection{Compilation}  \label{sec:compilation-ordinal}
\autoref{tab:make_stats_stat} and \autoref{tab:misc_stats_make} give more details about the compilation process.

\autoref{tab:make_stats_stat} is similar to the previous tables, but it is based on \emph{unpaired} tests, namely the Mann-Whitney $U$ test---a non-parametric ordinal test that can be applied to two samples of different size.
We first consider all solutions to tasks $T_{\textsc{comp}}$ marked for compilation (regardless of compilation outcome).
We then assign an ordinal value to each \emph{solution}:
\begin{enumerate}[1: ]
\setcounter{enumi}{-1}
\item if the solution compiles without errors (the compiler returns with exit status 0 and, if applicable, creates a non-empty binary) with the default compilation options;
\item if the solution compiles without errors (the compiler returns with exit status 0 and, if applicable, creates a non-empty binary), but it requires to set a compilation flag to specify where to find \emph{libraries};
\item if the solution compiles without errors (the compiler returns with exit status 0 and, if applicable, creates a non-empty binary), but it requires to specify how to \emph{merge} or otherwise process multiple input files;
\item if the solution compiles without errors (the compiler returns with exit status 0 and, if applicable, creates a non-empty binary), but only after applying a \emph{patch}, which deploys some settings (such as include directives);
\item if the solution compiles without errors (the compiler returns with exit status 0 and, if applicable, creates a non-empty binary), but only after \emph{fixing} some simple error (such as a type error, or a missing variable declaration);
\item \label{compile:worst} if the solution does not compile or compiles with errors (the compiler returns with exit status other than 1 or, if applicable, creates no non-empty binary), even after applying possible patches or fixing.
\end{enumerate}

To make the categories disjoint, we assign the highest possible value in each case, reflecting the fact that the lower the ordinal value the better.
For example, if a solution requires a patch and a merge, we classify it as a patch, which characterizes the most effort involved in making it compile.

The distinction between patch and fixing is somewhat subjective; it tries to reflect whether the error that had to be rectified was a trivial omission (patch) or a genuine error (fixing).
However, we stopped fixing at simple errors, dropping all programs that misinterpreted a task description, referenced obviously missing pieces of code, or required substantial structural modifications to work.
All solutions suffering from these problems these received an ordinal value of \ref{compile:worst}.

For each pair of languages $X$ and $Y$, a Mann-Whitney $U$ test assessed whether the two samples (ordinal values for language $X$ vs.\ ordinal values for language $Y$) come from the same population.
The test results appear in \autoref{tab:make_stats_stat}, under column labeled with language $X$ at a row labeled with language $Y$, and includes various statistics:

  \begin{enumerate}
   \item The $p$-value is the probability that the two samples come from the same population.

   \item The total sample size $N$ is the total number of solutions in language $X$ and $Y$ that received an ordinal value.

   \item The test statistics $U$ (see a description of the test for details).

   \item The related test statistics $Z$, derivable from $U$.


   \item The effect size---Cohen's~$d$.

   \item The difference $\Delta$ of the means, which gives a sign to the difference between the samples. Namely, if $p$ is small, a positive $\Delta$ indicates that language $Y$ is on average ``better'' (fewer compilation problems) than language $X$.
   \end{enumerate}

\autoref{tab:misc_stats_make} reports, for each language, 
the number of \textbf{tasks} and \textbf{solutions} considered for compilation;
in column \textbf{make ok}, the percentage of solutions that eventually compiled correctly (ordinal values in the range 0--4); 
in column \textbf{make ko}, the percentage of solutions that did not compile correctly (ordinal value \ref{compile:worst}); 
in columns \textbf{none} through \textbf{fix}, the percentage of solutions that eventually compiled correctly for each category corresponding to ordinal values in the range 0--4.

\subsection{Execution} \label{sec:execution-ordinal}
\autoref{tab:run_stats_stat} and \autoref{tab:misc_stats_run} give more details about the running process; they are the counterparts to \autoref{tab:make_stats_stat} and \autoref{tab:misc_stats_make}.

We first consider all solutions to tasks $T_{\textsc{exec}}$ marked for execution that we could run.
We then assign an ordinal value to each \emph{solution}:
\begin{enumerate}[1: ]
\setcounter{enumi}{-1}
\item if the solution runs without errors (it runs and terminates within the timeout, returns with exit status 0, and does not write anything to standard error);
\item if the solution runs with possible errors but \emph{terminates} successfully (it runs and terminates within the timeout, returns with exit status 0, but writes some messages to standard error);
\item if the solution \emph{times out} (it runs without errors, and it is still running when the time out elapses);
\item if the solution runs with visible \emph{error} (it runs and terminates within the timeout, returns with exit status other than 0, and writes some messages to standard error);
\item if the solution \emph{crashes} (it runs and terminates within the timeout, returns with exit status other than 0, and writes nothing to standard error, such as a \verb|Segfault|).
\end{enumerate}

The categories are disjoint and try to reflect increasing levels of problems.
We consider terminating without printing error (a crash) worse than printing some information.
Similarly, we consider nontermination without manifest error better than abrupt termination with error. (In fact, many solutions in this categories were either from correct solutions working on very large inputs, typically in the scalability tasks, or to correct solutions to \emph{interactive} tasks were termination is not to be expected.)
The distinctions are somewhat subjective; they try to reflect the difficulty of understanding and possibly debugging an error.

For each pair of languages $X$ and $Y$, a Mann-Whitney $U$ test assessed whether the two samples (ordinal values for language $X$ vs.\ ordinal values for language $Y$) come from the same population.
The test results appear in \autoref{tab:run_stats_stat}, under column labeled with language $X$ at a row labeled with language $Y$, and includes the same statistics as \autoref{tab:make_stats_stat}.

\autoref{tab:misc_stats_run} reports, for each language, 
the number of \textbf{tasks} and \textbf{solutions} considered for execution;
in columns \textbf{run ok} through \textbf{crash}, the percentage of solutions for each category corresponding to ordinal values in the range 0--4.

\subsection{Overall code quality (compilation + execution)}
\autoref{tab:corr_stats_stat} compares the \emph{sum} of ordinal values assigned to each \emph{solution} as described in \autoref{sec:compilation-ordinal} and \autoref{sec:execution-ordinal}, for all tasks $T_{\textsc{exec}}$ marked for execution (and compilation).
The overall score gives an idea of the code quality of \rc solutions based on how much we had to do for compilation and for execution.

\subsection{Fault proneness} \label{sec:faults}
\autoref{tab:fault_stats_stat} and \autoref{tab:misc_stats_faults} give an idea of the number of defects manifesting themselves as runtime failures; they draw on data similar to those presented in \autoref{sec:compilation-ordinal} and \autoref{sec:execution-ordinal}.

We first consider all solutions to tasks $T_{\textsc{exec}}$ marked for execution that we could compile without errors and that ran without timing out.
We then assign an ordinal value to each \emph{solution}:
\begin{enumerate}[1: ]
\setcounter{enumi}{-1}
\item if the solution runs without errors (it runs and terminates within the timeout with exit status 0);
\item if the solution runs with errors (it runs and terminates within the timeout with exit status other than 0).
\end{enumerate}
The categories are disjoint and do not include solutions that timed out.

For each pair of languages $X$ and $Y$, a Mann-Whitney $U$ test assessed whether the two samples (ordinal values for language $X$ vs.\ ordinal values for language $Y$) come from the same population.
The test results appear in \autoref{tab:fault_stats_stat}, under column labeled with language $X$ at a row labeled with language $Y$, and includes the same statistics as \autoref{tab:make_stats_stat}.

\autoref{tab:misc_stats_faults} reports, for each language, 
the number of \textbf{tasks} and \textbf{solutions} that compiled correctly and ran without timing out;
in columns \textbf{error} and \textbf{run ok}, the percentage of solutions for each category corresponding to ordinal values 1 (\textbf{error}) and 0 (\textbf{run ok}).

\subsection*{Visualizations of language comparisons}
To help visualize the results, a graph accompanies each table with the results of statistical significance tests between pairs of languages.
In such graphs, nodes correspond to programming languages.
Two nodes $\ell_1$ and $\ell_2$ are arranged so that their \emph{horizontal} distance is roughly proportional to how $\ell_1$ and $\ell_2$ compare according to the statistical test. In contrast, vertical distances are chosen only to improve readability and carry not meaning.

Precisely, there are two graphs for each measure: unnormalized and normalized.
In the unnormalized graph, let $p_M(\ell_1, \ell_2)$, $e_M(\ell_1, \ell_2)$, and $\Delta_M(\ell_1, \ell_2)$ be the $p$-value, effect size ($d$ or $r$ according to the metric), and difference of medians for the test that compares $\ell_1$ to $\ell_2$ on metric $M$.
If $p_M(\ell_1, \ell_2) > 0.05$ or $e_M(\ell_1, \ell_2) < 0.05$, then the horizontal distance between node $\ell_1$ and node $\ell_2$ is zero or very small (it may still be non-zero to improve the visual layout while satisfying the constraints as well as possible), and there is no edge between them.
Otherwise, their horizontal distance is proportional to $\sqrt{e_M(\ell_1, \ell_2) \cdot (1 - p_M(\ell_1, \ell_2))}$, and there is a directed edge to the node corresponding to the ``better'' language according to $M$ from the other node.
To improve the visual layout, edges that express an ordered pair that is subsumed by others are omitted, that is if $\ell_a \rightarrow \ell_b \rightarrow \ell_c$ the edge from $\ell_a$ to $\ell_c$ is omitted.
Paths on the graph following edges in the direction of the arrow list languages in approximate order from ``worse'' to ``better''.
Edges are dotted if they correspond to large significant $p$-values ($0.01 \leq p < 0.05$); they have a color corresponding to the color assigned to the effect size in the table: \textcolor{colorstrong}{large effect size}, \textcolor{colormedium}{medium effect size}, and \textcolor{colorweak}{small but non vanishing} effect size.

In the \emph{normalized} companion graphs (called with ``normalized horizontal distances'' in the captions), arrows and vertical distances are assigned as in the unnormalized graphs, whereas the horizontal distance are proportional to the mean $\overline{\ell_M^{\alpha}}$ for each language $\ell$ over all common tasks (that is where each language has at least one solution).
Namely, the language with the ``worst'' average measure (consistent with whether $M$ is such that ``smaller'' or ``larger'' is better) will be farthest on the left, the language with the ``best'' average measure will be farthest on the right, with other languages in between proportionally to their rank.
Since, to have a unique baseline, normalized average measures only refer to tasks common to all languages, the normalized horizontal distances may be inconsistent with the pairwise tests because they refer to a much smaller set of values (sensitive to noise). This is only visible in the case of performance and scalability tasks, which are often sufficiently many for pairwise comparisons, but become too few (for example, less than 10) when we only look at the tasks that have implementations in all languages.
In these cases, the unnormalized graphs may be more indicative (and, in any case, the data in the tables is the hard one).

For comparisons based on ordinal values, there is only one kind of graph whose horizontal distances do not have a significant quantitative meaning but mainly represent an ordering.

Remind that all graphs use approximations and heuristics to build their layout; hence they are mainly meant as qualitative visualization aids that cannot substitute a detailed analytical reading of the data.

\clearpage

\section{Appendix: Tables and graphs}

\subsection{Lines of code (tasks compiling successfully)}
\begin{table}[ht]
\begin{center}
{\scriptsize
\begin{tabular}{cc|rrrrrrr}
  \hline
\textsc{language} & \textsc{measure} & \multicolumn{1}{c}{C} & \multicolumn{1}{c}{C\#} & \multicolumn{1}{c}{F\#} & \multicolumn{1}{c}{Go} & \multicolumn{1}{c}{Haskell} & \multicolumn{1}{c}{Java} & \multicolumn{1}{c}{Python} \\ 
  \hline
C\# & $p$ & 5.433E-01 &  &  &  &  &  &  \\ 
   & $N$ & 4.640E+02 &  &  &  &  &  &  \\ 
   & $W$ & 1.245E+04 &  &  &  &  &  &  \\ 
   & $Z$ & -6.078E-01 &  &  &  &  &  &  \\ 
   & $d$ & 3.918E-03 &  &  &  &  &  &  \\ 
   & $\Delta$ & -1.500E+00 &  &  &  &  &  &  \\ 
   & $R$ & -1.054E+00 &  &  &  &  &  &  \\
\hline
\end{tabular}
}
\caption{Comparison of conciseness (by min) for tasks compiling successfully}
\label{tab:loc_exit0_min}
\end{center}
\end{table}

\newpage
\NetworkFig{loc_exit0_min}{Lines of code (min) of tasks compiling successfully}
\clearpage
\begin{table}[ht]
\begin{center}
{\scriptsize
\begin{tabular}{cc|rrrrrrr}
  \hline
\textsc{language} & \textsc{measure} & \multicolumn{1}{c}{C} & \multicolumn{1}{c}{C\#} & \multicolumn{1}{c}{F\#} & \multicolumn{1}{c}{Go} & \multicolumn{1}{c}{Haskell} & \multicolumn{1}{c}{Java} & \multicolumn{1}{c}{Python} \\ 
  \hline
C\# & $p$ & 5.898E-01 &  &  &  &  &  &  \\ 
   & $N$ & 4.640E+02 &  &  &  &  &  &  \\ 
   & $W$ & 1.359E+04 &  &  &  &  &  &  \\ 
   & $Z$ & 5.391E-01 &  &  &  &  &  &  \\ 
   & $d$ & \textcolor{colorweak}{1.146E-01} &  &  &  &  &  &  \\ 
   & $\Delta$ & -1.000E+00 &  &  &  &  &  &  \\ 
   & $R$ & -1.032E+00 &  &  &  &  &  &  \\
\hline
\end{tabular}
}
\caption{Comparison of conciseness (by mean) for tasks compiling successfully}
\label{tab:loc_exit0_mean}
\end{center}
\end{table}

\newpage
\NetworkFig{loc_exit0_mean}{Lines of code (mean) of tasks compiling successfully}

\clearpage
\subsection{Lines of code (all tasks)}
\begin{table}[ht]
\begin{center}
{\scriptsize
\begin{tabular}{cc|rrrrrrr}
  \hline
\textsc{language} & \textsc{measure} & \multicolumn{1}{c}{C} & \multicolumn{1}{c}{C\#} & \multicolumn{1}{c}{F\#} & \multicolumn{1}{c}{Go} & \multicolumn{1}{c}{Haskell} & \multicolumn{1}{c}{Java} & \multicolumn{1}{c}{Python} \\ 
  \hline
C\# & $p$ & 4.401E-01 &  &  &  &  &  &  \\ 
   & $N$ & 5.460E+02 &  &  &  &  &  &  \\ 
   & $W$ & 1.704E+04 &  &  &  &  &  &  \\ 
   & $Z$ & -7.720E-01 &  &  &  &  &  &  \\ 
   & $d$ & \textcolor{colorweak}{1.075E-01} &  &  &  &  &  &  \\ 
   & $\Delta$ & -3.223E-01 &  &  &  &  &  &  \\ 
   & $R$ & -1.009E+00 &  &  &  &  &  &  \\
\hline
\end{tabular}
}
\caption{Comparison of conciseness (by min) for all tasks}
\label{tab:loc_any_min}
\end{center}
\end{table}

\newpage
\NetworkFig{loc_any_min}{Lines of code (min) of all tasks}
\clearpage
\begin{table}[ht]
\begin{center}
{\scriptsize
\begin{tabular}{cc|rrrrrrr}
  \hline
\textsc{language} & \textsc{measure} & \multicolumn{1}{c}{C} & \multicolumn{1}{c}{C\#} & \multicolumn{1}{c}{F\#} & \multicolumn{1}{c}{Go} & \multicolumn{1}{c}{Haskell} & \multicolumn{1}{c}{Java} & \multicolumn{1}{c}{Python} \\ 
  \hline
C\# & $p$ & 2.617E-01 &  &  &  &  &  &  \\ 
   & $N$ & 5.460E+02 &  &  &  &  &  &  \\ 
   & $W$ & 1.959E+04 &  &  &  &  &  &  \\ 
   & $Z$ & 1.122E+00 &  &  &  &  &  &  \\ 
   & $d$ & \textcolor{colorweak}{6.772E-02} &  &  &  &  &  &  \\ 
   & $\Delta$ & 1.839E+00 &  &  &  &  &  &  \\ 
   & $R$ & 1.048E+00 &  &  &  &  &  &  \\
\hline
\end{tabular}
}
\caption{Comparison of conciseness (by mean) for all tasks}
\label{tab:loc_any_mean}
\end{center}
\end{table}

\newpage
\NetworkFig{loc_any_mean}{Lines of code (mean) of all tasks}

\clearpage
\subsection{Comments per line of code}
\begin{table}[ht]
\begin{center}
{\scriptsize
\begin{tabular}{cc|rrrrrrr}
  \hline
\textsc{language} & \textsc{measure} & \multicolumn{1}{c}{C} & \multicolumn{1}{c}{C\#} & \multicolumn{1}{c}{F\#} & \multicolumn{1}{c}{Go} & \multicolumn{1}{c}{Haskell} & \multicolumn{1}{c}{Java} & \multicolumn{1}{c}{Python} \\ 
  \hline
C\# & $p$ & \textcolor{colorweak}{1.280E-02} &  &  &  &  &  &  \\ 
   & $N$ & 5.460E+02 &  &  &  &  &  &  \\ 
   & $W$ & 5.485E+03 &  &  &  &  &  &  \\ 
   & $Z$ & 2.489E+00 &  &  &  &  &  &  \\ 
   & $d$ & \textcolor{colorweak}{1.196E-01} &  &  &  &  &  &  \\ 
   & $\Delta$ & 7.850E-03 &  &  &  &  &  &  \\ 
   & $R$ & 1.273E+00 &  &  &  &  &  &  \\
\hline
\end{tabular}
}
\caption{Comparison of comments per line of code (by min) for all tasks}
\label{tab:commentstocode_any_min}
\end{center}
\end{table}

\newpage
\NetworkFig{commentstocode_any_min}{Comments per line of code (min) of all tasks}
\clearpage
\begin{table}[ht]
\begin{center}
{\scriptsize
\begin{tabular}{cc|rrrrrrr}
  \hline
\textsc{language} & \textsc{measure} & \multicolumn{1}{c}{C} & \multicolumn{1}{c}{C\#} & \multicolumn{1}{c}{F\#} & \multicolumn{1}{c}{Go} & \multicolumn{1}{c}{Haskell} & \multicolumn{1}{c}{Java} & \multicolumn{1}{c}{Python} \\ 
  \hline
C\# & $p$ & \textcolor{colorstrong}{1.126E-04} &  &  &  &  &  &  \\ 
   & $N$ & 5.460E+02 &  &  &  &  &  &  \\ 
   & $W$ & 9.117E+03 &  &  &  &  &  &  \\ 
   & $Z$ & 3.862E+00 &  &  &  &  &  &  \\ 
   & $d$ & \textcolor{colorweak}{1.451E-01} &  &  &  &  &  &  \\ 
   & $\Delta$ & 1.888E-02 &  &  &  &  &  &  \\ 
   & $R$ & 1.602E+00 &  &  &  &  &  &  \\
\hline
\end{tabular}
}
\caption{Comparison of comments per line of code (by mean) for all tasks}
\label{tab:commentstocode_any_mean}
\end{center}
\end{table}

\newpage
\NetworkFig{commentstocode_any_mean}{Comments per line of code (mean) of all tasks}

\clearpage
\subsection{Size of binaries}
\begin{table}[ht]
\begin{center}
{\scriptsize
\begin{tabular}{cc|rrrrrr}
  \hline
\textsc{language} & \textsc{measure} & \multicolumn{1}{c}{C} & \multicolumn{1}{c}{C\#} & \multicolumn{1}{c}{F\#} & \multicolumn{1}{c}{Go} & \multicolumn{1}{c}{Haskell} & \multicolumn{1}{c}{Java} \\ 
  \hline
C\# & $p$ & \textcolor{colorstrong}{0.000E+00} &  &  &  &  &  \\ 
   & $N$ & 4.660E+02 &  &  &  &  &  \\ 
   & $W$ & 2.611E+04 &  &  &  &  &  \\ 
   & $Z$ & 1.349E+01 &  &  &  &  &  \\ 
   & $d$ & \textcolor{colorstrong}{2.669E+00} &  &  &  &  &  \\ 
   & $\Delta$ & 6.575E+00 &  &  &  &  &  \\ 
   & $R$ & 2.140E+00 &  &  &  &  &  \\
\hline
\end{tabular}
}
\caption{Comparison of binary size (by min) for tasks compiling successfully}
\label{tab:binsize_exit0_min}
\end{center}
\end{table}

\newpage
\NetworkFig{binsize_exit0_min}{Size of binaries (min) of tasks compiling successfully}
\clearpage
\begin{table}[ht]
\begin{center}
{\scriptsize
\begin{tabular}{cc|rrrrrr}
  \hline
\textsc{language} & \textsc{measure} & \multicolumn{1}{c}{C} & \multicolumn{1}{c}{C\#} & \multicolumn{1}{c}{F\#} & \multicolumn{1}{c}{Go} & \multicolumn{1}{c}{Haskell} & \multicolumn{1}{c}{Java} \\ 
  \hline
C\# & $p$ & \textcolor{colorstrong}{0.000E+00} &  &  &  &  &  \\ 
   & $N$ & 4.660E+02 &  &  &  &  &  \\ 
   & $W$ & 2.634E+04 &  &  &  &  &  \\ 
   & $Z$ & 1.343E+01 &  &  &  &  &  \\ 
   & $d$ & \textcolor{colorstrong}{2.633E+00} &  &  &  &  &  \\ 
   & $\Delta$ & 6.595E+00 &  &  &  &  &  \\ 
   & $R$ & 2.130E+00 &  &  &  &  &  \\
\hline
\end{tabular}
}
\caption{Comparison of binary size (by mean) for tasks compiling successfully}
\label{tab:binsize_exit0_mean}
\end{center}
\end{table}

\newpage
\NetworkFig{binsize_exit0_mean}{Size of binaries (mean) of tasks compiling successfully}

\clearpage
\subsection{Performance}
\begin{table}[ht]
\begin{center}
{\scriptsize
\begin{tabular}{cc|rrrrrrr}
  \hline
\textsc{language} & \textsc{measure} & \multicolumn{1}{c}{C} & \multicolumn{1}{c}{C\#} & \multicolumn{1}{c}{F\#} & \multicolumn{1}{c}{Go} & \multicolumn{1}{c}{Haskell} & \multicolumn{1}{c}{Java} & \multicolumn{1}{c}{Python} \\ 
  \hline
C\# & $p$ & \textcolor{colorstrong}{9.150E-04} &  &  &  &  &  &  \\ 
   & $N$ & 2.800E+01 &  &  &  &  &  &  \\ 
   & $W$ & 0.000E+00 &  &  &  &  &  &  \\ 
   & $Z$ & -3.315E+00 &  &  &  &  &  &  \\ 
   & $d$ & \textcolor{colorstrong}{2.009E+00} &  &  &  &  &  &  \\ 
   & $\Delta$ & -3.000E-02 &  &  &  &  &  &  \\ 
   & $R$ & -Inf &  &  &  &  &  &  \\
\hline
\end{tabular}
}
\caption{Comparison of performance (by min) for tasks running successfully}
\label{tab:performance_exit0_min}
\end{center}
\end{table}

\newpage
\NetworkFig{performance_exit0_min}{Performance (min) of tasks running successfully}
\clearpage
\begin{table}[ht]
\begin{center}
{\scriptsize
\begin{tabular}{cc|rrrrrrr}
  \hline
\textsc{language} & \textsc{measure} & \multicolumn{1}{c}{C} & \multicolumn{1}{c}{C\#} & \multicolumn{1}{c}{F\#} & \multicolumn{1}{c}{Go} & \multicolumn{1}{c}{Haskell} & \multicolumn{1}{c}{Java} & \multicolumn{1}{c}{Python} \\ 
  \hline
C\# & $p$ & \textcolor{colorstrong}{9.259E-04} &  &  &  &  &  &  \\ 
   & $N$ & 2.800E+01 &  &  &  &  &  &  \\ 
   & $W$ & 0.000E+00 &  &  &  &  &  &  \\ 
   & $Z$ & -3.312E+00 &  &  &  &  &  &  \\ 
   & $d$ & \textcolor{colorstrong}{2.104E+00} &  &  &  &  &  &  \\ 
   & $\Delta$ & -3.000E-02 &  &  &  &  &  &  \\ 
   & $R$ & -Inf &  &  &  &  &  &  \\
\hline
\end{tabular}
}
\caption{Comparison of performance (by mean) for tasks running successfully}
\label{tab:performance_exit0_mean}
\end{center}
\end{table}

\newpage
\NetworkFig{performance_exit0_mean}{Performance (mean) of tasks running successfully}

\clearpage
\subsection{Scalability}
\begin{table}[ht]
\begin{center}
{\scriptsize
\begin{tabular}{cc|rrrrrrr}
  \hline
\textsc{language} & \textsc{measure} & \multicolumn{1}{c}{C} & \multicolumn{1}{c}{C\#} & \multicolumn{1}{c}{F\#} & \multicolumn{1}{c}{Go} & \multicolumn{1}{c}{Haskell} & \multicolumn{1}{c}{Java} & \multicolumn{1}{c}{Python} \\ 
  \hline
C\# & $p$ & \textcolor{colorstrong}{1.020E-03} &  &  &  &  &  &  \\ 
   & $N$ & 4.400E+01 &  &  &  &  &  &  \\ 
   & $W$ & 2.100E+01 &  &  &  &  &  &  \\ 
   & $Z$ & -3.285E+00 &  &  &  &  &  &  \\ 
   & $d$ & \textcolor{colormedium}{3.279E-01} &  &  &  &  &  &  \\ 
   & $\Delta$ & -1.700E+00 &  &  &  &  &  &  \\ 
   & $R$ & -7.538E+00 &  &  &  &  &  &  \\
\hline
\end{tabular}
}
\caption{Comparison of scalability (by min) for tasks running successfully}
\label{tab:scalability_exit0_min}
\end{center}
\end{table}

\newpage
\NetworkFig[scalability_exit0_noFsharp_min]{scalability_exit0_min}{Scalability (min) of tasks running successfully}
\clearpage
\begin{table}[ht]
\begin{center}
{\scriptsize
\begin{tabular}{cc|rrrrrrr}
  \hline
\textsc{language} & \textsc{measure} & \multicolumn{1}{c}{C} & \multicolumn{1}{c}{C\#} & \multicolumn{1}{c}{F\#} & \multicolumn{1}{c}{Go} & \multicolumn{1}{c}{Haskell} & \multicolumn{1}{c}{Java} & \multicolumn{1}{c}{Python} \\ 
  \hline
C\# & $p$ & \textcolor{colorstrong}{4.061E-03} &  &  &  &  &  &  \\ 
   & $N$ & 4.400E+01 &  &  &  &  &  &  \\ 
   & $W$ & 3.800E+01 &  &  &  &  &  &  \\ 
   & $Z$ & -2.873E+00 &  &  &  &  &  &  \\ 
   & $d$ & \textcolor{colormedium}{3.274E-01} &  &  &  &  &  &  \\ 
   & $\Delta$ & -1.820E+00 &  &  &  &  &  &  \\ 
   & $R$ & -6.600E+00 &  &  &  &  &  &  \\
\hline
\end{tabular}
}
\caption{Comparison of scalability (by mean) for tasks running successfully}
\label{tab:scalability_exit0_mean}
\end{center}
\end{table}

\newpage
\NetworkFig[scalability_exit0_noFsharp_mean]{scalability_exit0_mean}{Scalability (mean) of tasks running successfully}

\clearpage
\subsection{Maximum RAM}
\begin{table}[ht]
\begin{center}
{\scriptsize
\begin{tabular}{cc|rrrrrrr}
  \hline
\textsc{language} & \textsc{measure} & \multicolumn{1}{c}{C} & \multicolumn{1}{c}{C\#} & \multicolumn{1}{c}{F\#} & \multicolumn{1}{c}{Go} & \multicolumn{1}{c}{Haskell} & \multicolumn{1}{c}{Java} & \multicolumn{1}{c}{Python} \\ 
  \hline
C\# & $p$ & \textcolor{colorstrong}{7.994E-05} &  &  &  &  &  &  \\ 
   & $N$ & 4.400E+01 &  &  &  &  &  &  \\ 
   & $W$ & 5.000E+00 &  &  &  &  &  &  \\ 
   & $Z$ & -3.945E+00 &  &  &  &  &  &  \\ 
   & $d$ & \textcolor{colorstrong}{2.022E+00} &  &  &  &  &  &  \\ 
   & $\Delta$ & -7.837E+04 &  &  &  &  &  &  \\ 
   & $R$ & -2.479E+00 &  &  &  &  &  &  \\
\hline
\end{tabular}
}
\caption{Comparison of RAM used (by min) for tasks running successfully}
\label{tab:maxram_exit0_min}
\end{center}
\end{table}

\newpage
\NetworkFig[maxram_exit0_noFsharp_min]{maxram_exit0_min}{Maximum RAM usage (min) of scalability tasks}
\clearpage
\begin{table}[ht]
\begin{center}
{\scriptsize
\begin{tabular}{cc|rrrrrrr}
  \hline
\textsc{language} & \textsc{measure} & \multicolumn{1}{c}{C} & \multicolumn{1}{c}{C\#} & \multicolumn{1}{c}{F\#} & \multicolumn{1}{c}{Go} & \multicolumn{1}{c}{Haskell} & \multicolumn{1}{c}{Java} & \multicolumn{1}{c}{Python} \\ 
  \hline
C\# & $p$ & \textcolor{colorstrong}{4.010E-05} &  &  &  &  &  &  \\ 
   & $N$ & 4.400E+01 &  &  &  &  &  &  \\ 
   & $W$ & 0.000E+00 &  &  &  &  &  &  \\ 
   & $Z$ & -4.107E+00 &  &  &  &  &  &  \\ 
   & $d$ & \textcolor{colorstrong}{2.239E+00} &  &  &  &  &  &  \\ 
   & $\Delta$ & -1.850E+05 &  &  &  &  &  &  \\ 
   & $R$ & -4.076E+00 &  &  &  &  &  &  \\
\hline
\end{tabular}
}
\caption{Comparison of RAM used (by mean) for tasks running successfully}
\label{tab:maxram_exit0_mean}
\end{center}
\end{table}

\newpage
\NetworkFig[maxram_exit0_noFsharp_mean]{maxram_exit0_mean}{Maximum RAM usage (mean) of scalability tasks}

\clearpage
\subsection{Page faults}
\begin{table}[ht]
\begin{center}
{\scriptsize
\begin{tabular}{cc|rrrrrrr}
  \hline
\textsc{language} & \textsc{measure} & \multicolumn{1}{c}{C} & \multicolumn{1}{c}{C\#} & \multicolumn{1}{c}{F\#} & \multicolumn{1}{c}{Go} & \multicolumn{1}{c}{Haskell} & \multicolumn{1}{c}{Java} & \multicolumn{1}{c}{Python} \\ 
  \hline
C\# & $p$ &  NaN &  &  &  &  &  &  \\ 
   & $N$ & 4.400E+01 &  &  &  &  &  &  \\ 
   & $W$ &  NaN &  &  &  &  &  &  \\ 
   & $Z$ &  NaN &  &  &  &  &  &  \\ 
   & $d$ &  NaN &  &  &  &  &  &  \\ 
   & $\Delta$ &  NaN &  &  &  &  &  &  \\ 
   & $R$ &  NaN &  &  &  &  &  &  \\
\hline
\end{tabular}
}
\caption{Comparison of page faults (by min) for tasks running successfully}
\label{tab:pfs_exit0_min}
\end{center}
\end{table}

\newpage
\NetworkFig{pfs_exit0_min}{Page faults (min) of scalability tasks}
\clearpage
\begin{table}[ht]
\begin{center}
{\scriptsize
\begin{tabular}{cc|rrrrrrr}
  \hline
\textsc{language} & \textsc{measure} & \multicolumn{1}{c}{C} & \multicolumn{1}{c}{C\#} & \multicolumn{1}{c}{F\#} & \multicolumn{1}{c}{Go} & \multicolumn{1}{c}{Haskell} & \multicolumn{1}{c}{Java} & \multicolumn{1}{c}{Python} \\ 
  \hline
C\# & $p$ & 3.173E-01 &  &  &  &  &  &  \\ 
   & $N$ & 4.400E+01 &  &  &  &  &  &  \\ 
   & $W$ & 1.000E+00 &  &  &  &  &  &  \\ 
   & $Z$ & 1.000E+00 &  &  &  &  &  &  \\ 
   & $d$ & \textcolor{colorweak}{2.194E-01} &  &  &  &  &  &  \\ 
   & $\Delta$ & 3.409E-02 &  &  &  &  &  &  \\ 
   & $R$ & 4.000E+00 &  &  &  &  &  &  \\
\hline
\end{tabular}
}
\caption{Comparison of page faults (by mean) for tasks running successfully}
\label{tab:pfs_exit0_mean}
\end{center}
\end{table}

\newpage
\NetworkFig{pfs_exit0_mean}{Page faults (mean) of scalability tasks}

\clearpage
\subsection{Timeout analysis}
\begin{table}[ht]
\begin{center}
{\scriptsize
\begin{tabular}{cc|rrrrrrr}
  \hline
\textsc{language} & \textsc{measure} & \multicolumn{1}{c}{C} & \multicolumn{1}{c}{C\#} & \multicolumn{1}{c}{F\#} & \multicolumn{1}{c}{Go} & \multicolumn{1}{c}{Haskell} & \multicolumn{1}{c}{Java} & \multicolumn{1}{c}{Python} \\ 
  \hline
C\# & $p$ & \textcolor{colorweak}{4.550E-02} &  &  &  &  &  &  \\ 
   & $N$ & 5.400E+01 &  &  &  &  &  &  \\ 
   & $W$ & 0.000E+00 &  &  &  &  &  &  \\ 
   & $Z$ & -2.000E+00 &  &  &  &  &  &  \\ 
   & $d$ & \textcolor{colormedium}{4.760E-01} &  &  &  &  &  &  \\ 
   & $\Delta$ & -1.481E-01 &  &  &  &  &  &  \\ 
   & $R$ & -5.000E+00 &  &  &  &  &  &  \\
\hline
\end{tabular}
}
\caption{Comparison of time-outs (3 minutes) on scalability tasks}
\label{tab:timeout_binary.timeout.selector}
\end{center}
\end{table}

\newpage
\NetworkFig{timeout_binary.timeout.selector}{Timeout analysis of scalability tasks}

\clearpage
\subsection{Number of solutions}
\begin{table}[ht]
\begin{center}
{\scriptsize
\begin{tabular}{cc|rrrrrrr}
  \hline
\textsc{language} & \textsc{measure} & \multicolumn{1}{c}{C} & \multicolumn{1}{c}{C\#} & \multicolumn{1}{c}{F\#} & \multicolumn{1}{c}{Go} & \multicolumn{1}{c}{Haskell} & \multicolumn{1}{c}{Java} & \multicolumn{1}{c}{Python} \\ 
  \hline
C\# & $p$ & \textcolor{colorstrong}{0.000E+00} &  &  &  &  &  &  \\ 
   & $N$ & 1.488E+03 &  &  &  &  &  &  \\ 
   & $W$ & 2.168E+04 &  &  &  &  &  &  \\ 
   & $Z$ & 9.277E+00 &  &  &  &  &  &  \\ 
   & $d$ & \textcolor{colorweak}{2.975E-01} &  &  &  &  &  &  \\ 
   & $\Delta$ & 2.285E-01 &  &  &  &  &  &  \\ 
   & $R$ & 1.480E+00 &  &  &  &  &  &  \\
\hline
\end{tabular}
}
\caption{Comparison of number of solutions per task}
\label{tab:solutions_count_length.na.zero}
\end{center}
\end{table}

\newpage
\NetworkFig{solutions_count_length.na.zero}{Number of solutions per task}

\clearpage
\subsection{Compilation and execution statistics}
\begin{table}[ht]
\begin{center}
{\scriptsize
\begin{tabular}{cc|rrrrrrr}
  \hline
\textsc{language} & \textsc{measure} & \multicolumn{1}{c}{C} & \multicolumn{1}{c}{C\#} & \multicolumn{1}{c}{F\#} & \multicolumn{1}{c}{Go} & \multicolumn{1}{c}{Haskell} & \multicolumn{1}{c}{Java} & \multicolumn{1}{c}{Python} \\ 
  \hline
C\# & $p$ & \textcolor{colorweak}{2.210E-02} &  &  &  &  &  &  \\ 
   & $N$ & 8.780E+02 &  &  &  &  &  &  \\ 
   & $U$ & 8.511E+04 &  &  &  &  &  &  \\ 
   & $Z$ & -2.289E+00 &  &  &  &  &  &  \\ 
   & $d$ & \textcolor{colorweak}{1.714E-01} &  &  &  &  &  &  \\ 
   & $\Delta$ & -3.091E-01 &  &  &  &  &  &  \\ 
   \hline
F\# & $p$ & \textcolor{colorstrong}{3.302E-06} & \textcolor{colorstrong}{3.089E-09} &  &  &  &  &  \\
\hline
\end{tabular}
}
\caption{Unpaired comparisons of compilation status}
\label{tab:make_stats_stat}
\end{center}
\end{table}

\NetworkFig{make_stats_stat}{Comparisons of compilation status}
\clearpage
\begin{table}[ht]
\begin{center}
{\scriptsize
\begin{tabular}{cc|rrrrrrr}
  \hline
\textsc{language} & \textsc{measure} & \multicolumn{1}{c}{C} & \multicolumn{1}{c}{C\#} & \multicolumn{1}{c}{F\#} & \multicolumn{1}{c}{Go} & \multicolumn{1}{c}{Haskell} & \multicolumn{1}{c}{Java} & \multicolumn{1}{c}{Python} \\ 
  \hline
C\# & $p$ & 7.612E-01 &  &  &  &  &  &  \\ 
   & $N$ & 7.490E+02 &  &  &  &  &  &  \\ 
   & $U$ & 6.718E+04 &  &  &  &  &  &  \\ 
   & $Z$ & -3.040E-01 &  &  &  &  &  &  \\ 
   & $d$ & 4.167E-02 &  &  &  &  &  &  \\ 
   & $\Delta$ & 4.716E-02 &  &  &  &  &  &  \\ 
   \hline
F\# & $p$ & 1.522E-01 & 1.134E-01 &  &  &  &  &  \\
\hline
\end{tabular}
}
\caption{Unpaired comparisons of running status}
\label{tab:run_stats_stat}
\end{center}
\end{table}

\NetworkFig{run_stats_stat}{Comparisons of running status}
\clearpage
\begin{table}[ht]
\begin{center}
{\scriptsize
\begin{tabular}{cc|rrrrrrr}
  \hline
\textsc{language} & \textsc{measure} & \multicolumn{1}{c}{C} & \multicolumn{1}{c}{C\#} & \multicolumn{1}{c}{F\#} & \multicolumn{1}{c}{Go} & \multicolumn{1}{c}{Haskell} & \multicolumn{1}{c}{Java} & \multicolumn{1}{c}{Python} \\ 
  \hline
C\# & $p$ & \textcolor{colorstrong}{5.525E-05} &  &  &  &  &  &  \\ 
   & $N$ & 7.490E+02 &  &  &  &  &  &  \\ 
   & $U$ & 5.690E+04 &  &  &  &  &  &  \\ 
   & $Z$ & -4.032E+00 &  &  &  &  &  &  \\ 
   & $d$ & \textcolor{colormedium}{3.231E-01} &  &  &  &  &  &  \\ 
   & $\Delta$ & -5.362E-01 &  &  &  &  &  &  \\ 
   \hline
F\# & $p$ & \textcolor{colorstrong}{2.097E-04} & \textcolor{colorstrong}{1.676E-11} &  &  &  &  &  \\
\hline
\end{tabular}
}
\caption{Unpaired comparisons of combined compilation and running}
\label{tab:corr_stats_stat}
\end{center}
\end{table}

\NetworkFig{corr_stats_stat}{Comparisons of combined compilation and running status}
\clearpage
\begin{table}[ht]
\begin{center}
{\scriptsize
\begin{tabular}{cc|rrrrrrr}
  \hline
\textsc{language} & \textsc{measure} & \multicolumn{1}{c}{C} & \multicolumn{1}{c}{C\#} & \multicolumn{1}{c}{F\#} & \multicolumn{1}{c}{Go} & \multicolumn{1}{c}{Haskell} & \multicolumn{1}{c}{Java} & \multicolumn{1}{c}{Python} \\ 
  \hline
C\# & $p$ & \textcolor{colorweak}{3.687E-02} &  &  &  &  &  &  \\ 
   & $N$ & 6.370E+02 &  &  &  &  &  &  \\ 
   & $U$ & 5.060E+04 &  &  &  &  &  &  \\ 
   & $Z$ & 2.087E+00 &  &  &  &  &  &  \\ 
   & $d$ & \textcolor{colorweak}{1.703E-01} &  &  &  &  &  &  \\ 
   & $\Delta$ & 5.215E-02 &  &  &  &  &  &  \\ 
   \hline
F\# & $p$ & 5.047E-01 & 2.038E-01 &  &  &  &  &  \\
\hline
\end{tabular}
}
\caption{Unpaired comparisons of runtime fault proneness}
\label{tab:fault_stats_stat}
\end{center}
\end{table}

\NetworkFig{fault_stats_stat}{Comparisons of fault proneness (based on exit status) of solutions that compile correctly and do not timeout}

\clearpage
\begin{table}[ht]
\begin{center}
{\scriptsize
\begin{tabular}{crrr|rrrrr|r}
  \hline
 & \multicolumn{1}{c}{\textbf{tasks}} & \multicolumn{1}{c}{\textbf{solutions}} & \multicolumn{1}{c}{\textbf{make ok}} & \multicolumn{1}{c}{\textbf{none}} & \multicolumn{1}{c}{\textbf{libs}} & \multicolumn{1}{c}{\textbf{merge}} & \multicolumn{1}{c}{\textbf{patch}} & \multicolumn{1}{c}{\textbf{fix}} & \multicolumn{1}{c}{\textbf{make ko}} \\ 
  \hline
C & 398 & 524 & 0.85 & 0.68 & 0.21 & 0.03 & 0.06 & 0.02 & 0.15 \\ 
  C\# & 288 & 354 & 0.90 & 0.58 & 0.08 & 0.00 & 0.34 & 0.00 & 0.10 \\ 
  F\# & 216 & 254 & 0.95 & 0.82 & 0.00 & 0.00 & 0.18 & 0.00 & 0.05 \\ 
  Go & 392 & 497 & 0.89 & 0.82 & 0.00 & 0.01 & 0.17 & 0.00 & 0.11 \\ 
  Haskell & 358 & 519 & 0.84 & 0.47 & 0.00 & 0.06 & 0.47 & 0.00 & 0.16 \\ 
  Java & 344 & 446 & 0.78 & 0.82 & 0.01 & 0.02 & 0.10 & 0.04 & 0.22 \\ 
  Python & 415 & 775 & 1.00 & 0.67 & 0.00 & 0.05 & 0.28 & 0.00 & 0.00 \\ 
  Ruby & 391 & 581 & 1.00 & 0.80 & 0.00 & 0.03 & 0.17 & 0.00 & 0.00 \\ 
   \hline
\textsc{total} & 442 & 3950 & 0.91 & 0.70 & 0.03 & 0.03 & 0.22 & 0.01 & 0.09 \\ 
   \hline
\end{tabular}
}
\caption{Statistics about the compilation process: columns \textbf{make ok} and \textbf{make ko} report percentages relative to \textbf{solutions} for each language; the columns in between report percentages relative to \textbf{make ok} for each language}
\label{tab:misc_stats_make}
\end{center}
\end{table}

\begin{table}[ht]
\begin{center}
{\scriptsize
\begin{tabular}{crr|rrrrr}
  \hline
 & \multicolumn{1}{c}{\textbf{tasks}} & \multicolumn{1}{c}{\textbf{solutions}} & \multicolumn{1}{c}{\textbf{run ok}} & \multicolumn{1}{c}{\textbf{stderr}} & \multicolumn{1}{c}{\textbf{timeout}} & \multicolumn{1}{c}{\textbf{error}} & \multicolumn{1}{c}{\textbf{crash}} \\ 
  \hline
C & 351 & 442 & 0.76 & 0.01 & 0.11 & 0.06 & 0.05 \\ 
  C\# & 253 & 307 & 0.73 & 0.01 & 0.20 & 0.06 & 0.00 \\ 
  F\# & 201 & 236 & 0.81 & 0.01 & 0.09 & 0.10 & 0.00 \\ 
  Go & 349 & 427 & 0.89 & 0.01 & 0.09 & 0.01 & 0.00 \\ 
  Haskell & 298 & 426 & 0.82 & 0.00 & 0.12 & 0.06 & 0.00 \\ 
  Java & 275 & 339 & 0.75 & 0.02 & 0.11 & 0.13 & 0.00 \\ 
  Python & 399 & 751 & 0.76 & 0.01 & 0.10 & 0.13 & 0.01 \\ 
  Ruby & 378 & 563 & 0.79 & 0.01 & 0.08 & 0.12 & 0.00 \\ 
   \hline
\textsc{total} & 422 & 3491 & 0.79 & 0.01 & 0.11 & 0.09 & 0.01 \\ 
   \hline
\end{tabular}
}
\caption{Statistics about the running process: all columns \emph{other than} \textbf{tasks} and \textbf{solutions} report percentages relative to \textbf{solutions} for each language}
\label{tab:misc_stats_run}
\end{center}
\end{table}

\begin{table}[ht]
\begin{center}
{\scriptsize
\begin{tabular}{crr|rr}
  \hline
 & \multicolumn{1}{c}{\textbf{tasks}} & \multicolumn{1}{c}{\textbf{solutions}} & \multicolumn{1}{c}{\textbf{error}} & \multicolumn{1}{c}{\textbf{run ok}} \\ 
  \hline
C & 313 & 391 & 0.13 & 0.87 \\ 
  C\# & 203 & 246 & 0.07 & 0.93 \\ 
  F\# & 182 & 215 & 0.11 & 0.89 \\ 
  Go & 317 & 389 & 0.02 & 0.98 \\ 
  Haskell & 264 & 376 & 0.07 & 0.93 \\ 
  Java & 247 & 297 & 0.15 & 0.85 \\ 
  Python & 354 & 675 & 0.15 & 0.85 \\ 
  Ruby & 349 & 516 & 0.14 & 0.86 \\ 
   \hline
\textsc{total} & 422 & 3105 & 0.11 & 0.89 \\ 
   \hline
\end{tabular}
}
\caption{Statistics about fault proneness: columns \textbf{error} and \textbf{run ok} report percentages relative to \textbf{solutions} for each language}
\label{tab:misc_stats_faults}
\end{center}
\end{table}

C: incomplete (27 solutions), snippet (21 solutions), nomain (21 solutions), nonstandard libs (17 solutions), other (12 solutions), unsupported dialect (2 solutions)
C\#: nomain (24 solutions), nonstandard libs (17 solutions), snippet (14 solutions), bug (3 solutions), filename (1 solutions), other (1 solutions)
F\#: nonstandard libs (10 solutions), bug (2 solutions), other (1 solutions)
Go: nonstandard libs (41 solutions), snippet (9 solutions), other (5 solutions), bug (2 solutions), nomain (0 solutions)
Haskell: nomain (45 solutions), nonstandard libs (30 solutions), snippet (20 solutions), bug (18 solutions), other (16 solutions)
Java: nomain (60 solutions), snippet (56 solutions), other (24 solutions), nonstandard libs (13 solutions), incomplete (6 solutions)
Python: other (2 solutions)
Ruby: other (519 solutions), nonstandard libs (35 solutions), merge (9 solutions), package (7 solutions), bug (6 solutions), snippet (2 solutions), abort (1 solutions), incomplete (1 solutions), require (1 solutions)

\clearpage
\changepage{}{}{}{}{}{-15mm}{}{}{}

\section{Appendix: Plots}

\clearpage\newpage
\subsection{Lines of code (tasks compiling successfully)}
\GraphMeasure{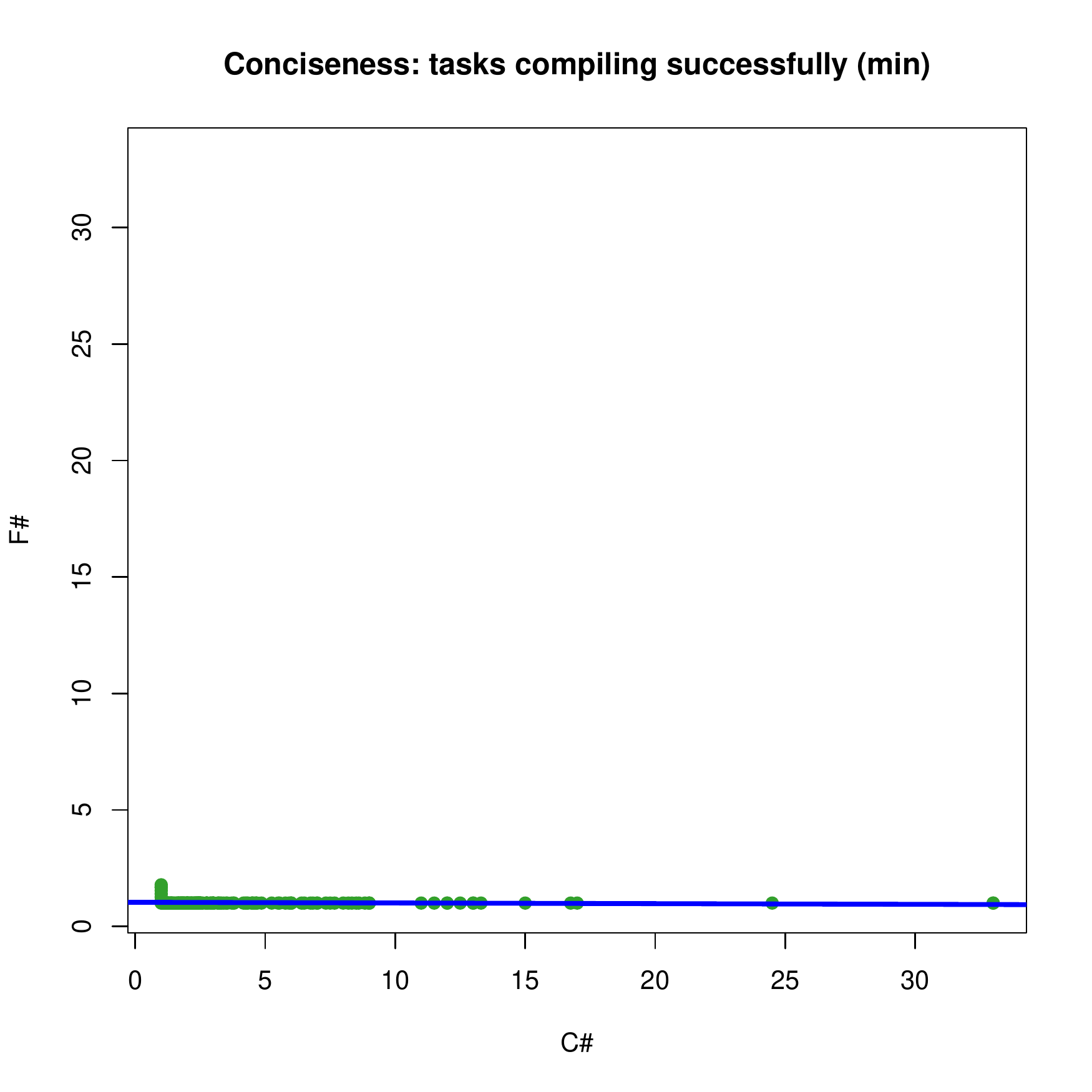}{Lines of code (min) of tasks compiling successfully}
\clearpage\newpage
\GraphMeasure{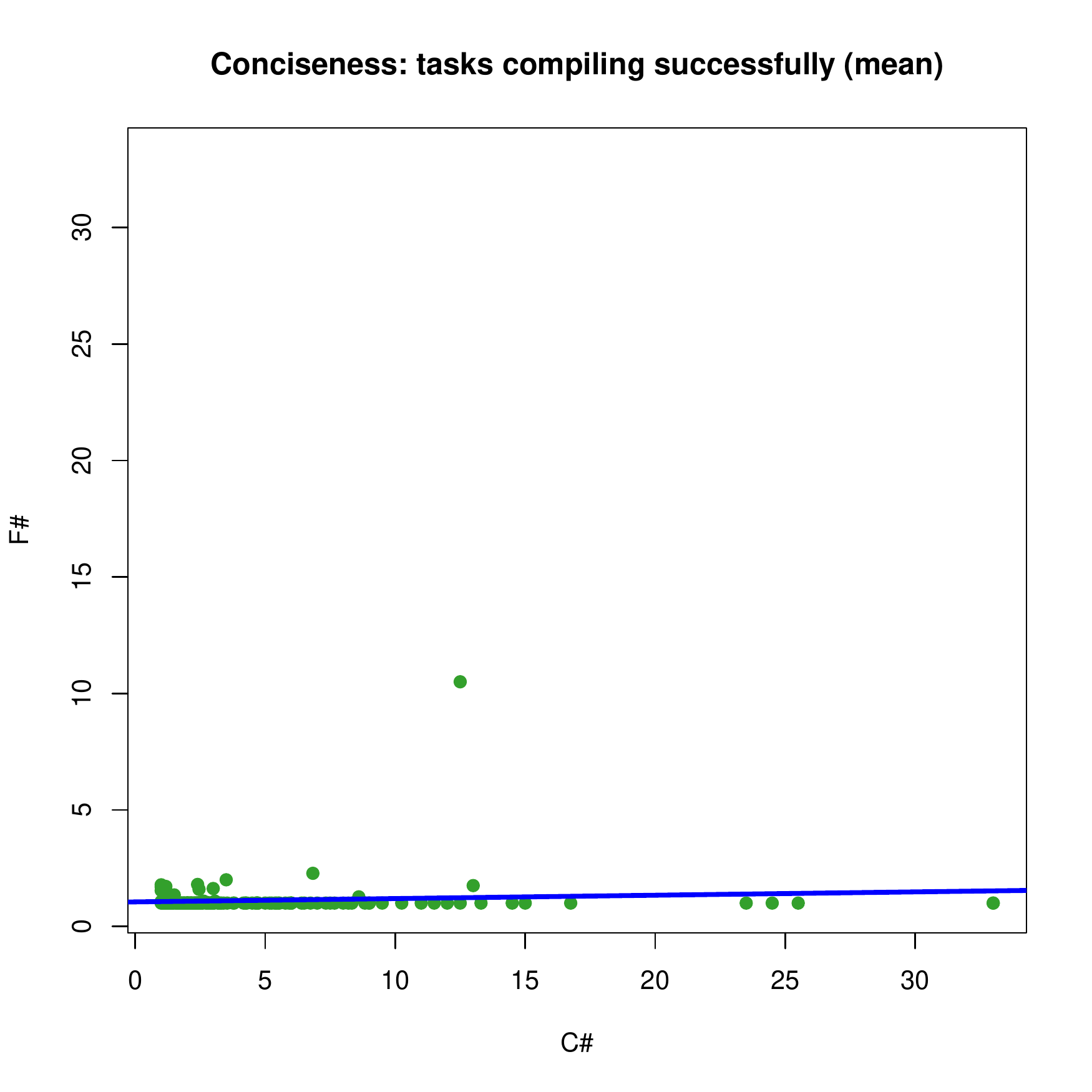}{Lines of code (mean) of tasks compiling successfully}

\clearpage\newpage
\subsection{Lines of code (all tasks)}
\GraphMeasure{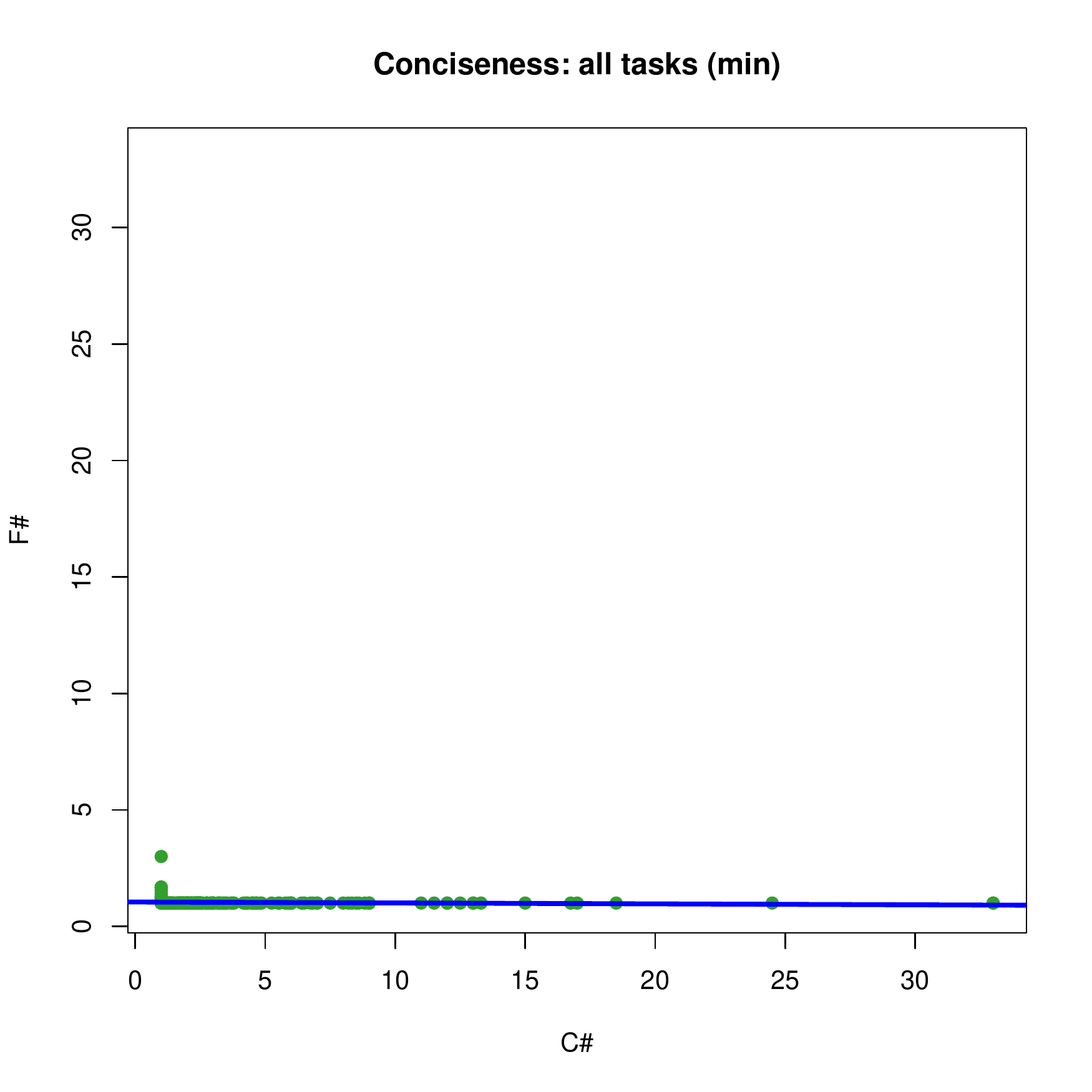}{Lines of code (min) of all tasks}
\clearpage\newpage
\GraphMeasure{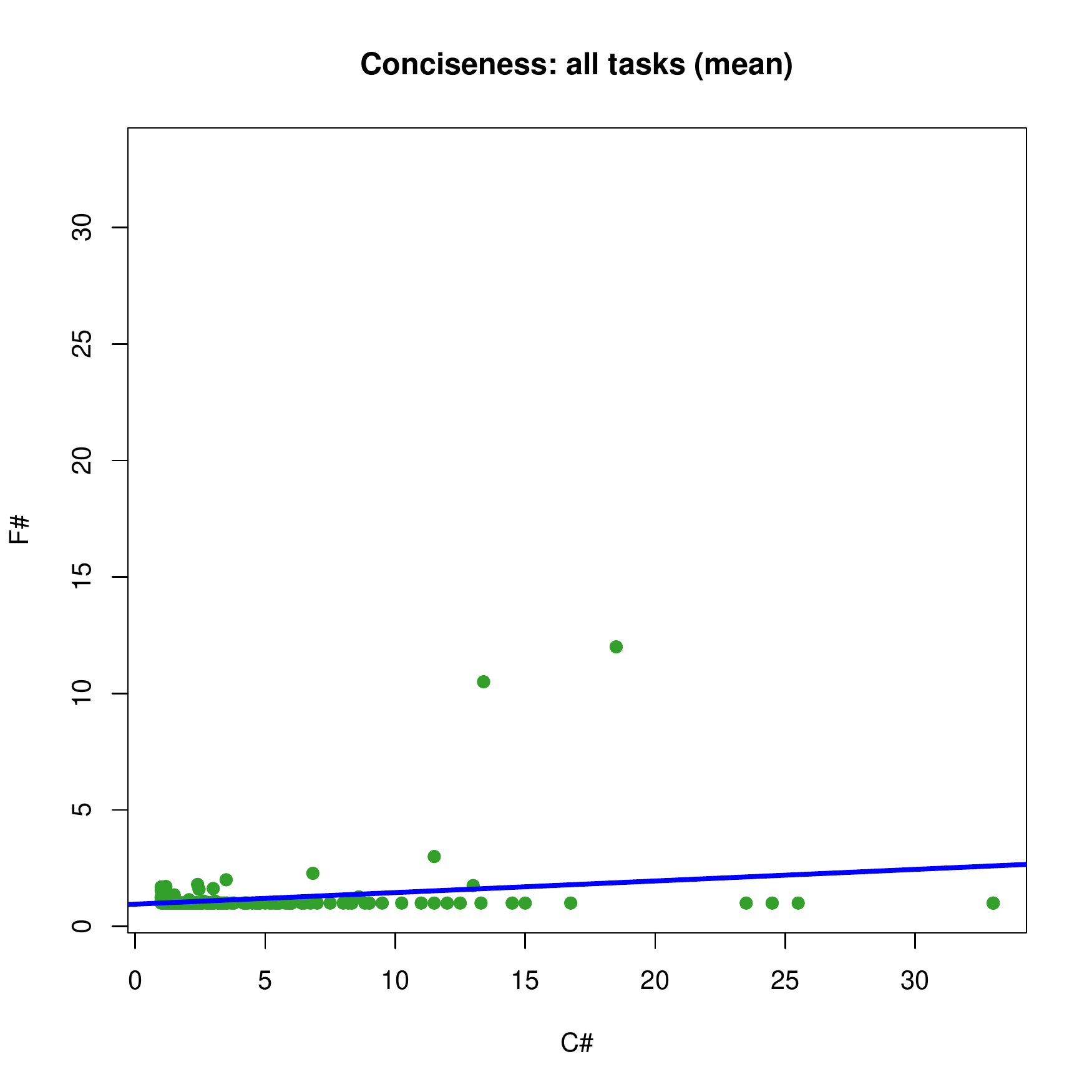}{Lines of code (mean) of all tasks}

\clearpage\newpage
\subsection{Comments per line of code}
\GraphMeasure{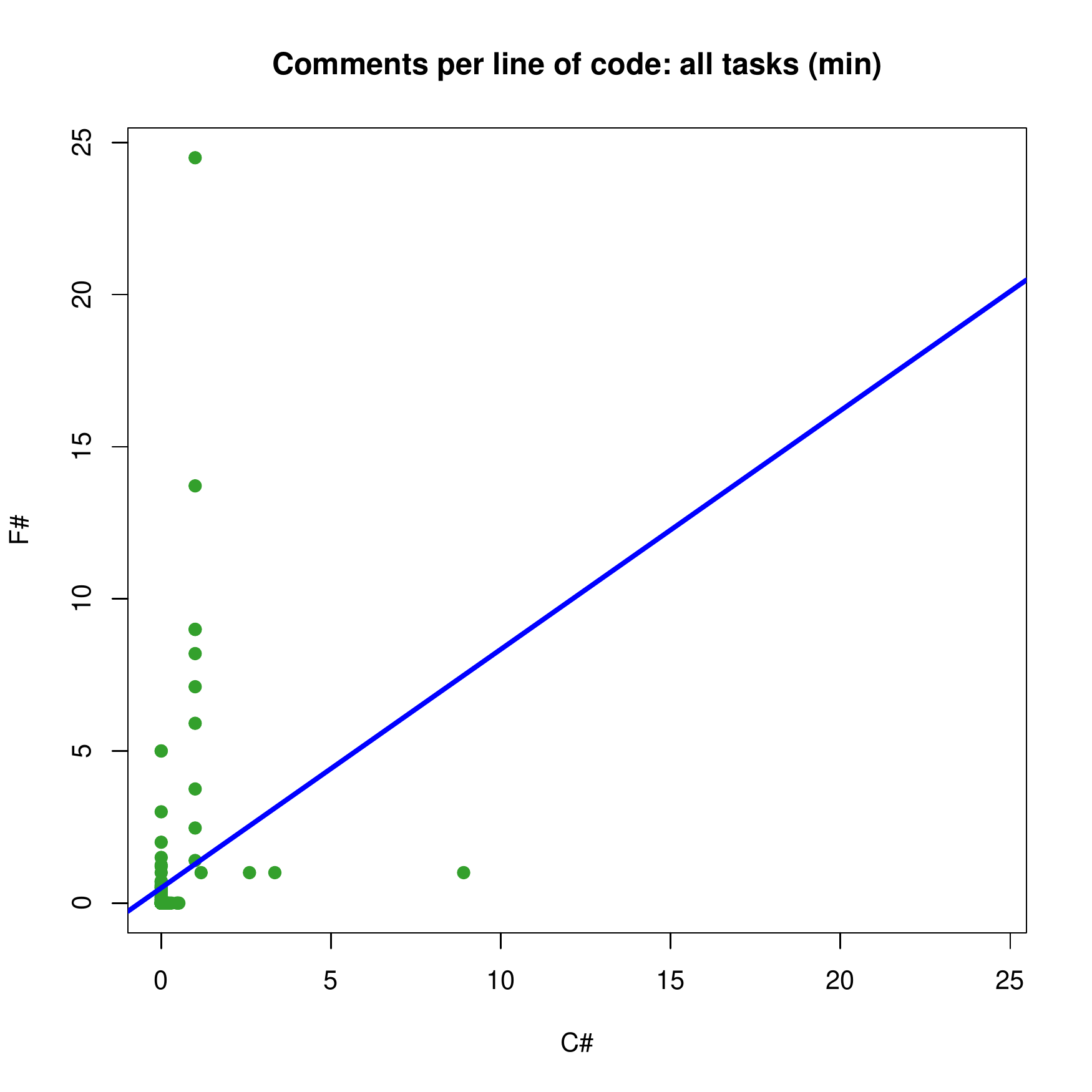}{Comments per line of code (min) of all tasks}
\clearpage\newpage
\GraphMeasure{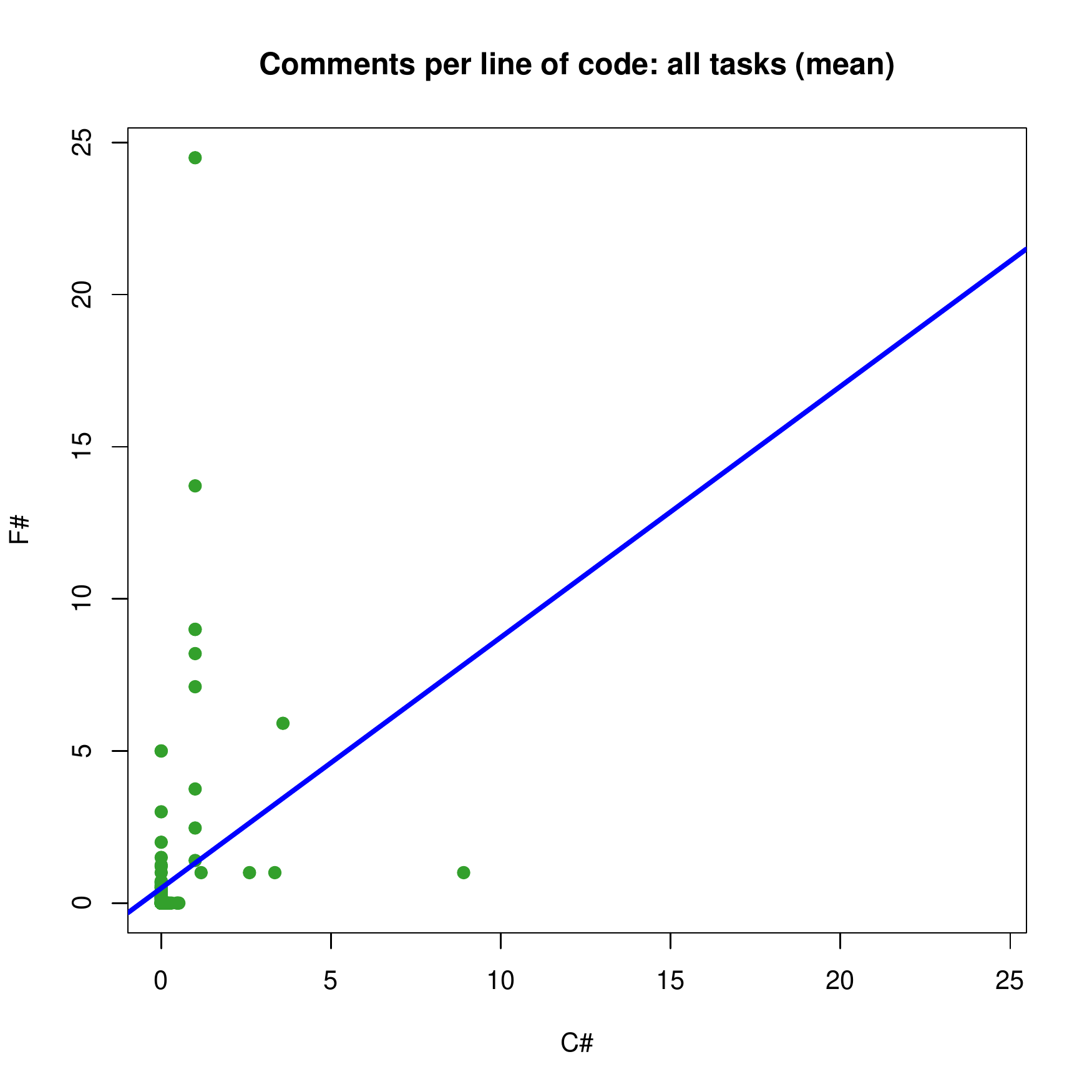}{Comments per line of code (mean) of all tasks}

\clearpage\newpage
\subsection{Size of binaries}
\GraphMeasure{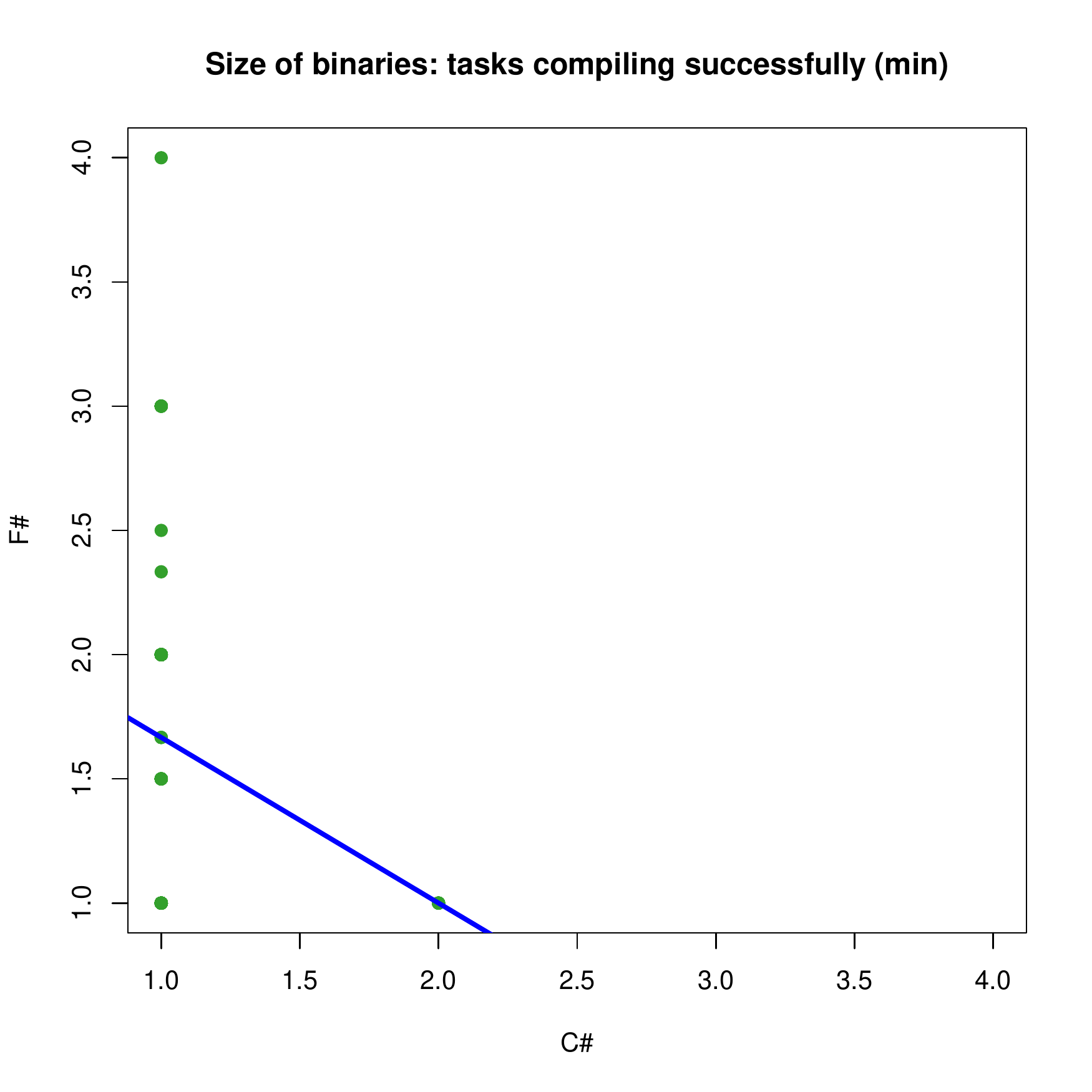}{Size of binaries (min) of tasks compiling successfully}
\clearpage\newpage
\GraphMeasure{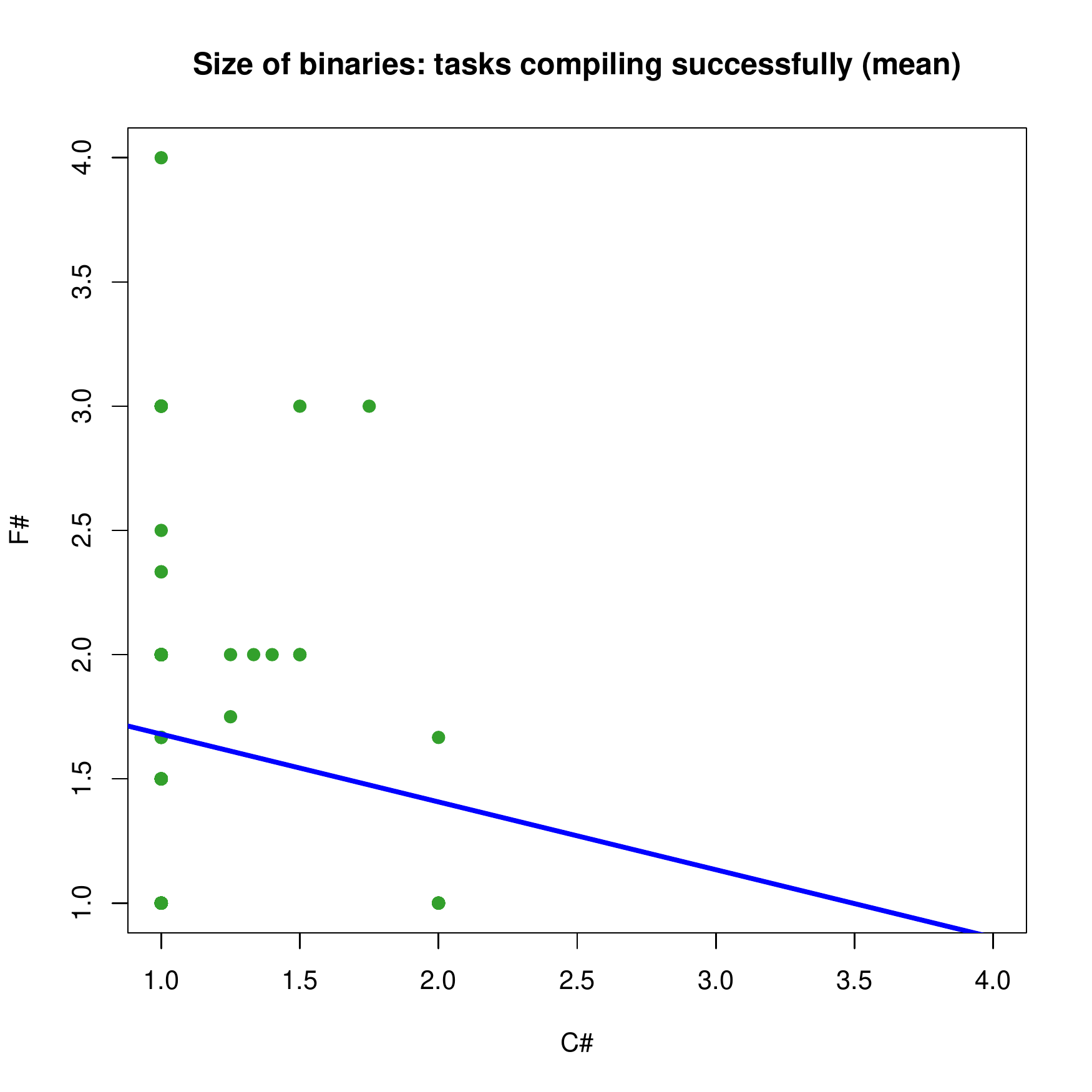}{Size of binaries (mean) of tasks compiling successfully}

\clearpage\newpage
\subsection{Performance}
\GraphMeasure{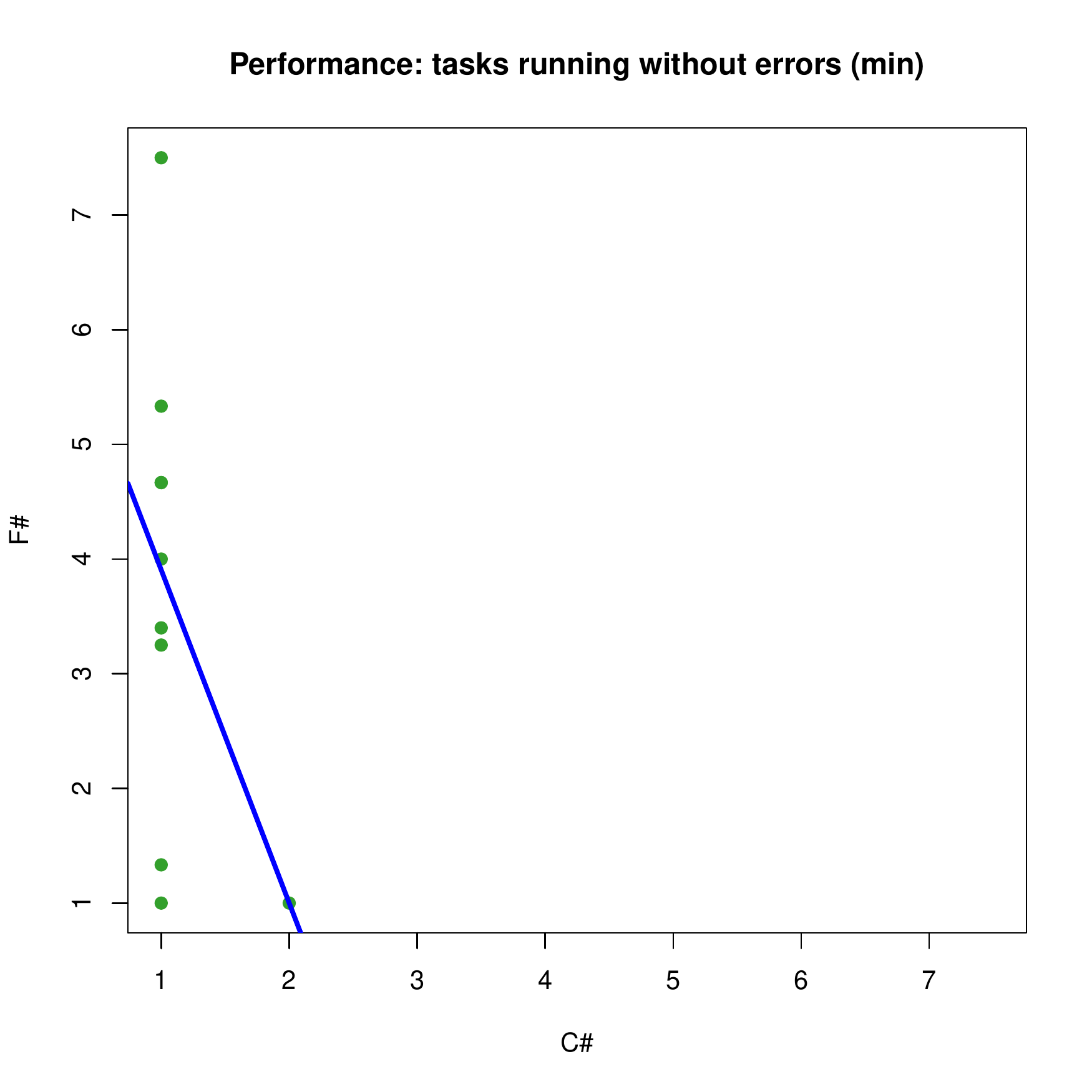}{Performance (min) of tasks running successfully}
\clearpage\newpage
\GraphMeasure{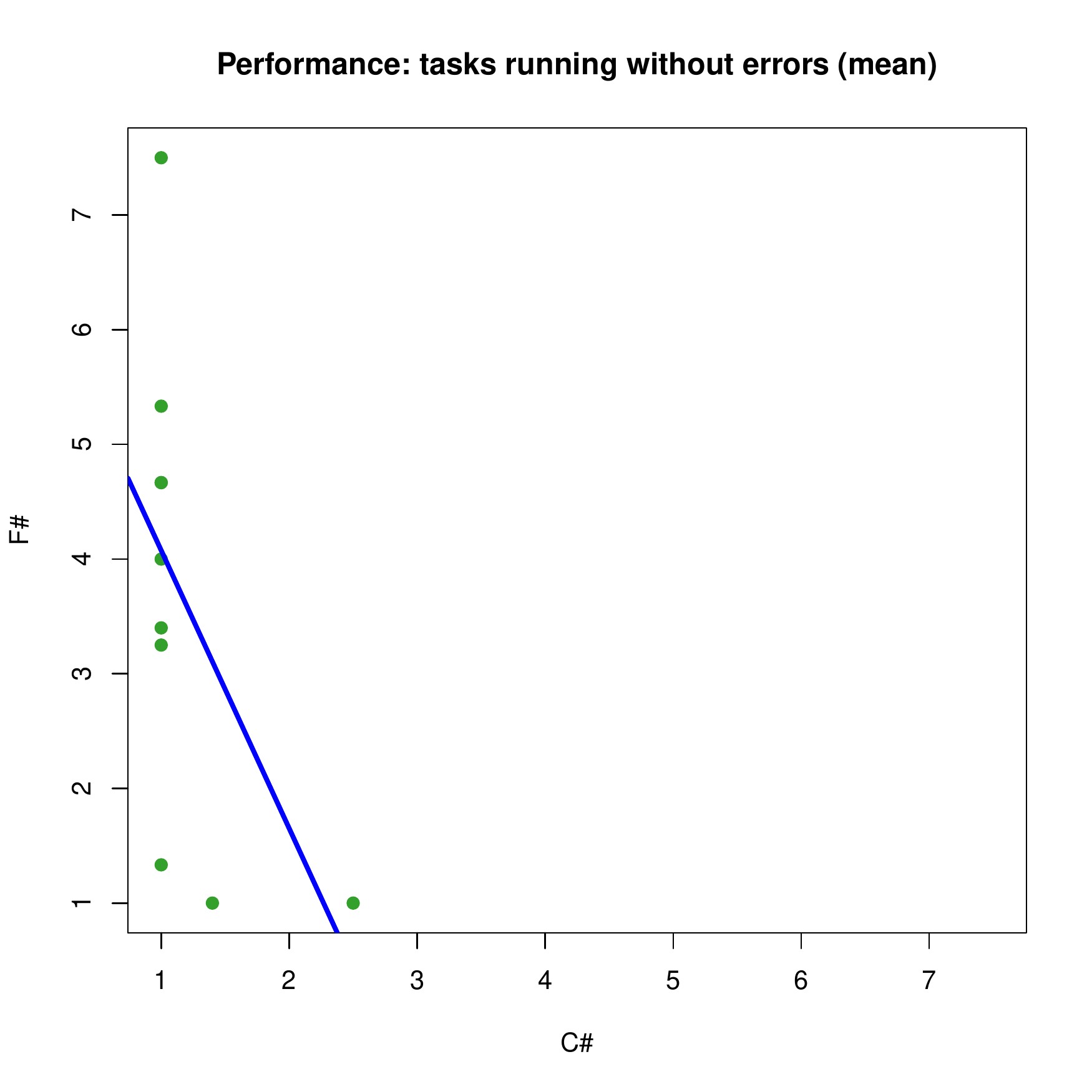}{Performance (mean) of tasks running successfully}

\clearpage\newpage
\subsection{Scalability}
\GraphMeasure{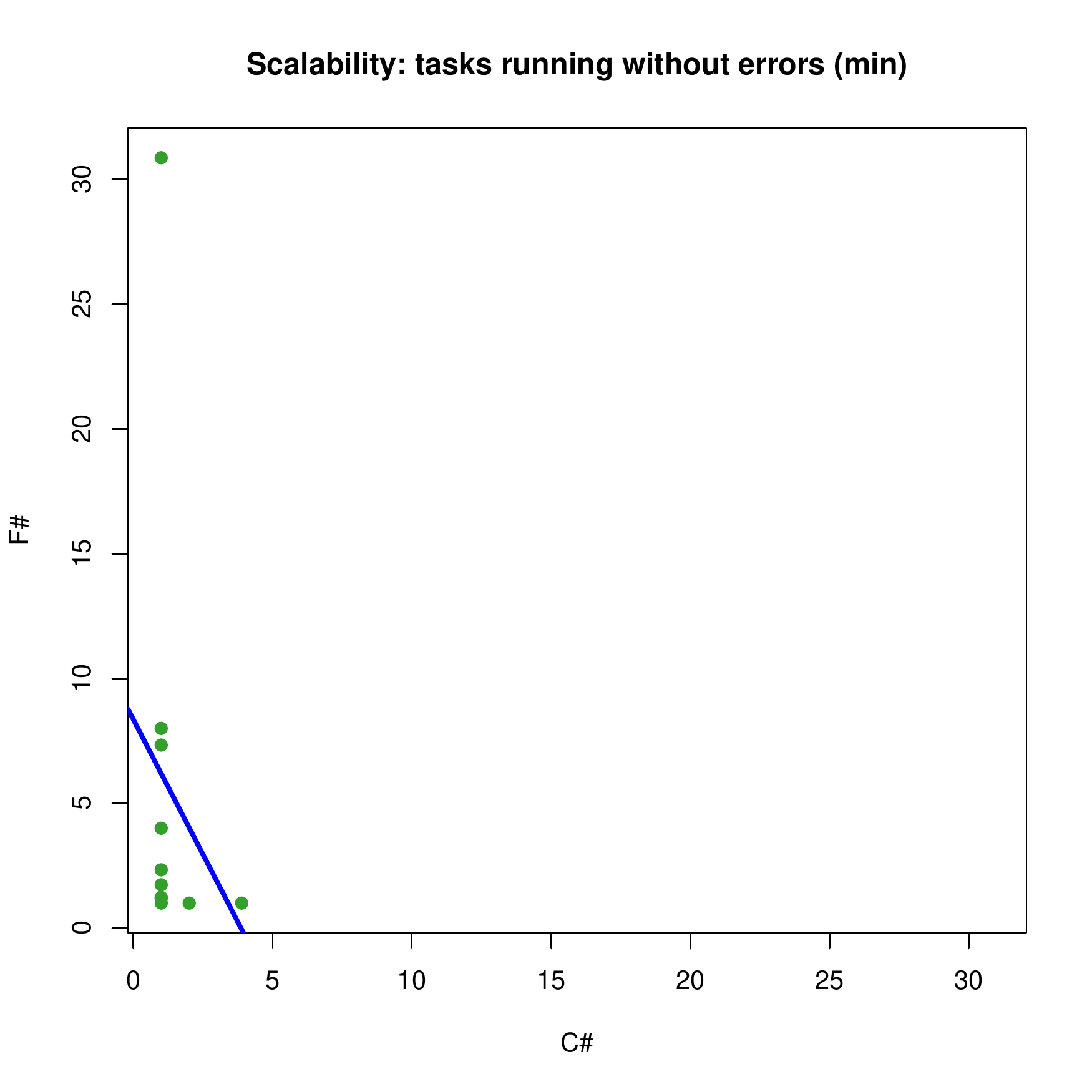}{Scalability (min) of tasks running successfully}
\clearpage\newpage
\GraphMeasure{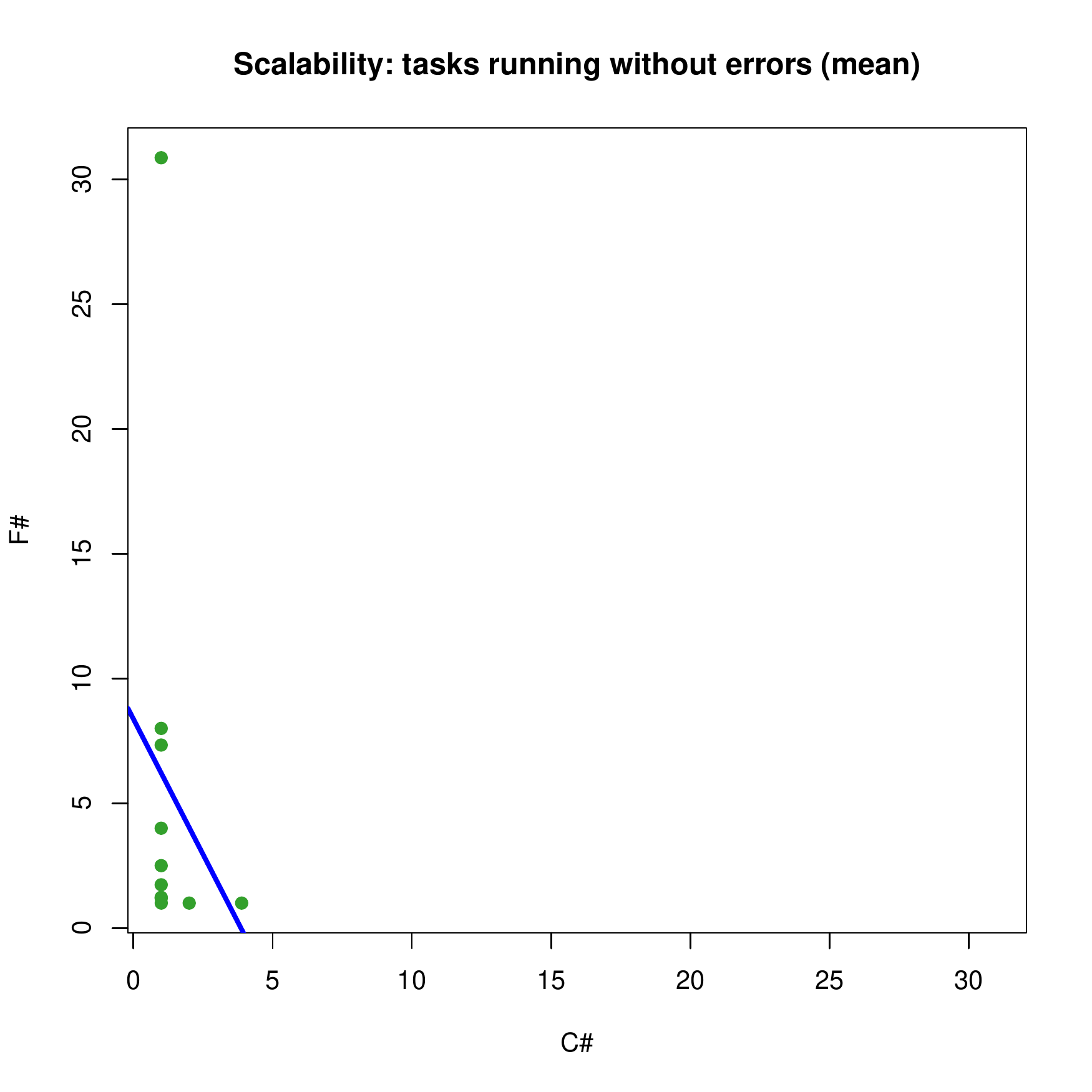}{Scalability (mean) of tasks running successfully}

\clearpage\newpage
\subsection{Maximum RAM}
\GraphMeasure{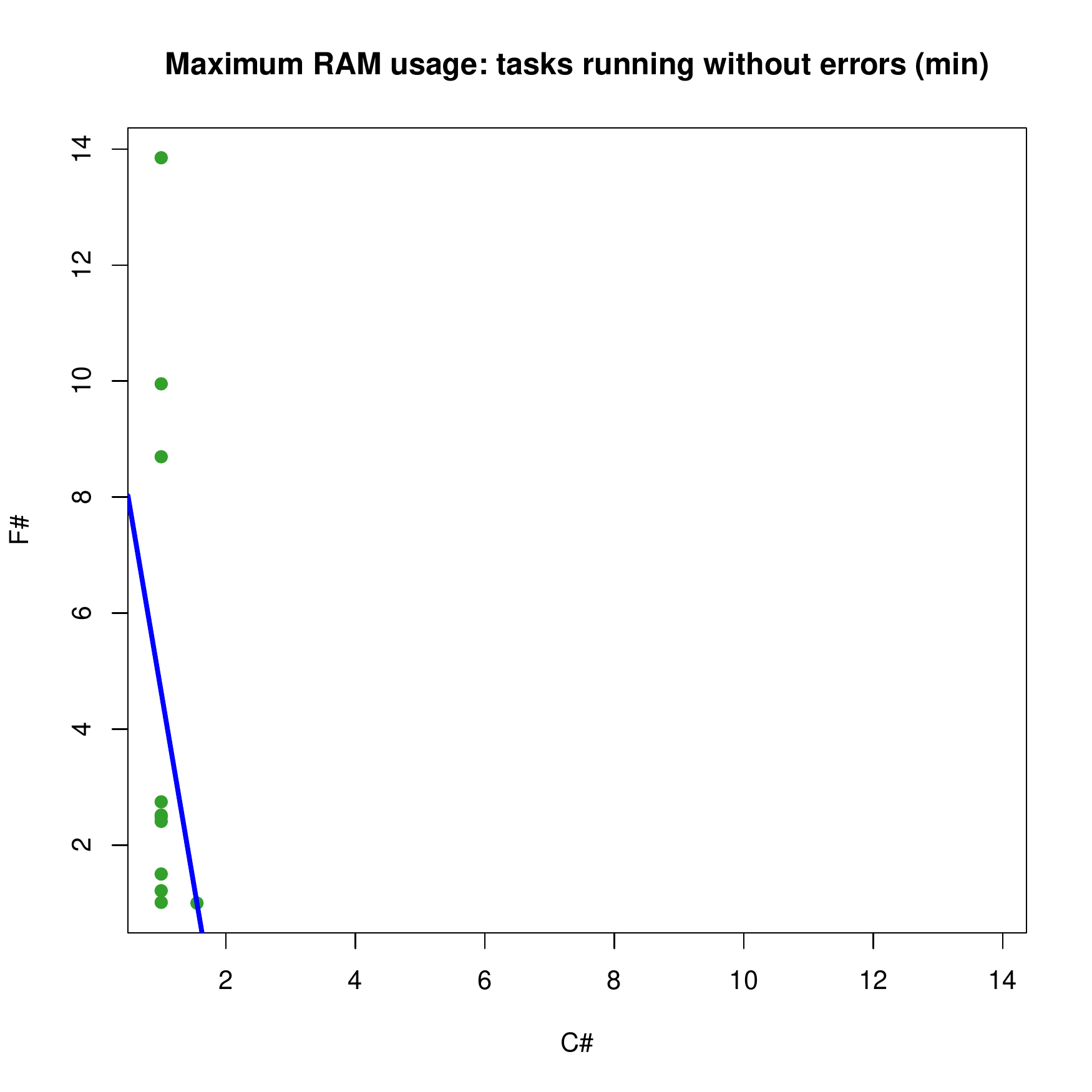}{Maximum RAM usage (min) of tasks running successfully}
\clearpage\newpage
\GraphMeasure{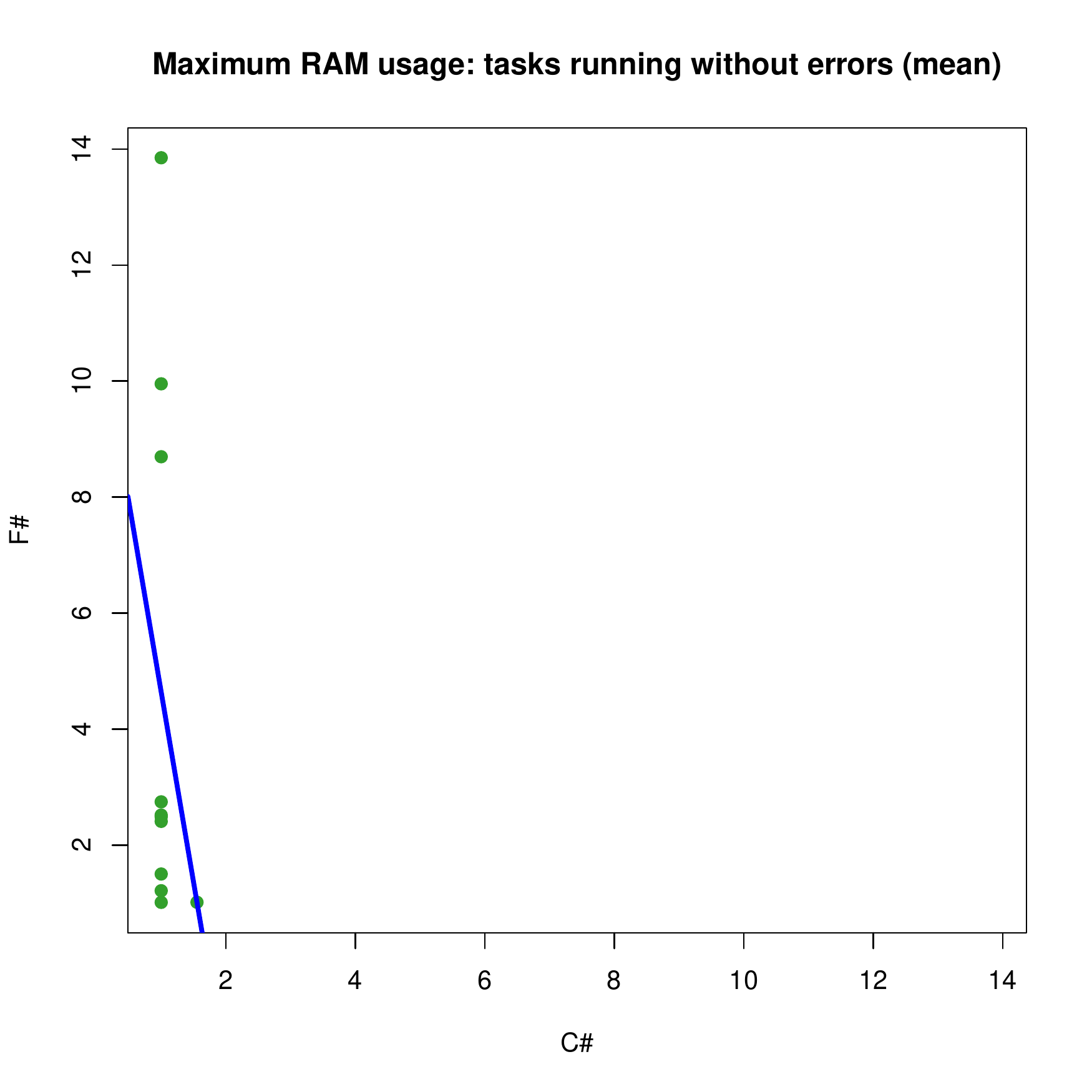}{Maximum RAM usage (mean) of tasks running successfully}

\clearpage\newpage
\subsection{Page faults}
\GraphMeasure{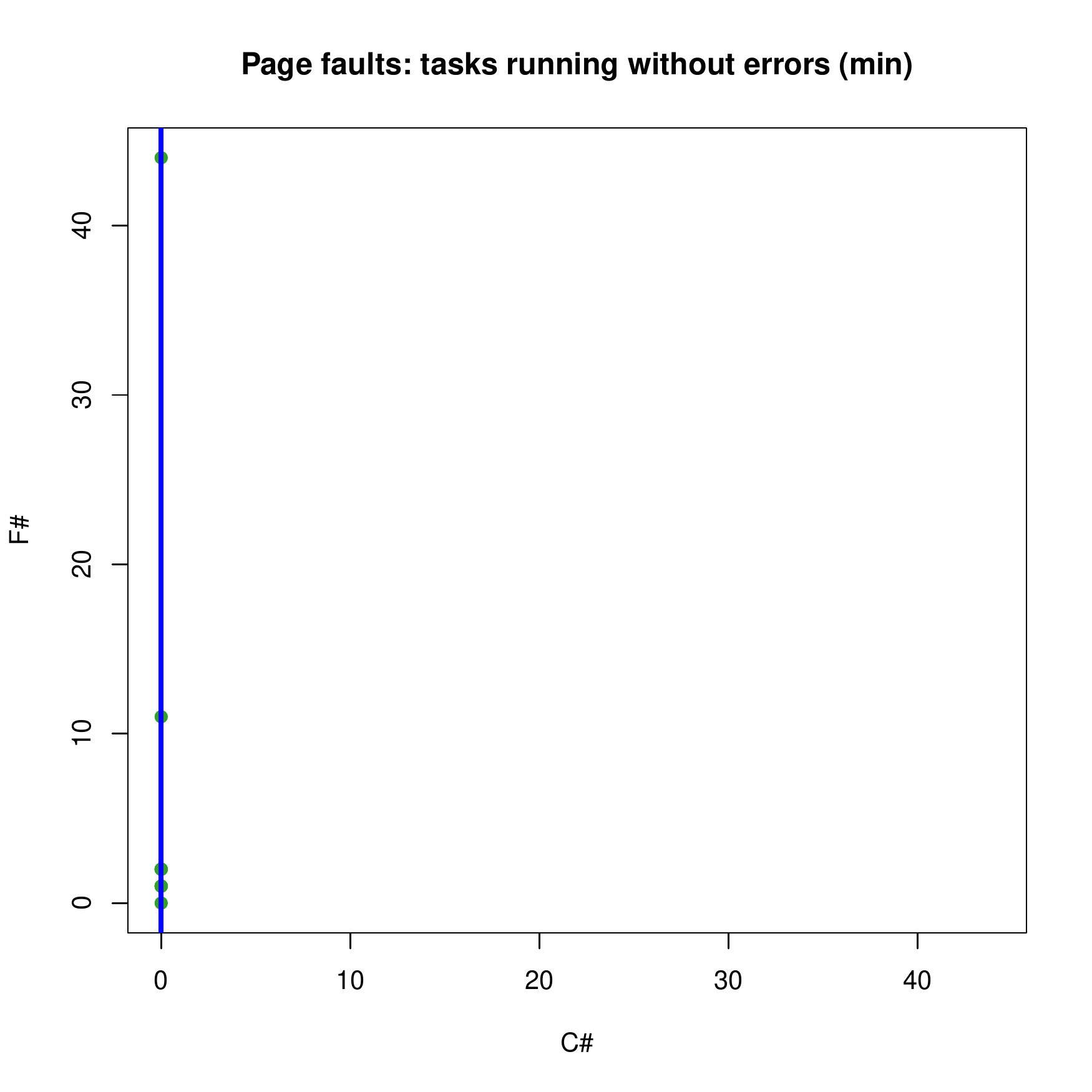}{Page faults (min) of tasks running successfully}
\clearpage\newpage
\GraphMeasure{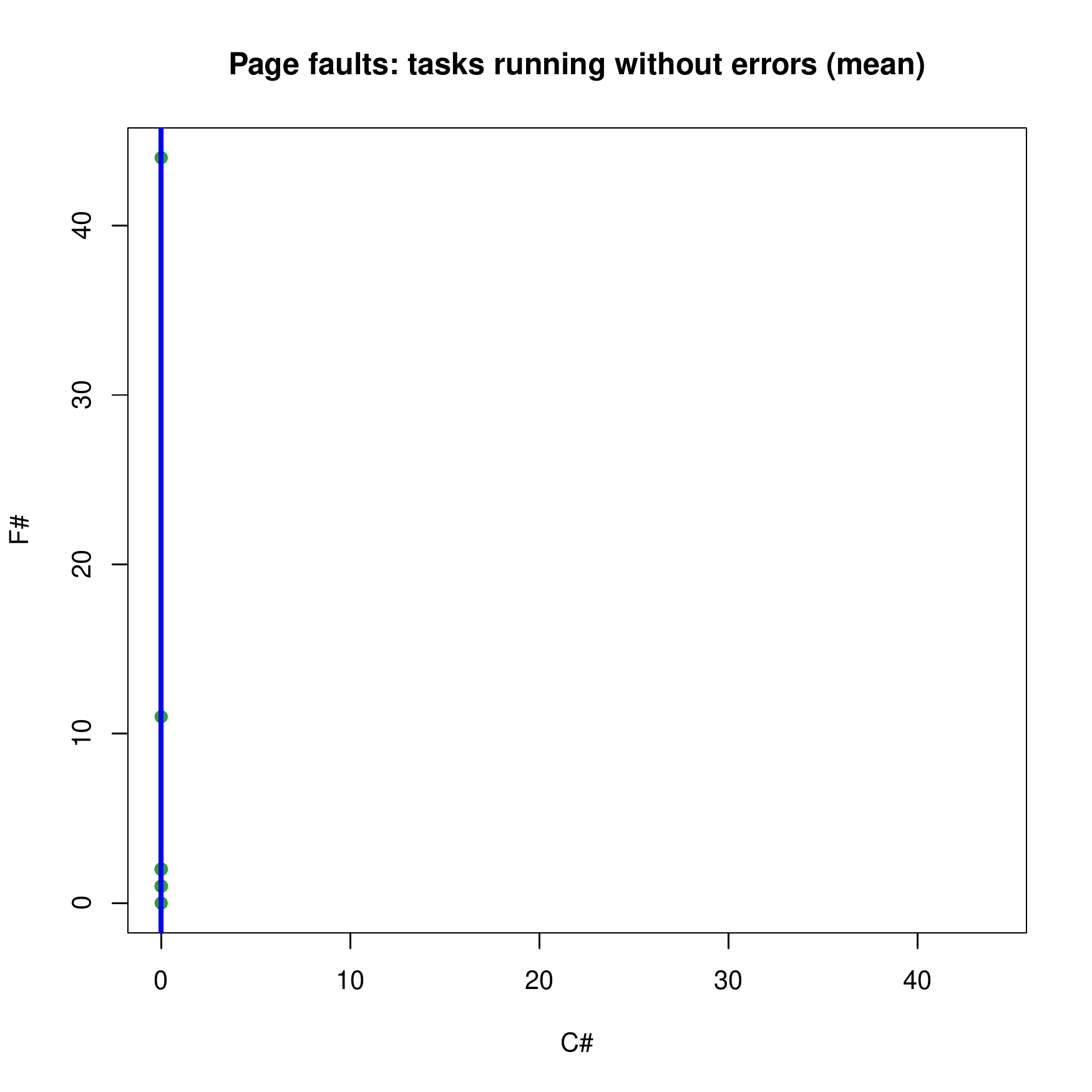}{Page faults (mean) of tasks running successfully}

\clearpage\newpage
\subsection{Timeout analysis}
\GraphMeasure{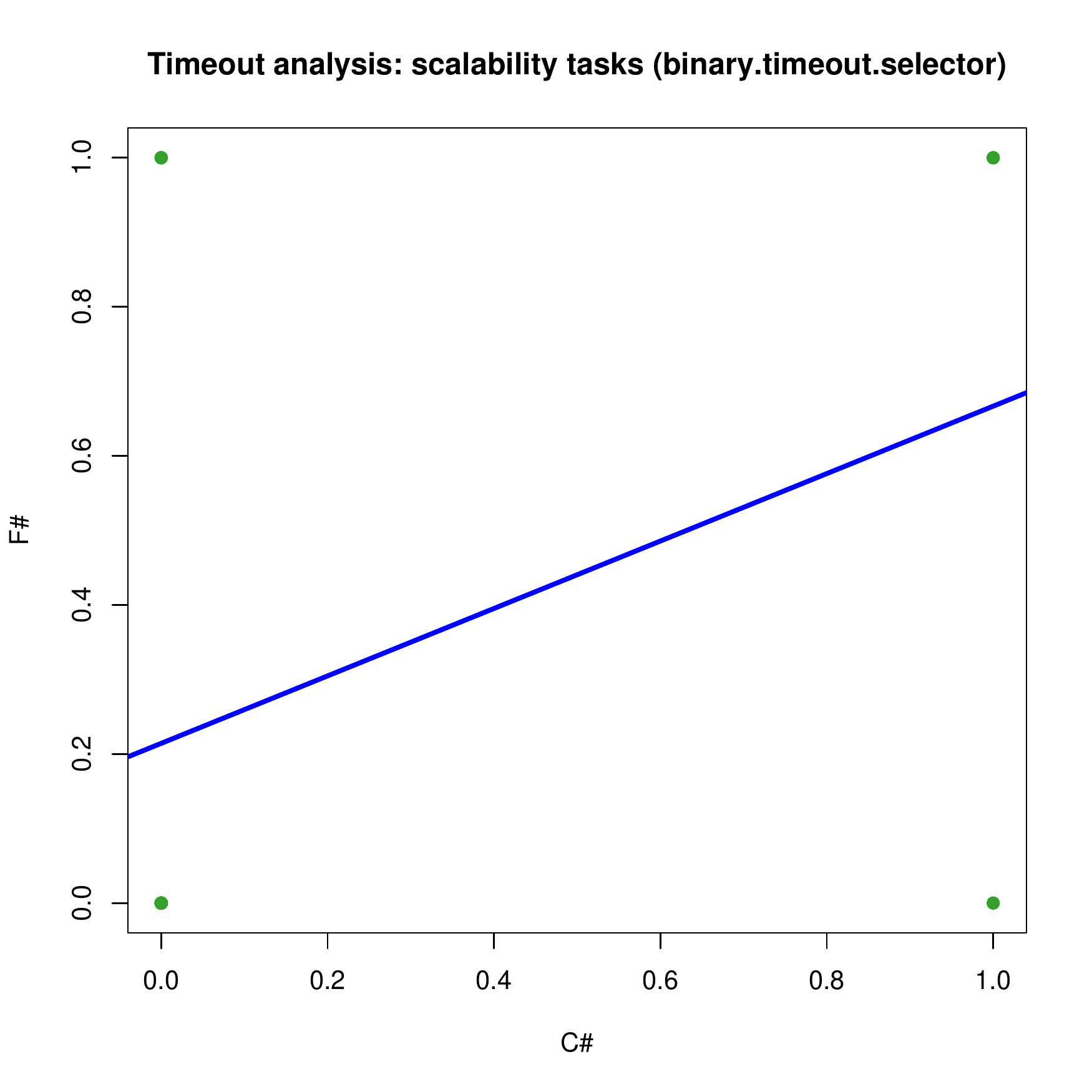}{Timeout analysis of scalability tasks}

\clearpage\newpage
\subsection{Number of solutions}
\GraphMeasure{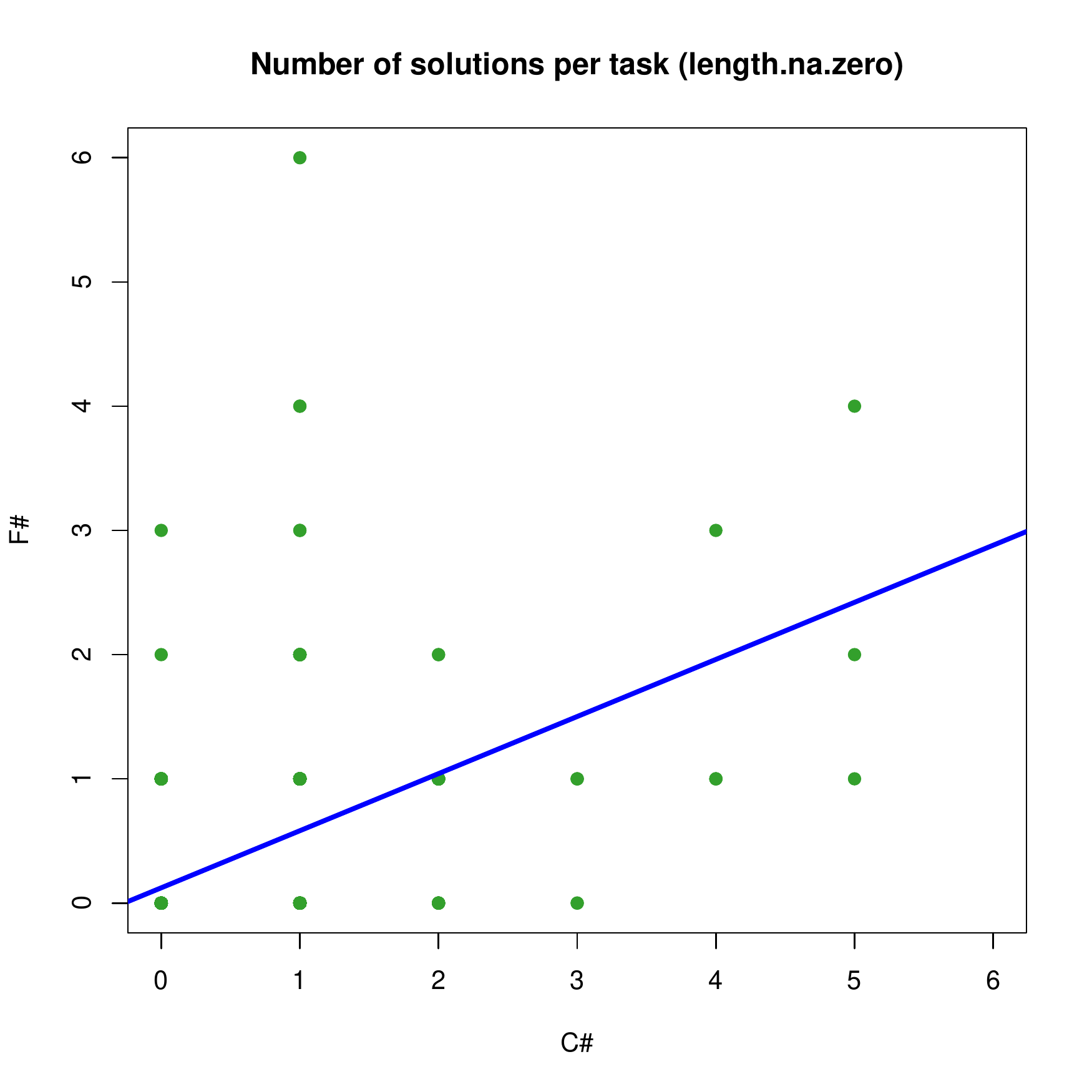}{Number of solutions per task}


\fi

\end{document}